\documentclass[pdftex,twocolumn,epjc3]{svjour3}          

\usepackage[margin=0.9cm,top=10mm,left=15mm]{geometry} 

\RequirePackage[T1]{fontenc}
\smartqed  
\RequirePackage{mathptmx}      
\RequirePackage{caption}
\RequirePackage[numbers,sort&compress]{natbib}
\RequirePackage[colorlinks,citecolor=blue,urlcolor=blue,linkcolor=blue]{hyperref}

\journalname{Eur. Phys. J. C}

\usepackage{pgfplotstable}
\usepackage{longtable}
\usepackage{supertabular}

\usepackage{xspace}
\usepackage{color}\definecolor{darkblue}{rgb}{0,0,0.5}
\usepackage{hyperref}
\hypersetup{
    bookmarks=true,         
    unicode=false,          
    pdftoolbar=true,        
    pdfmenubar=true,        
    pdffitwindow=true,      
    pdftitle={My title},    
    pdfauthor={Author},     
    pdfsubject={Subject},   
    pdfnewwindow=true,      
    pdfkeywords={keywords}, 
    colorlinks=true,       
    linkcolor=darkblue,          
    citecolor=darkblue,        
    filecolor=magenta,      
    urlcolor=darkblue,         
}

\newcommand{\NASixtyOne}{NA61\slash SHINE\xspace}

\usepackage[latin2]{inputenc}
\usepackage{indentfirst}
\usepackage{enumerate}

\usepackage{amsmath}
\usepackage{bm}
\usepackage{amssymb}
\usepackage[english]{babel}
\usepackage{url}
\usepackage{natbib}

\usepackage{multirow}
\usepackage[abs]{overpic}

\usepackage{subcaption} 

\newif\ifpdf

\ifx\pdfoutput\undefined

\pdffalse

\else

\pdfoutput=1

\pdftrue

\fi

\ifpdf

\else

\usepackage{graphicx}

\fi

\usepackage{epstopdf}

\usepackage{stmaryrd}

\usepackage[switch]{lineno}

\journalname{Eur. Phys. J. C}

\begin{document}
\newcommand{\ra}[1]{\renewcommand{\arraystretch}{#1}}
\DeclareGraphicsExtensions{.pdf,.png,.eps,.jpg,.ps}
\cleardoublepage

\pagenumbering{roman}

\title{\large \bf \vspace{-1.3cm} \, \\
  Measurements
  of $\pi^\pm$ differential yields from the surface of the T2K replica target \\
  for incoming 31~GeV/$c$ protons 
  with the NA61/SHINE spectrometer at the CERN SPS
}

\clearpage

\institute{
{National Nuclear Research Center, Baku, Azerbaijan}\label{inst0}
\and{Faculty of Physics, University of Sofia, Sofia, Bulgaria}\label{inst1}
\and{Ru{\dj}er Bo\v{s}kovi\'c Institute, Zagreb, Croatia}\label{inst2}
\and{LPNHE, University of Paris VI and VII, Paris, France}\label{inst3}
\and{Karlsruhe Institute of Technology, Karlsruhe, Germany}\label{inst4}
\and{Fachhochschule Frankfurt, Frankfurt, Germany}\label{inst5}
\and{University of Frankfurt, Frankfurt, Germany}\label{inst6}
\and{University of Athens, Athens, Greece}\label{inst7}
\and{Wigner Research Centre for Physics of the Hungarian Academy of Sciences, Budapest, Hungary}\label{inst8}
\and{Institute for Particle and Nuclear Studies, Tsukuba, Japan}\label{inst9}
\and{University of Bergen, Bergen, Norway}\label{inst10}
\and{Jan Kochanowski University in Kielce, Poland}\label{inst11}
\and{National Centre for Nuclear Research, Warsaw, Poland}\label{inst12}
\and{Jagiellonian University, Cracow, Poland}\label{inst13}
\and{University of Silesia, Katowice, Poland}\label{inst14}
\and{University of Warsaw, Warsaw, Poland}\label{inst15}
\and{University of Wroc{\l}aw,  Wroc{\l}aw, Poland}\label{inst16}
\and{Warsaw University of Technology, Warsaw, Poland}\label{inst17}
\and{Institute for Nuclear Research, Moscow, Russia}\label{inst18}
\and{Joint Institute for Nuclear Research, Dubna, Russia}\label{inst19}
\and{National Research Nuclear University ``MEPhI'' (Moscow Engineering Physics Institute), Moscow, Russia}\label{inst20}
\and{St. Petersburg State University, St. Petersburg, Russia}\label{inst21}
\and{University of Belgrade, Belgrade, Serbia}\label{inst22}
\and{ETH Z\"urich, Z\"urich, Switzerland}\label{inst23}
\and{University of Bern, Bern, Switzerland}\label{inst24}
\and{University of Geneva, Geneva, Switzerland}\label{inst25}
\and{Los Alamos National Laboratory, Los Alamos, USA}\label{inst27}
\and{University of Colorado, Boulder, USA}\label{inst28}
\and{University of Pittsburgh, Pittsburgh, USA}\label{inst29}
\and{IPNL, University of Lyon, Villeurbanne, France}\label{Lyon}
\and{Kavli Institute for the Physics and Mathematics of the Universe (WPI), The  University of Tokyo Institutes for Advanced Study, University of Tokyo, Kashiwa, Chiba, Japan}\label{IPMU}
\and{TRIUMF, Vancouver, British Columbia, Canada}\label{TRIUMF}
\and{Kyoto University, Department of Physics, Kyoto, Japan}\label{Kyoto}
\and{Louisiana State University, Department of Physics and Astronomy, Baton Rouge, Louisiana, USA}\label{Louisiana}
\and{York University, Department of Physics and Astronomy, Toronto, Ontario, Canada}\label{York}
}

\thankstext{e1}{present address: Department of Physics, COMSATS Institute of Information Technology, Islamabad 44000 Pakistan}

\author{
{N.~Abgrall}\thanksref{inst25}
\and{A.~Aduszkiewicz}\thanksref{inst15}
\and{M.~Ajaz}\thanksref{inst25}
\and{Y.~Ali}\thanksref{inst13,e1}
\and{E.~Andronov}\thanksref{inst21}
\and{T.~Anti\'ci\'c}\thanksref{inst2}
\and{N.~Antoniou}\thanksref{inst7}
\and{B.~Baatar}\thanksref{inst19}
\and{F.~Bay}\thanksref{inst23}
\and{A.~Blondel}\thanksref{inst25}
% ...
}


 \date{\today}

 \maketitle
 
\begin{abstract}
  Measurements of particle emission from a replica of the T2K 
  90~cm-long carbon target
  were performed in the \NASixtyOne experiment at CERN SPS,
  using data collected during a high-statistics run in 2009. 
  An efficient use of the long-target measurements for neutrino flux
  predictions in T2K requires dedicated reconstruction and analysis techniques.
  Fully-corrected differential yields of $\pi^\pm$-mesons from the surface of 
  the T2K replica target for incoming 31~GeV/$c$ protons are presented. 
  A possible strategy to implement these results into the T2K neutrino 
  beam predictions is discussed and the propagation of the uncertainties 
  of these results to the final neutrino flux is performed.
\end{abstract}


 \tableofcontents

 \pagenumbering{arabic}


\section{Introduction} \label{sec:Introduction}

The \NASixtyOne (SPS Heavy Ion and Neutrino Experiment) 
experiment~\cite{NA61detector_paper} at
the CERN Super Proton Synchrotron (SPS) is pursuing a rich physics program in
various fields.
Precise hadron production measurements are performed for the T2K long-baseline neutrino 
experiment~\cite{T2K,T2Knueapp_obs,T2Knumudisap_precise,T2Kcombined}
and for more reliable simulations of cosmic-ray air showers for the Pierre
Auger and KASCADE experiments~\cite{Auger, KASCADE}. 
The properties of the onset of deconfinement are studied with measurements 
of p+p~\cite{NA61_2009_pp_EPJC},
p+Pb and nucleus+nucleus collisions at the SPS energies~\cite{Status_Report_2014,Status_Report_2015}.

The on-going T2K 
experiment~\cite{T2K,T2Knueapp_obs,T2Knumudisap_precise,T2Kcombined} 
requires precise predictions of the expected neutrino fluxes 
at both near and far detectors.  

In T2K, a high-intensity
neutrino beam is produced at \mbox{J-PARC} by a 30~GeV (kinetic energy) proton beam impinging
on a 90~cm long graphite target (1.9 nuclear interaction length, $\lambda_{I}$). 
Positively or negatively
charged hadrons exiting the target (mainly $\pi$ and $K$ mesons) are
focused by a set of three magnetic horns and decay along a 96~m long
decay tunnel. The flavour content and energy spectrum of the neutrino interactions are
measured at the near detector complex located 280~m away from the
target station, and by the Super-Kamiokande (SK) detector at a
distance of 295~km.

Accelerator-based neutrino beams provide well defined and controlled sources of
neutrinos. However, intrinsic uncertainties on the fluxes predicted with Monte
Carlo (MC) simulations arise from the models employed to simulate hadron
emission from long nuclear targets used in accelerator-based
experiments. In these types of experiments, a non-negligible fraction
of the neutrino flux actually originates from particles which are produced
in hadronic re-interactions in the long target.  Up to now, neutrino
flux predictions have been constrained by using either
parametrizations based on existing hadron production data available in
the literature, 
or dedicated hadron
production measurements performed on thin nuclear targets.

This approach was also followed
by the T2K experiment for (anti)neutrino flux predictions~\cite{T2Kflux} 
with the \NASixtyOne
measurements~\cite{pion_paper,kaon_paper,V0_2007,thin2009paper} 
performed using a thin (0.04~$\lambda_{I}$) graphite target
and a 31~GeV/$c$ proton beam.
For the first time, the kinematical
phase space 
of pions and kaons exiting the target
and producing neutrinos in the direction of the near and far
detectors was fully covered by a single hadron production
experiment.

Such hadron production measurements performed with thin targets 
provide constraints on the production of secondary
particles in the primary interaction of beam protons in the
target. 
However, the lack of direct measurements of the production of tertiary particles
in re-interactions, and hence the use of sparse data sets to cover these
contributions, limits the achievable precision of the flux
prediction. The main motivation for measurements of hadron
emission from a replica of the T2K target is therefore to reduce the systematic
uncertainties on the prediction of the initial neutrino flux
originating from products of interactions in the target.

Thus, in addition to the published charged pion and kaon measurements  in p+C 
interactions at 31~GeV/$c$ on a thin target~\cite{pion_paper,kaon_paper,V0_2007},
already used for the T2K neutrino flux 
predictions~\cite{T2Kflux}, the
\NASixtyOne collaboration also performed studies of hadron 
emission
in interactions of a 31~GeV/$c$ proton beam with a full-size replica of the T2K
target using data taken in 2007~\cite{LTpaper}.
New dedicated reconstruction and analysis techniques were developed and  
described.
Uncorrected differential yields of positively charged pions at the surface
of the replica target and their ratios with respect to the predictions
of the model used to simulate hadronic interactions in the T2K target 
were published~\cite{LTpaper}.

In this article we present new measurements of fully-corrected 
differential yields 
of $\pi^\pm$-mesons from the surface of 
the T2K replica target for incoming 31~GeV/$c$ protons, 
denoted as p+(T2K RT)@31~GeV/$c$.
The results were obtained using data collected during 
a high-statistics run performed in 2009.

The paper is organized as follows:
Section 2 describes the T2K neutrino beam and discusses different sources of neutrinos.
Section 3 is devoted to the \NASixtyOne experimental setup.
Section~4 presents the analysis techniques used for the T2K replica target measurements, followed by Section 5 which describes the corresponding systematic uncertainties.
Section 6 shows and discusses the results on the fully-corrected 
differential yields of $\pi^\pm$-mesons from the surface of the T2K replica target.
A possible implementation of these results in the T2K neutrino flux predictions and the expected constraints on these predictions with the T2K replica target measurements are explained in Section 7.

\section{T2K neutrino flux predictions and hadron production measurements} \label{sec:T2KAndHadronProduction}

\begin{figure}[htb]
    \centering
    \includegraphics[width=0.45\textwidth]{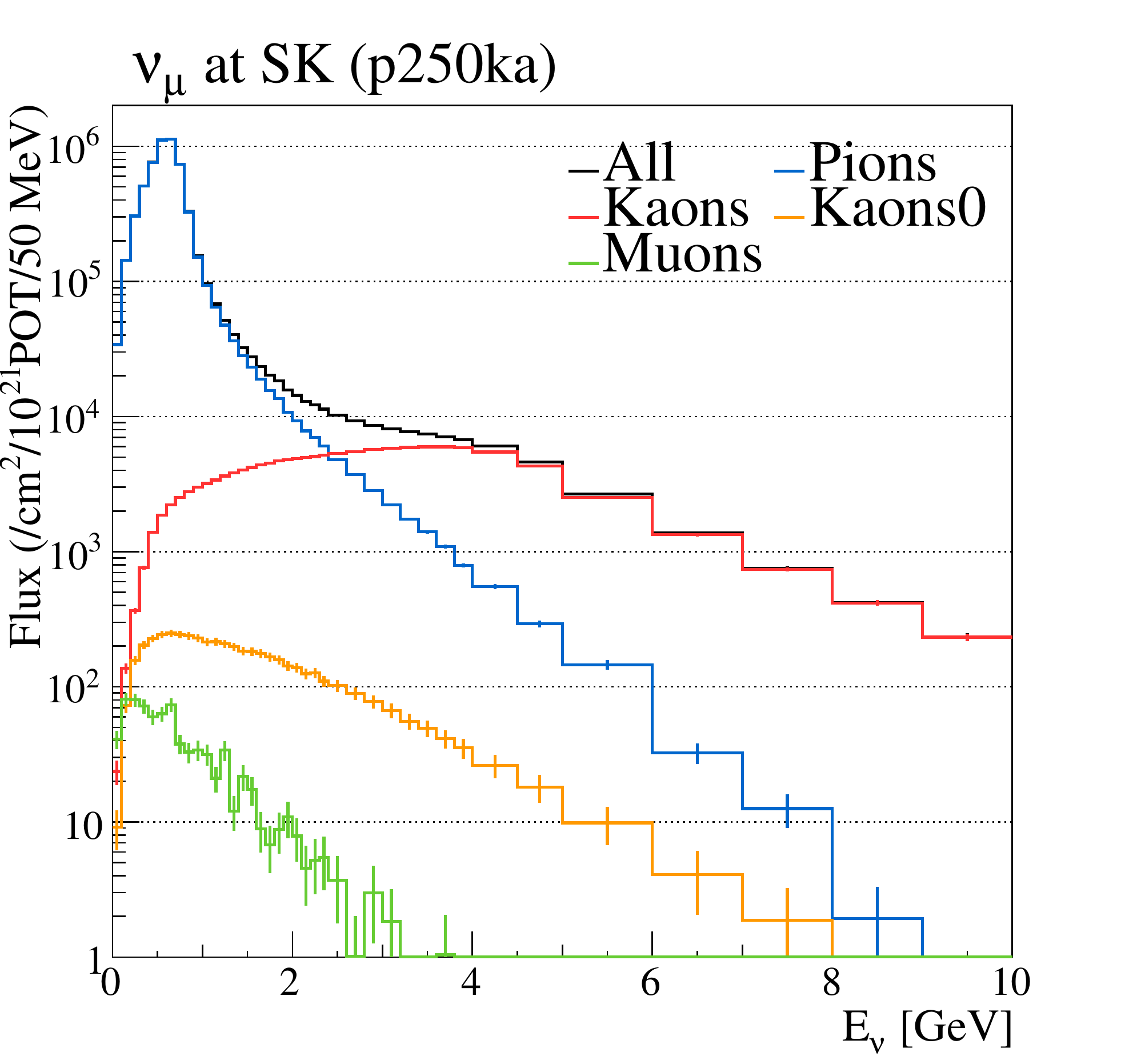}
    \includegraphics[width=0.45\textwidth]{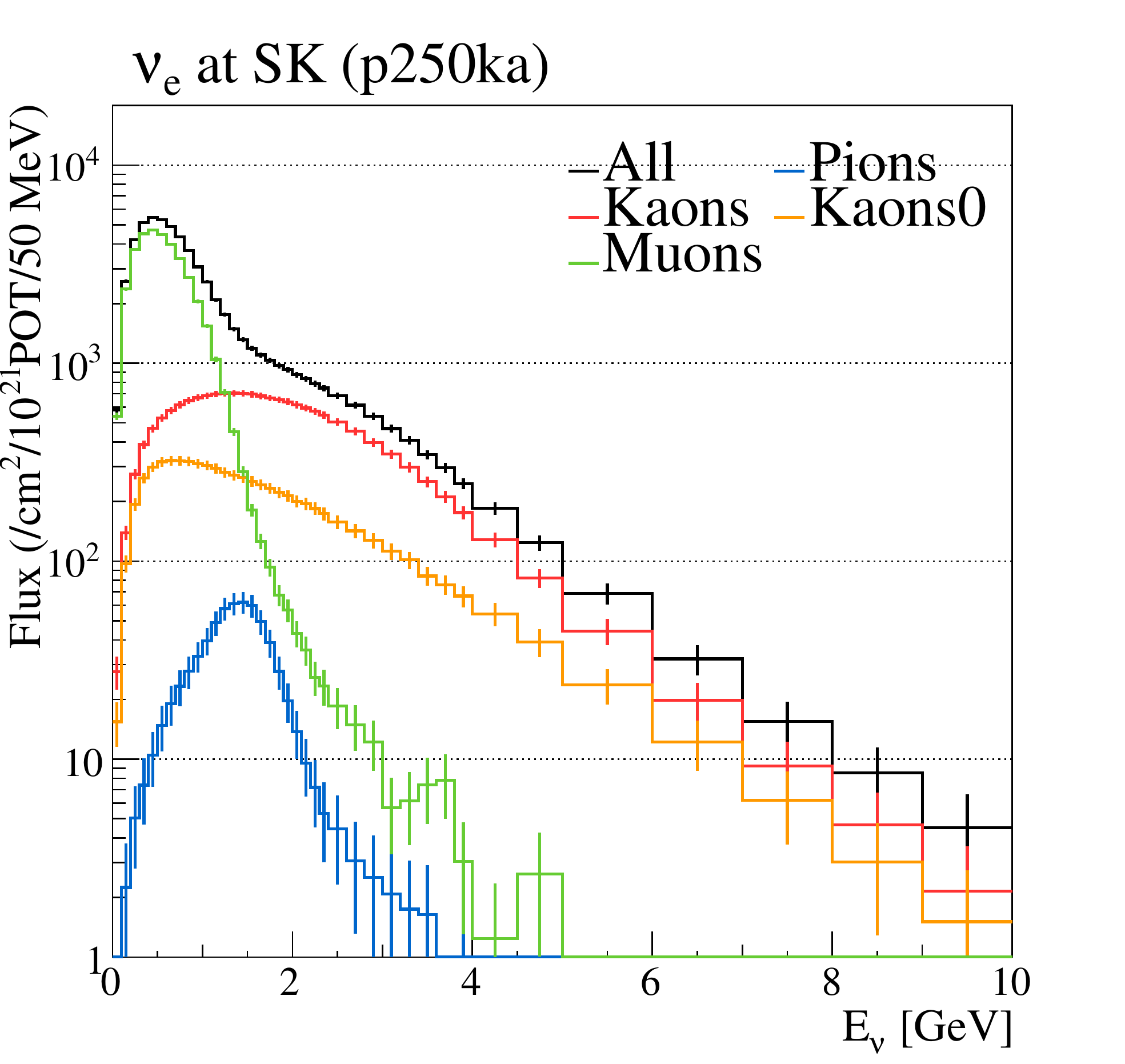}
    \caption{Contribution of different parent particles to the total neutrino flux at SK for $\nu_{\mu}$ ({\it{top}}) and $\nu_{e}$ ({\it{bottom}}), computed with the T2K beam Monte-Carlo simulation program~\cite{T2Kflux} for the positive focussing at 250~kA horn current ('p250ka' configuration).}
    \label{fig:NeutrinoParents}
\end{figure}

\subsection{T2K neutrino beams} \label{subsec:DescriptionNuBeam}
The T2K neutrino beam is generated at the J-PARC complex by 30~GeV protons impinging on a target which is a 90~cm long graphite rod.
The primary proton beam is monitored by a set of detectors which allows to precisely measure the characteristics of the beam.
The produced hadrons are focused by magnetic horns.
By choosing the polarity of the horn currents, it is possible to create either an enhanced neutrino beam or an enhanced antineutrino beam.
In this article we concentrate on the case of the enhanced neutrino beam
but the results of this paper can also be used for the prediction of 
the flux in the enhanced anti-neutrino configuration.
A detailed description of the beam and its properties 
can be found in Ref.~\cite{T2Kflux}.

The neutrino beam predictions are based on a detailed Monte-Carlo simulation.
The input parameters are given by the values describing the ellipsoid representing the primary proton beam impact points
on the target upstream face as measured by the beam position detectors placed along the proton beam line.
The FLUKA2011 \mbox{\cite{Fluka,Fluka_CERN,Fluka_new}} model is used to simulate the interactions of beam protons with the long graphite target.
The propagation of the particles emerging from the surface of the target is modeled by a GEANT3~\cite{Brun:1987ma} simulation using GCALOR~\cite{GCALOR} as hadronic model for re-interactions in the detector.

During the MC simulation,
information on particle production and decay is stored,
so the full history of neutrinos crossing either the near or far detector is available.
This allows to study different components of the neutrino beam and the origin of the neutrino species.
As shown in Fig.~\ref{fig:NeutrinoParents}, the $\nu_{\mu}$ flux around the beam peak energy at the SK far detector arises mainly from pion decays, while it is mainly due to kaons at higher energies.
This motivates the extraction of charged pion yields at the surface of the target, which is the subject of this paper.

\begin{figure}
    \centering
    \includegraphics[width=0.5\textwidth]{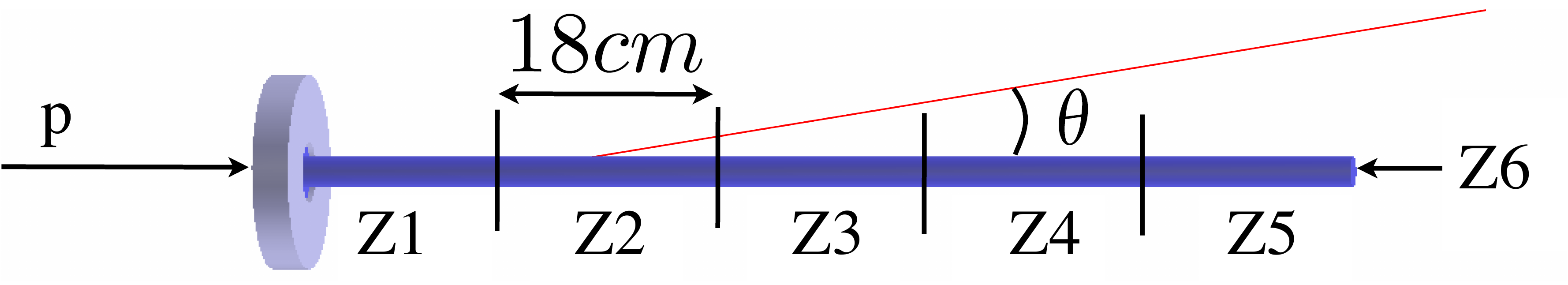}
    \caption{A sketch of the longitudinal binning of the T2K replica target.
      The aluminum flange at the upstream edge is used in \NASixtyOne to hold and align the target.
    }
    \label{fig:TargetZBinning}
\end{figure}

\begin{figure}
    \centering
    \includegraphics[width=0.5\textwidth]{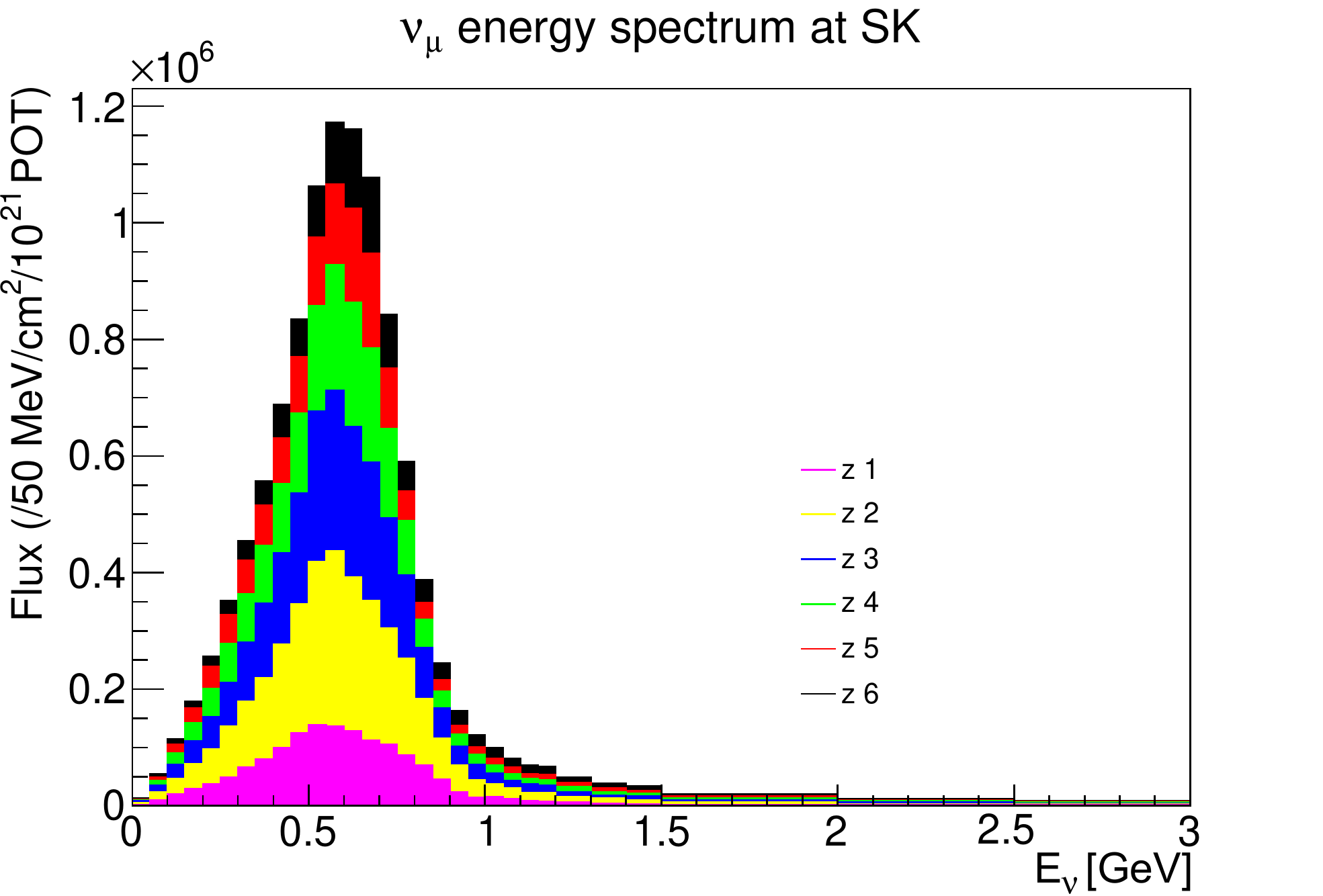}
    \caption{Stacked histograms showing the contribution of each of the 6 longitudinal target bins (see Fig.~\ref{fig:TargetZBinning}) to the muon neutrino flux at SK.}
    \label{fig:StackedNuFluxZBins}
\end{figure}

It is important to note that not only the pion angular and momentum spectra are of interest, but also the longitudinal position where they exit the target.
By dividing the 90~cm long graphite rod into 5 bins of 18~cm length each and considering the downstream face of the target as an additional sixth bin, as shown in Fig.~\ref{fig:TargetZBinning}, it is possible to study the contribution of each of these bins to the total neutrino flux.
Figure~\ref{fig:StackedNuFluxZBins} presents these different contributions as predicted at SK.

\subsection{Requirements on the T2K neutrino flux prediction} \label{subsec:Requirements}

The T2K experiment 
pursues three main physics goals~\cite{T2K} 
with an off-axis (essentially narrow band) neutrino or antineutrino beam peaked around the so-called atmospheric oscillation maximum (energy range from 0.2 to 1.2 GeV). These are:
\begin{enumerate}[(i)]
    \item the muon neutrino disappearance,
    \item the electron neutrino appearance ($\nu_\mu \to \nu_e$),
    \item neutrino cross section measurements. 
\end{enumerate}
The muon-neutrino flux in the region of interest for the oscillation analysis is mainly generated by pion decays.
For the oscillation measurements,
the ratio of the flux of neutrinos at the near detector to the one at the far detector is the most important quantity and a desirable level of uncertainty is about 1-2\%.
Another quantity of interest for the electron neutrino appearance is the ratio between electron and muon neutrino cross sections, whose measurement in the near detector will require a knowledge of the electron to muon neutrino  fluxes to better than about 2\%.
Failing to match this required precision might limit the precision of the results for the full expected T2K exposure. 
Existing data on (anti)neutrino cross sections in the energy range of interest are very limited,
the precision of measurements ranging typically between 10\% and 20\%.
A precision on the T2K neutrino flux with a 5\% absolute normalization error would allow considerable improvement in the understanding of low energy neutrino interactions.

\subsection{T2K flux predictions and the T2K replica target measurements in NA61/SHINE} \label{subsec:CoverageT2KbyNA61}

As already described in Ref.~\cite{LTpaper}, the neutrino fluxes can be split into secondary and tertiary components.
The secondary component originates from neutrino parents produced in the primary interaction of the beam protons in the target.
The tertiary component refers to neutrino parents produced in interactions of secondary particles.
The latter component is due to re-interactions in the target and re-interactions taking place in the elements of the beamline.
Secondary and tertiary interactions occurring in the target are constrained by the measurements of identified hadron spectra from the surface of the T2K replica target.

\begin{figure}[ht]
    \centering
    \centering
    \includegraphics[width=0.5\textwidth]{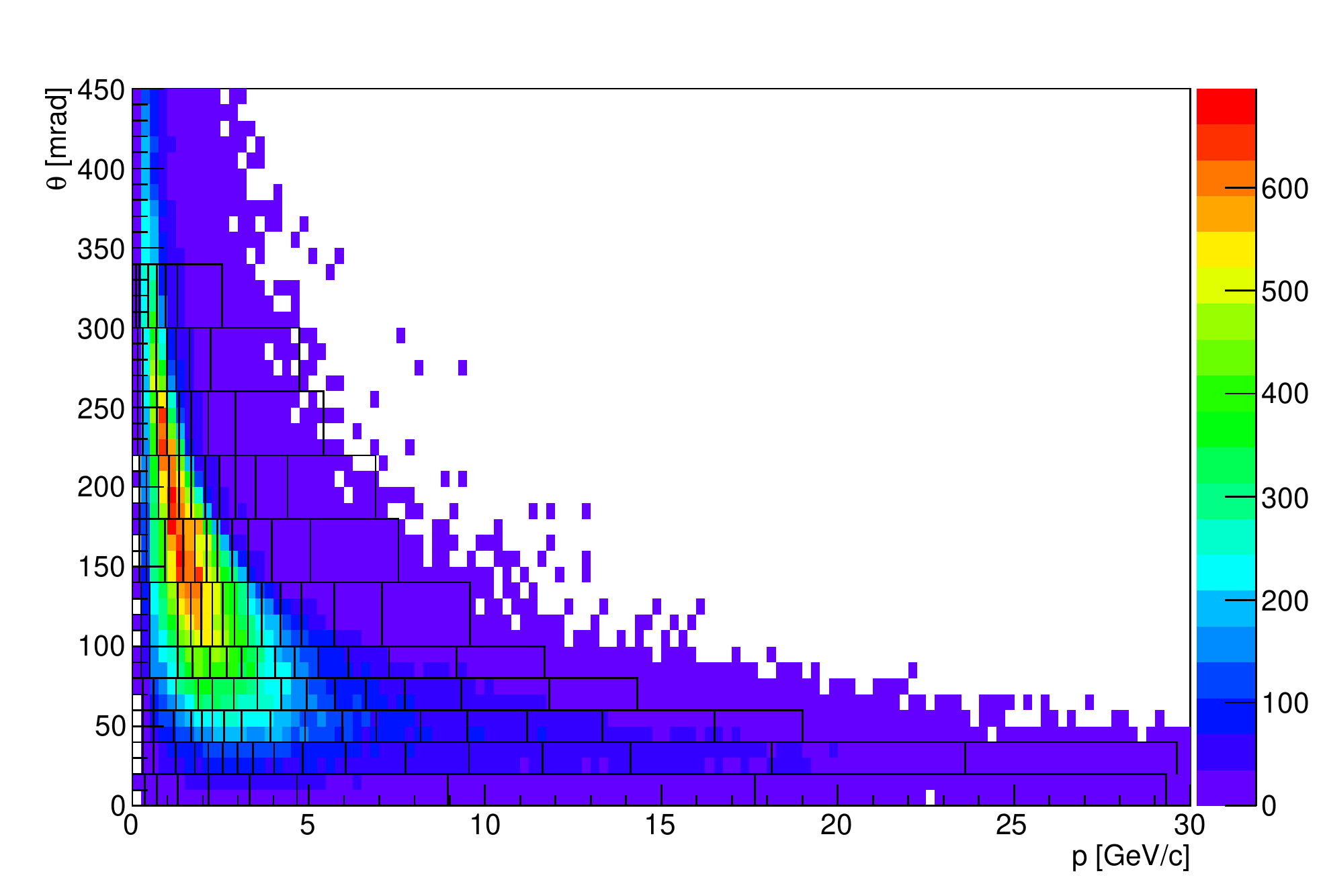}
    \caption{\NASixtyOne analysis binning overlaid on the $(p,\theta)$ distribution of pion parent particles exiting the target surface and producing $\nu_{\mu}$ at SK.}
    \label{fig:AnalysisBinning}
\end{figure}

The $\nu_{\mu}$ and $\nu_{e}$ spectra around the most probable neutrino energy in T2K are predominantly produced by pions (see Fig.~\ref{fig:NeutrinoParents}).
Figure~\ref{fig:AnalysisBinning} shows the phase space 
(in the kinematic variables $p$ and $\theta$ -- the momentum and polar 
angle of particles in the laboratory frame) of the pions exiting 
the target surface and contributing to the $\nu_{\mu}$ flux at SK.
The binning of the T2K replica target analysis is overlaid.
The bins in polar angle and in $z$ along the target were defined 
to ensure adequate sampling of the T2K beam focusing efficiency. 
The binning in momentum was then chosen to obtain roughly 
equipopulated bins.
As can be seen, the T2K replica target analysis region 
covers most of the phase space of interest for T2K.

\begin{figure*}[t]
    \centering
    \includegraphics[width=0.45\textwidth]{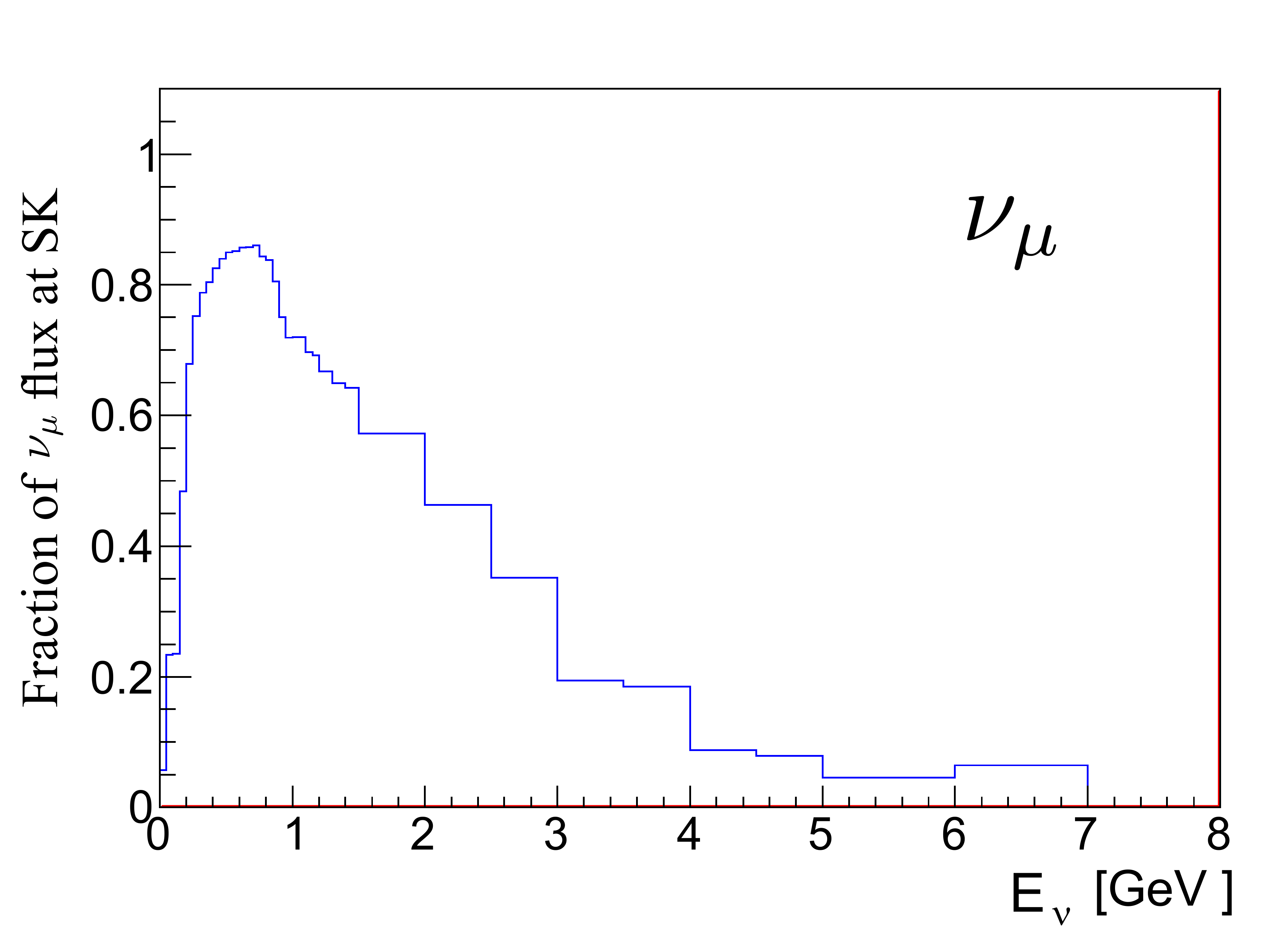}
    \includegraphics[width=0.45\textwidth]{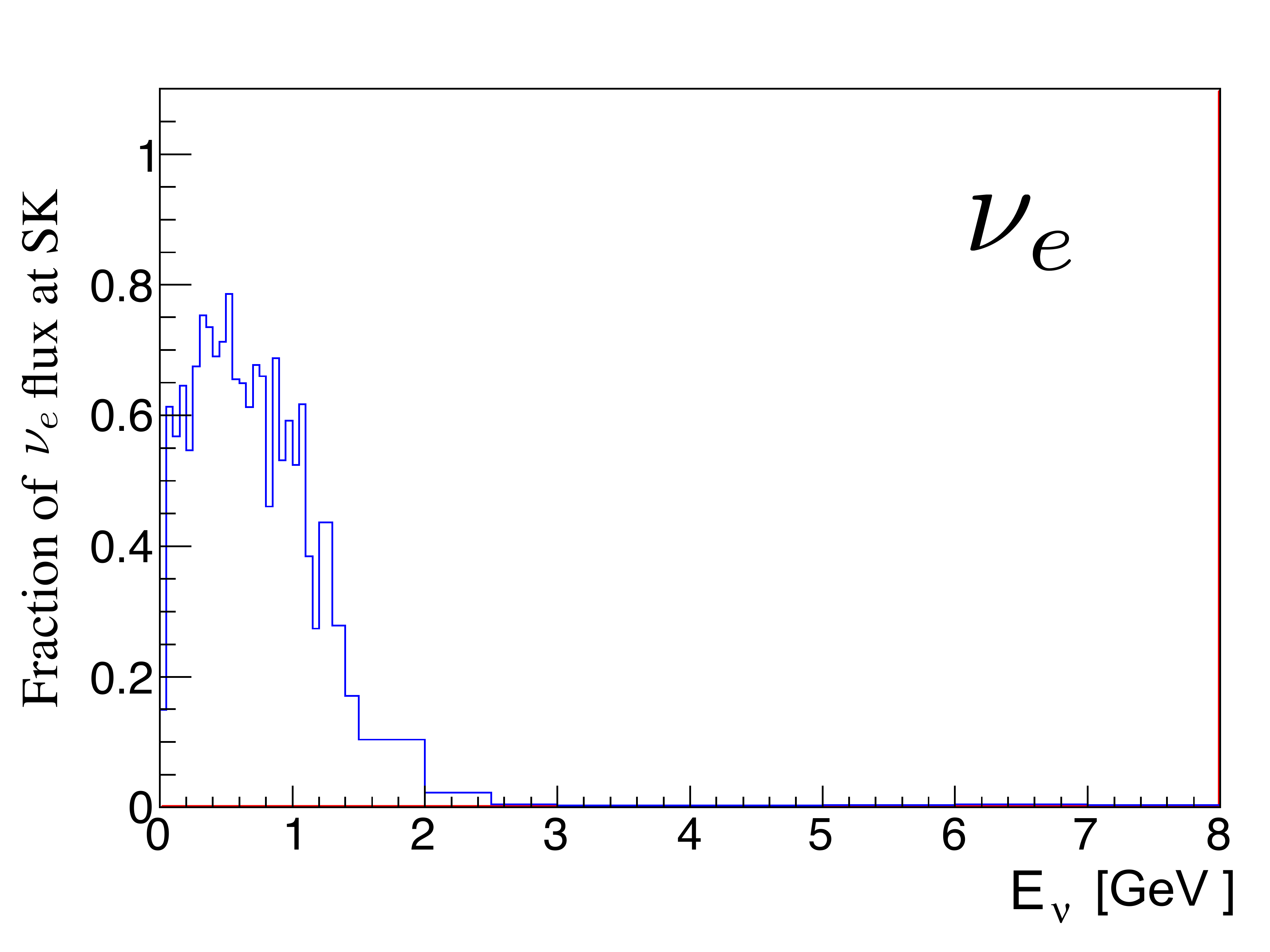}
    \caption{Fraction of $\nu_{\mu}$ ({\it{left}}) and $\nu_{e}$ ({\it{right}}) fluxes at SK
which can be constrained directly with the results presented in this article.
}
    \label{fig:FractionNonTargetSK}
\end{figure*}

Figure~\ref{fig:FractionNonTargetSK} presents the fractions of the $\nu_{\mu}$ and $\nu_{e}$ fluxes at SK that 
can be constrained by the T2K replica target measurements presented 
in this article.
The remaining flux originates from particles produced in interactions of
primary protons or secondaries with the beam line elements, or by other
particle species such as kaon decays, which are not included in the
present analysis.

\section{NA61/SHINE experimental setup} \label{sec:ExpSetup}

\begin{figure}[ht]
    \centering
    \includegraphics[width=0.5\textwidth]{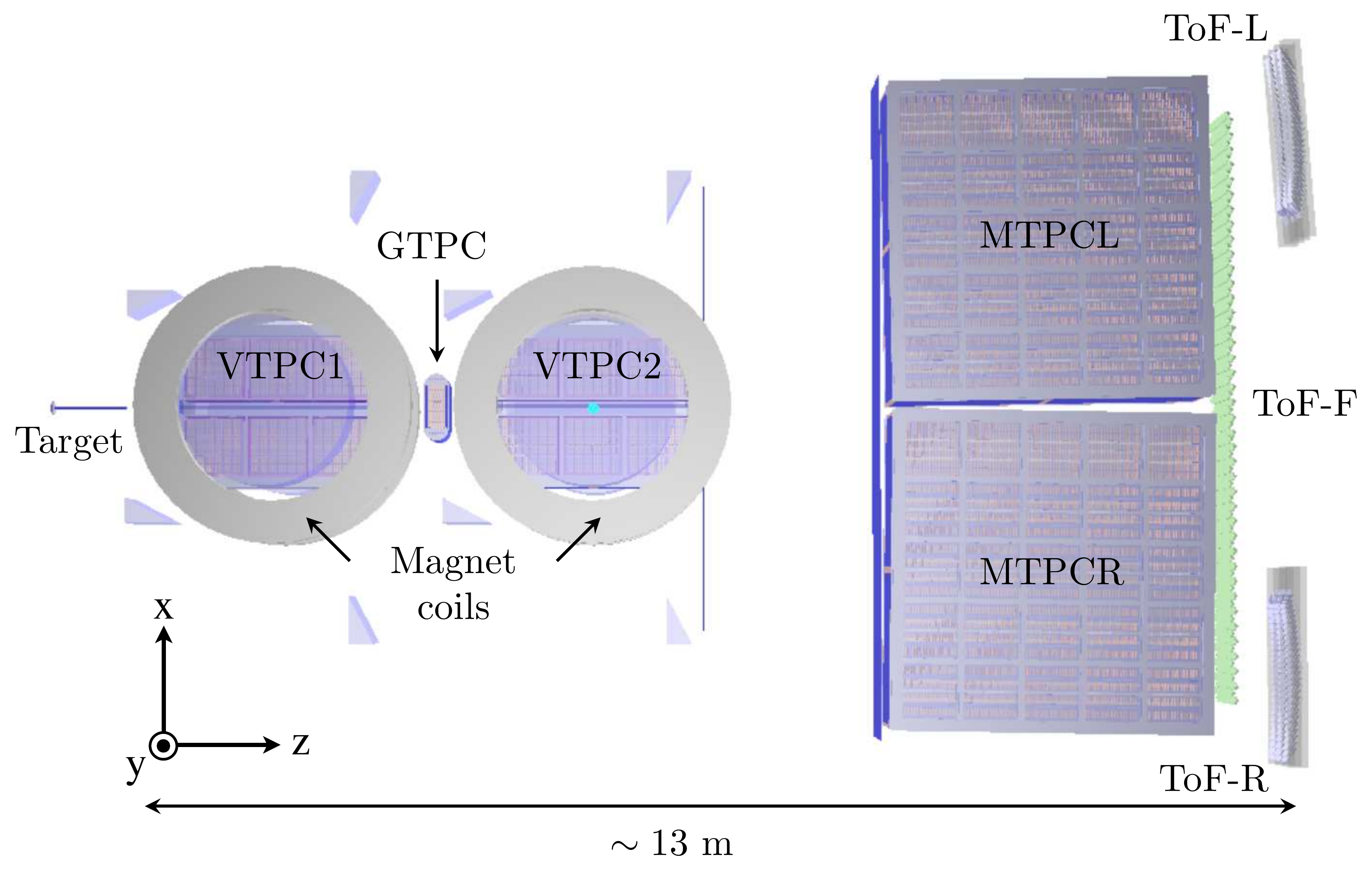}
    \caption{The \NASixtyOne experimental setup (horizontal cut). 
    The beam is coming from the left, 
    impinging on the T2K replica target shown in this figure.
    The chosen coordinate system is as follows: 
    its origin lies in the middle of the VTPC-2, on the beam axis.
    The nominal beam direction is along the $z$ axis.
    The magnetic field bends charged particle trajectories
    in the $x$--$z$ (horizontal) plane. Positively charged particles are bent
    towards the top of the plot.
    The drift direction in the TPCs is along the $y$ (vertical) axis.
    }
    \label{fig:NA61ExpSetUp}
\end{figure}

The \NASixtyOne apparatus is a wide acceptance spectrometer 
at the CERN SPS. Most of the detector components
were inherited from the NA49 experiment and are described in
Refs~\cite{NA49-NIM,NA61detector_paper}. 
A more detailed analysis-oriented description of the \NASixtyOne setup 
can also be found in Ref.~\cite{pion_paper}-Sec.~II.
Only some features relevant for the 2009 running
period are briefly mentioned here.
The general layout of the detector is displayed in Fig.~\ref{fig:NA61ExpSetUp}.
The \NASixtyOne right-handed coordinate system is displayed in the figure 
with the $z$ axis along the beam line, the $x$ axis in the horizontal plane 
and the $y$ axis pointing upwards.
The origin of the coordinate system is placed in the center of the VTPC-2.

The spectrometer is built around five
Time Projection Chambers (TPCs): two Vertex TPCs
(VTPC-1 and VTPC-2) placed in the magnetic field produced by two
superconducting dipole magnets and two Main-TPCs (MTPC) located
downstream symmetrically with respect to the beam line. 
A small additional TPC is placed between VTPC-1 and VTPC-2,
covering the very-forward region, and is referred to as the GAP-TPC (GTPC).

The experimental setup is complemented by time-of-flight detectors.
The \mbox{ToF-F} is located in the forward region, downstream of the MTPCs.
It was used in the analysis presented in this paper.
The detector consists of 80 scintillator bars read out at both ends
by photo-multipliers. 
The time resolution of each scintillator is
about 115~ps \cite{Sebastien}.

For the study presented here the
magnetic field was set to a bending power of 1.14~Tm. This leads to a
momentum resolution $\sigma(p)/p^2$ in the track reconstruction of
about $5 \times 10^{-3}$~(GeV/$c$)$^{-1}$.

The T2K replica target is 
an isotropic graphite rod of density 1.83~g/cm$^3$ with a thickness along 
the beam axis of 90~cm, equivalent to about 1.9~$\lambda_{\mathrm{I}}$ 
and a radius of 1.3~cm. 
Detailed descriptions of the minor differences between the target mounted in the T2K beam line and the T2K replica target are given in Ref.~\cite{LTpaper}.
The downstream face of the target was placed
52~cm upstream of the VTPC-1.
Aluminum flanges were mounted at the most upstream part of the target in order to fix it in the \NASixtyOne experimental set-up.
The rod and the flanges can be seen in Fig.~\ref{fig:TargetZBinning}.
Alignment screws, specially added for the 2009 data taking period and mounted on the flanges, allowed to precisely place and align the target with respect to the beamline.

\section{Analysis} \label{sec:Analysis}

The analysis described in this paper is based on 2.8$\times 10^6$ 
reconstructed events collected during the 2009 data-taking period.

\subsection{NA61/SHINE beam line} \label{subsec:NA61BeamLine}

\begin{figure*}
    \centering
    \includegraphics[width=0.9\textwidth]{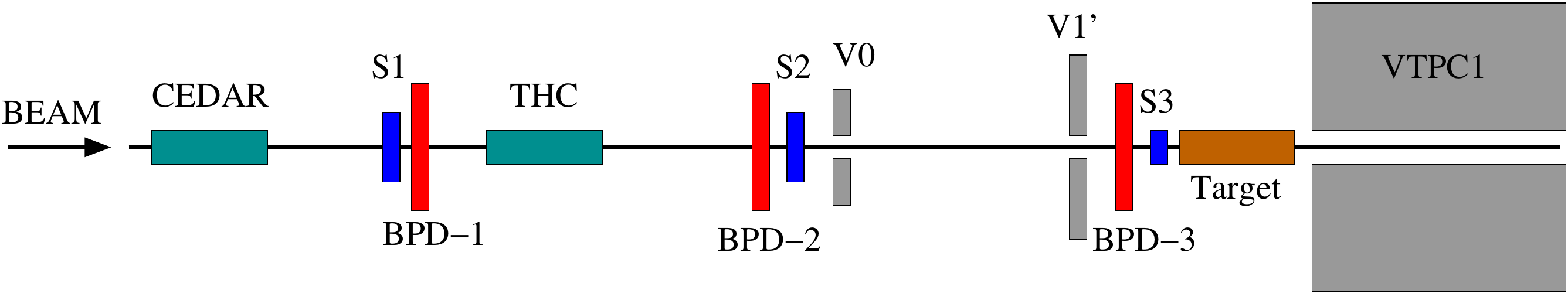}
    \caption{A schematic layout of the \NASixtyOne beam line 
    instrumentation~\cite{NA61detector_paper} 
    for the T2K replica target configuration.
    Horizontal cut in the beam plane is not to scale.
    }
    \label{fig:NA61BeamLine}
\end{figure*}

The \NASixtyOne beam line with  different counters and positioning detectors used 
in the 2009 T2K replica target run is presented in Fig.~\ref{fig:NA61BeamLine}.
The  beam line is instrumented with counters, S1--S3, 
and veto counters, V0 and V1$'$, which provide the beam definition.
Furthermore S1 also sets the start time for all other detectors. 
The S3 scintillator counter has the same radius as the target and 
is positioned 5~mm upstream of the target upstream face.
Hence, it ensures that the incident beam proton is impinging on the target.
Two Cherenkov counters, CEDAR and THC, allow to trigger on protons as beam particles.
The trigger used in the analysis includes the following combination of counters:
\begin{equation}
S1 \land S2 \land S3  \land CEDAR \land \overline{THC}.
\end{equation}

With the measurements of the three beam position detectors (BPDs), tracks of beam particles are reconstructed by 
fitting two straight lines in the $x-z$ and $y-z$ planes. Least square fits use the measured charge clusters in the BPDs.

The distribution of outgoing hadrons along the T2K replica target depends on 
the impact point of a primary proton at the target upstream face.
This point is reconstructed using information from the BPD detectors.
To ensure the quality of the beam track
strict cuts on the BPD measurements are applied.
All three BPDs must have properly reconstructed clusters in the $x$ and $y$ coordinates.
Furthermore, a cut on the $\chi^{2}$ of the fit of the reconstructed tracks allows to reach a resolution of 300~$\mu$m at the BPD-3 which is positioned 7~cm upstream of the target 
front face.

\begin{figure}
    \centering
    \includegraphics[width=7.5cm, height=7cm]{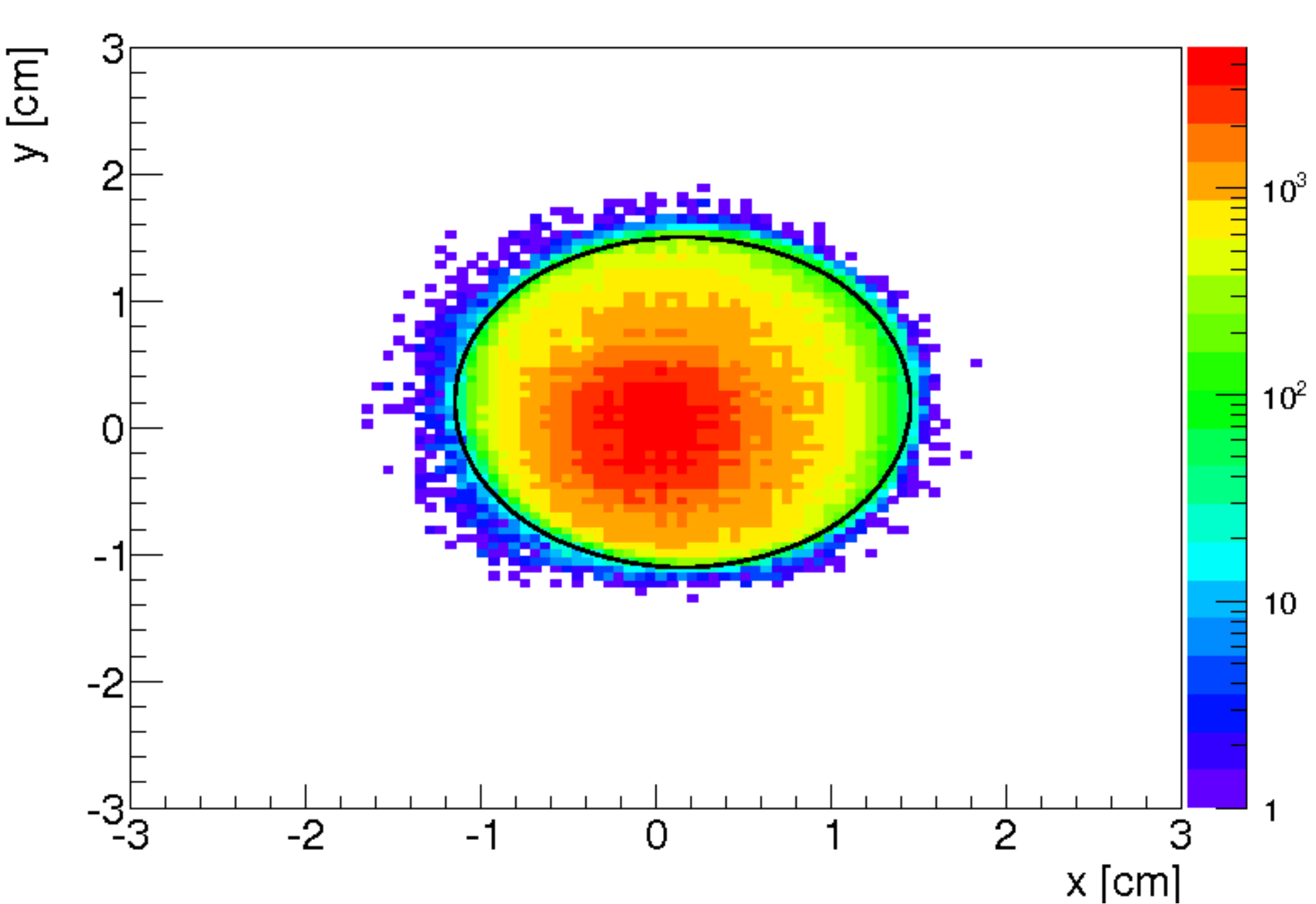}
    \caption{Proton beam profile at the upstream face of the target as reconstructed with the BPD information. The black circle represents the edge of the target surface.}
    \label{fig:BeamProfile}
\end{figure}

Figure~\ref{fig:BeamProfile} shows the beam profile on the target upstream face as reconstructed with the BPD information.


\subsection{Reconstruction of particle parameters at the target surface}

The final neutrino flux depends on the longitudinal distribution of the hadrons exiting the target surface.
Hence, 
the exit position of the produced particles and their momentum vectors at 
the target surface were reconstructed.

The reconstruction procedure was similar to the one described in Ref.\,\cite{thin2009paper}.
First, charge deposits by tracks traversing the TPCs measured by neighbouring readout pads are joined to form clusters.
Local track segments are then reconstructed from the clusters in each TPC separately.
The matching of the track segments from different TPCs allows to reconstruct global tracks.
The track fitting through the magnetic field allows the determination of the track parameters at the first measured TPC cluster.

In order to finally determine the track parameters at the surface of the target, a backward extrapolation is performed from the first measured TPC cluster.
A Runge-Kutta method is used to propagate track parameters and their uncertainties in the non-uniform magnetic field.
An exit point of the track from the target surface is found when the backward extrapolated track intersects the target volume.
If no intersection point can be found, the point of closest approach between the backward extrapolated track and the surface of the target is recorded and assigned as the exit position.
The track is then considered as originating from the target 
if the distance between the point of closest approach and the target surface is
within the one standard deviation uncertainty.
Most particles point to the inside of the target, both in data and Monte-Carlo. 
The cut at 1 standard deviation keeps 96.6\% of the selected tracks in the data 
and 96.5\% in the Monte-Carlo. The resulting systematic uncertainty 
is considered negligible.


\subsection{Determination of the target position and tilt} \label{subsec:TargetPosAlign}

A precise knowledge of the position of the target and its alignment along the beam line is mandatory in order to reconstruct the exit point of the produced hadrons at the surface of the target.
The alignment of the target with respect to the experimental setup is determined in three consecutive steps based on the data.

The first step consists of determining the angular alignment of the target.
This is done by dividing the 90~cm long target into 17 slices.
The reconstructed tracks are extrapolated backward from the TPCs to the center of each of the 17 slices.
For each slice, a mean of the $x$ (respectively $y$) position of the backward extrapolated tracks is determined and this mean is assigned as the central $x$ (respectively $y$) position of the target.
Fitting all the mean positions with a straight line allows to determine a tilt in the $x-z$ (respectively $y-z$) planes.
For the 2009 T2K replica target dataset, these tilts were found to be negligible as expected from the target positioning precision of better than 2.2~mrad.

The second step consists of determining the transverse position of the target with respect to the beam line.
This position is determined with the help of the reconstructed beam tracks from the BPDs.
Requiring a hit in the S3 scintillator counter ensures that the incident protons have reached the target upstream face and by drawing the $x-y$ distribution of the beam particles under this requirement, the center of the target can be determined as the mean of the $x-y$ distribution.

\begin{figure}[t]
    \centering
    \includegraphics[width=0.5\textwidth]{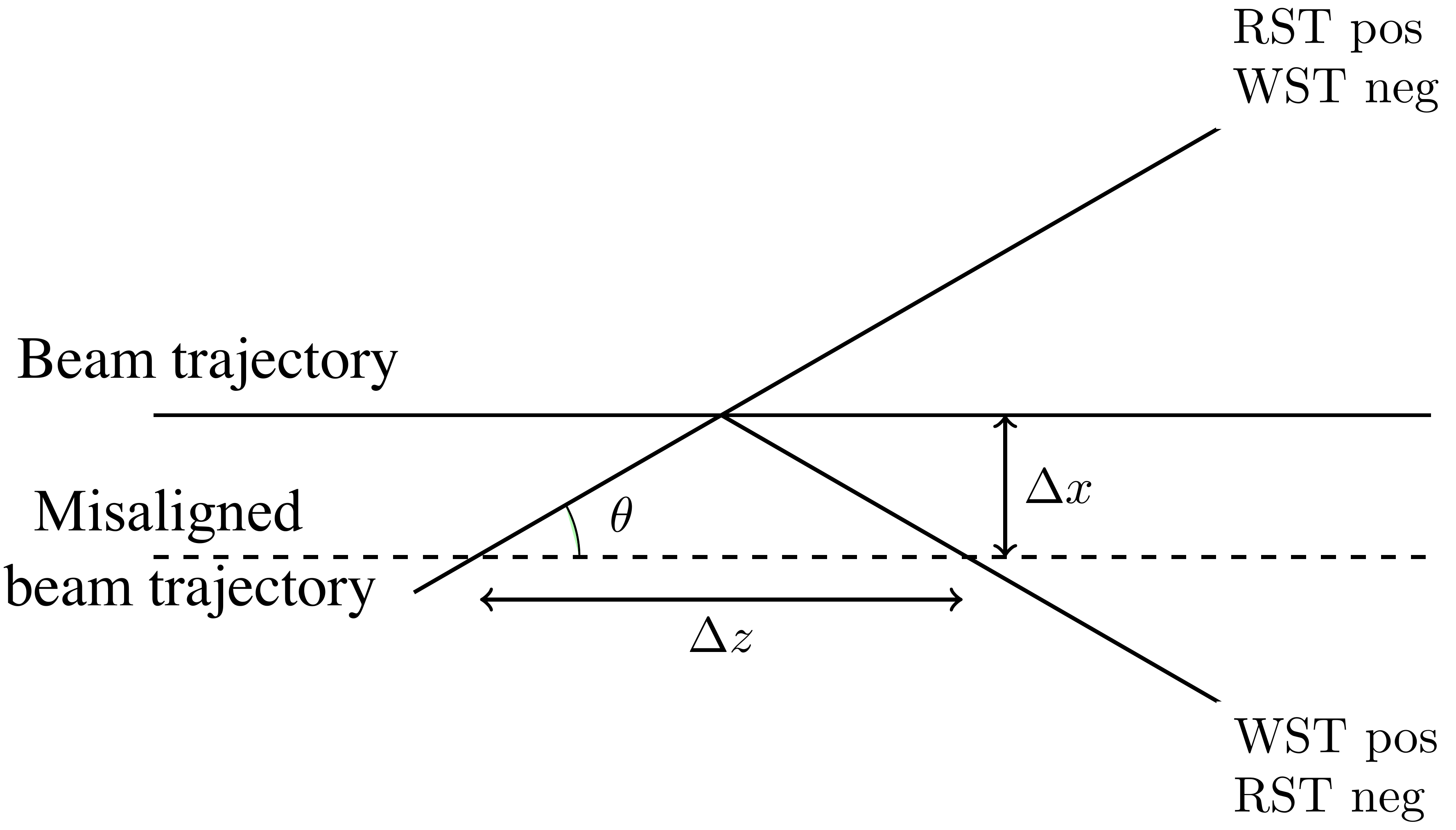}
    \caption{A transverse misalignment, $\Delta x$, between the beam particles and the tracks extrapolated backward from the TPCs is translated into a longitudinal shift, $\Delta z$, of the vertex distributions when comparing the two sets of tracks composed of different topologies: positively charged RST and negatively charged WST versus positively charged WST and negatively charged RST.}
    \label{fig:BPDTPCAlignment}
\end{figure}

The third step consists of cross-checking the alignment between the beam line and the spectrometer and subsequently extracting the longitudinal position of the target.
As this position is determined by using the beam tracks reconstructed from the BPD information, and as the position of the particles exiting the target surface is determined by using tracks from the TPCs, it is important to know the precision of the alignment between the BPDs and the TPCs.
This is done by checking the consistency between the reconstructed vertices from the beam tracks and the reconstructed tracks from the TPCs for two independent track topologies.
These topologies are defined as
\begin{itemize}
    \item Right Side Tracks (RSTs): ~
	tracks of particles which are emitted in the direction of bending in the magnetic field, i.e. \mbox{$p_{x}/Q > 0$}
    \item Wrong Side Tracks (WSTs): ~
	tracks of particles which are emitted with an angle opposite to the direction of bending, i.e. $p_{x}/Q < 0$
\end{itemize}
Then a potential misalignment in the transverse $x-y$ plane 
can be reconstructed from a shift in the  distribution of the longitudinal $z$ coordinate.
Figure~\ref{fig:BPDTPCAlignment} shows a schematic of this procedure.

To reconstruct coordinates of the primary interaction point inside the long target, specific cuts have to be applied in order to reduce the number of tracks that originate from re-interactions inside the target 
and hence have trajectories not intersecting with the beam particle.
The following cuts are  applied to the events used for the construction of the distribution of the vertices:
\begin{enumerate}[(i)]
    \item the beam particle hits the target at least 0.5~cm inside the target surface
    \item the associated track in the TPCs has to be on the same side in the $y-z$ plane as the beam particle, i.e. if the beam particle hits the target upstream face at the positive $x$, then the tracks have to exit the target surface at positive $x$ as well.
\end{enumerate}

In order to get sufficient precision only tracks with  $100< \theta < 180$~mrad were used.
The procedure was tested and validated by applying it to a Monte-Carlo simulation for which the $z$ position of the target is well known and the alignment of the beam is perfect with respect to the spectrometer.

\begin{figure*}[ht]
    \centering
    \begin{subfigure}{.5\textwidth}
	\centering
	\includegraphics[width=.98\linewidth]{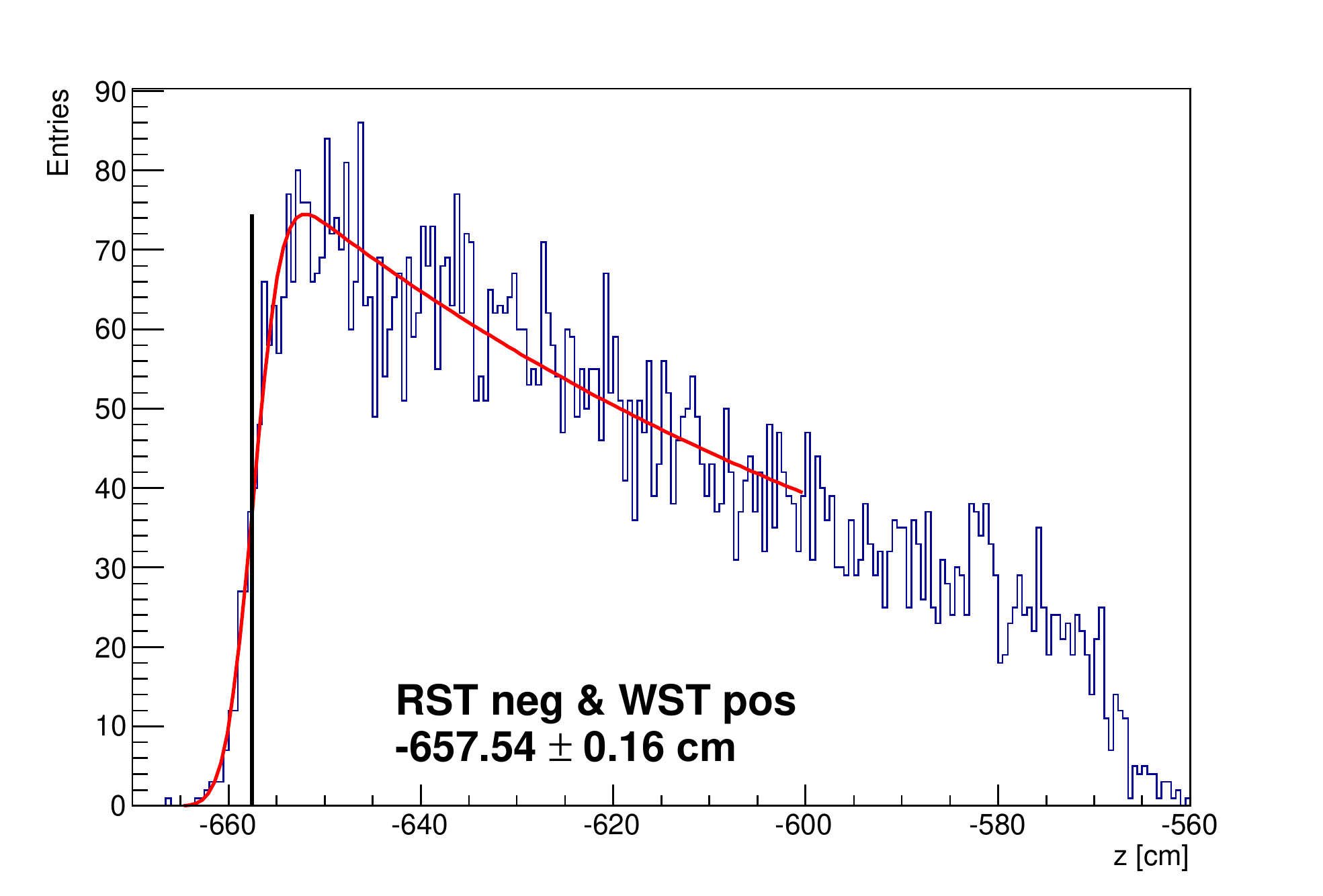}
    \end{subfigure}%
    \begin{subfigure}{.5\textwidth}
	\centering
	\includegraphics[width=.98\linewidth]{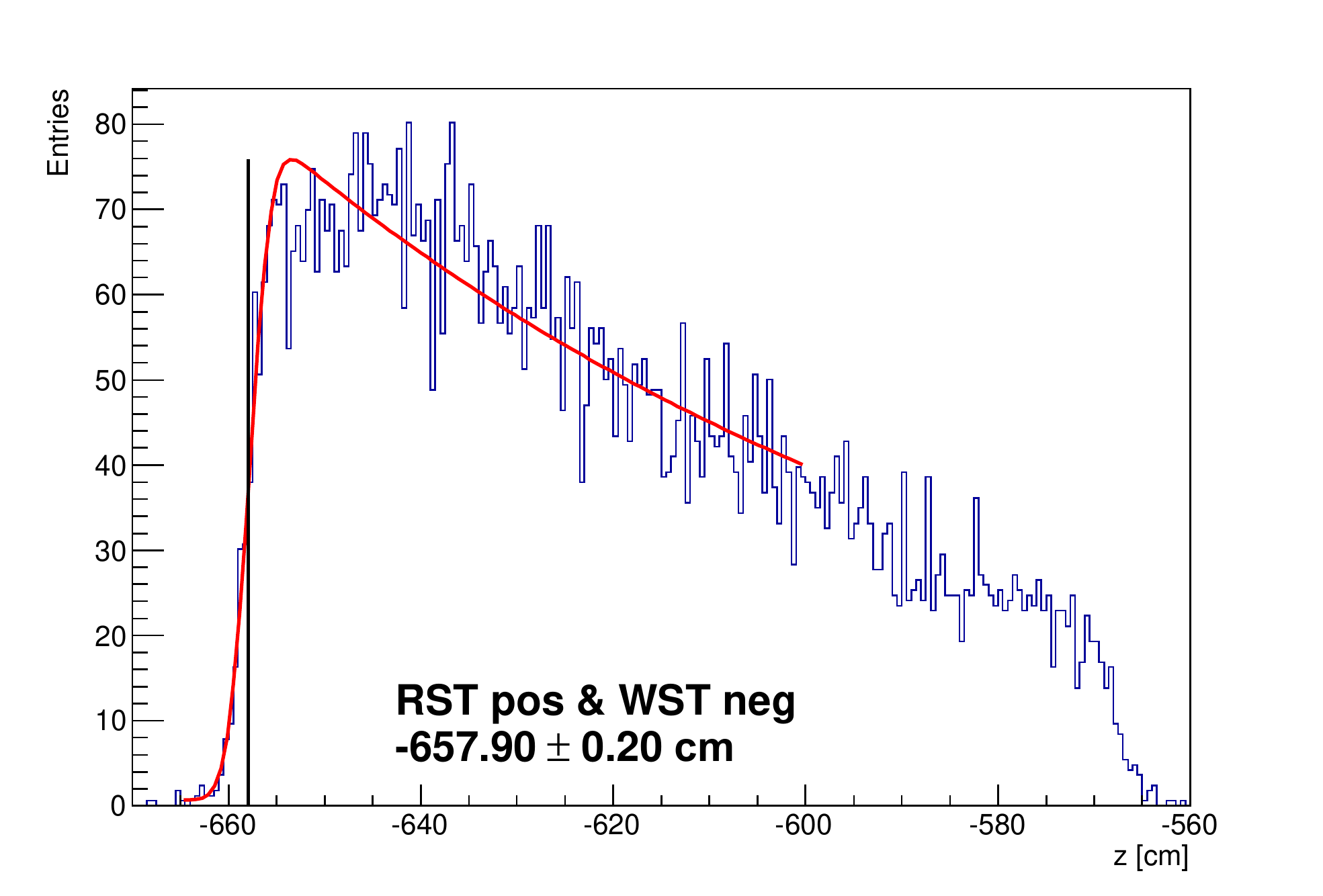}
    \end{subfigure}
    \caption{The fitted $z$ coordinate of primary interactions obtained using 
      negatively charged RST and positively charged WST ({\it{left}}) and
      positively charged WST and negatively charged RST ({\it{right}}).
      The vertical line shows the position of the target upstream face obtained with the fit.}
    \label{fig:ZVertexDATA}
\end{figure*}

Figure~\ref{fig:ZVertexDATA} shows the results of the procedure when applied to the 2009 T2K replica target dataset.
The $z$ distribution is fitted with an exponentially modified Gaussian function and the upstream face of the target is determined as the position of the half maximum of the distribution (see vertical line in Fig.~\ref{fig:ZVertexDATA}).
It can be seen that the distributions for the two samples of tracks do agree within statistical uncertainty. Thus there is no significant misalignment between the beam and the spectrometer.
The statistical uncertainties on the $z$ positions can actually be translated into uncertainties of the $x$ and $y$ positions.

The final position of the upstream face of the T2K replica target obtained for the 2009 dataset is:
\begin{align}
    x &= ~~ 0.16 \pm 0.04~\mbox{cm} \nonumber \\
    y &= ~~ 0.21 \pm 0.04~\mbox{cm} \\
    z &= -657.62 \pm 0.36~\mbox{cm} \nonumber
\end{align}


\subsection{Data analysis} \label{subsec:YieldExtr}

There are several analysis techniques applied in \NASixtyOne for the extraction
of raw yields of charged pions (see Refs.\,\cite{pion_paper,thin2009paper}).
The analysis presented in this article, utilizes the TPC measurements of 
the specific energy loss ($dE/dx$) and the time-of-flight measurements ($tof$) 
using the 
ToF-F detector. 

In this section, we will concentrate on specific details related to the extraction of 
charged pion yields at the surface of the T2K replica target.


\subsubsection{Event and track selection} \label{subsec:EventTrackSelection}

As mentioned in Section~\ref{subsec:NA61BeamLine}, the events of interest for this analysis are selected based on two requirements:
\begin{enumerate}[(i)]
    \item 
      a hit in the S3 counter is required to make sure that the incoming proton is hitting the target upstream face.
    \item a strict cut on the $\chi^{2}$ of the fitted beam track.
\end{enumerate}
After applying these two cuts, a sample of $1.6\times 10^{6}$ events remains available for the analysis.

The requirements on the reconstructed tracks 
depend on the analysis technique to be used (as explained later) 
but they are based on the same criteria:
\begin{enumerate}[(i)]
    
    \item The reconstructed tracks must have a certain number of measured points 
    through the TPCs.
    
    \item Tracks must have a properly measured energy loss $dE/dx$ in the TPCs.
    
    \item In addition, for the $tof-dE/dx$ analysis, they must have 
    a properly measured time of flight from the ToF-F detector.
    
    \item The reconstructed tracks have to be within a given azimuthal $\phi$ angle range 
    in order to be well within the acceptance of the spectrometer.
    
    \item Tracks have to originate from the surface of the target.
    
\end{enumerate}
These criteria were also used for the target alignment procedure described in Section~\ref{subsec:TargetPosAlign} as they select tracks with 
well fitted parameters.


\subsubsection{Data binning} \label{subsec:AnalysisBinning}

As mentioned in Section~\ref{subsec:DescriptionNuBeam}, different longitudinal sections of the target contribute differently to the final neutrino flux while the focusing of the horns will affect the particles depending on their momentum and polar angles.
Hence, the analysis of the T2K replica target will be conducted in $(p,\theta,z)$ bins.
The longitudinal $z$ binning was determined by a study performed together with the T2K beam group.
It was found that five longitudinal bins are sufficient to obtain a neutrino flux
prediction that matches the non-binned case, both in terms of shape and overall normalization, within a known and correctable bias of less than 2\%.
Hence the target surface is divided into five bins of 18~cm length and the downstream face of the target is taken as a sixth longitudinal bin, as shown 
in Fig.~\ref{fig:TargetZBinning}.
The chosen $(p,\theta)$ binning scheme is illustrated 
in Fig.~\ref{fig:AnalysisBinning}.


\subsubsection{The $tof-dE/dx$ analysis for particle identification}
\label{subsec:Analysis-PID}

\begin{figure*}[ht]
    \centering
    \includegraphics[width=0.5\textwidth]{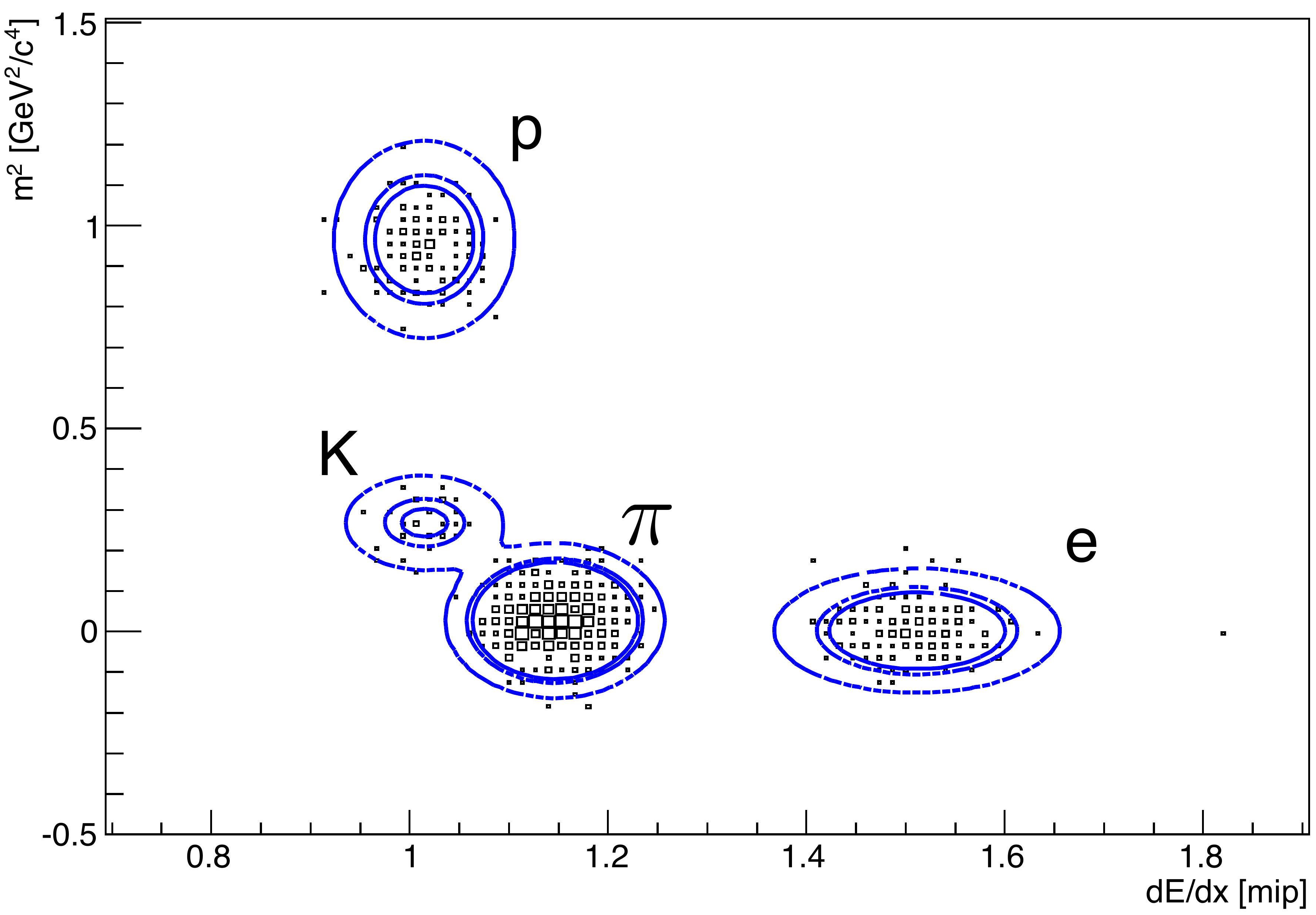}
    \begin{subfigure}{.5\textwidth}
	\centering
	\includegraphics[width=.98\linewidth]{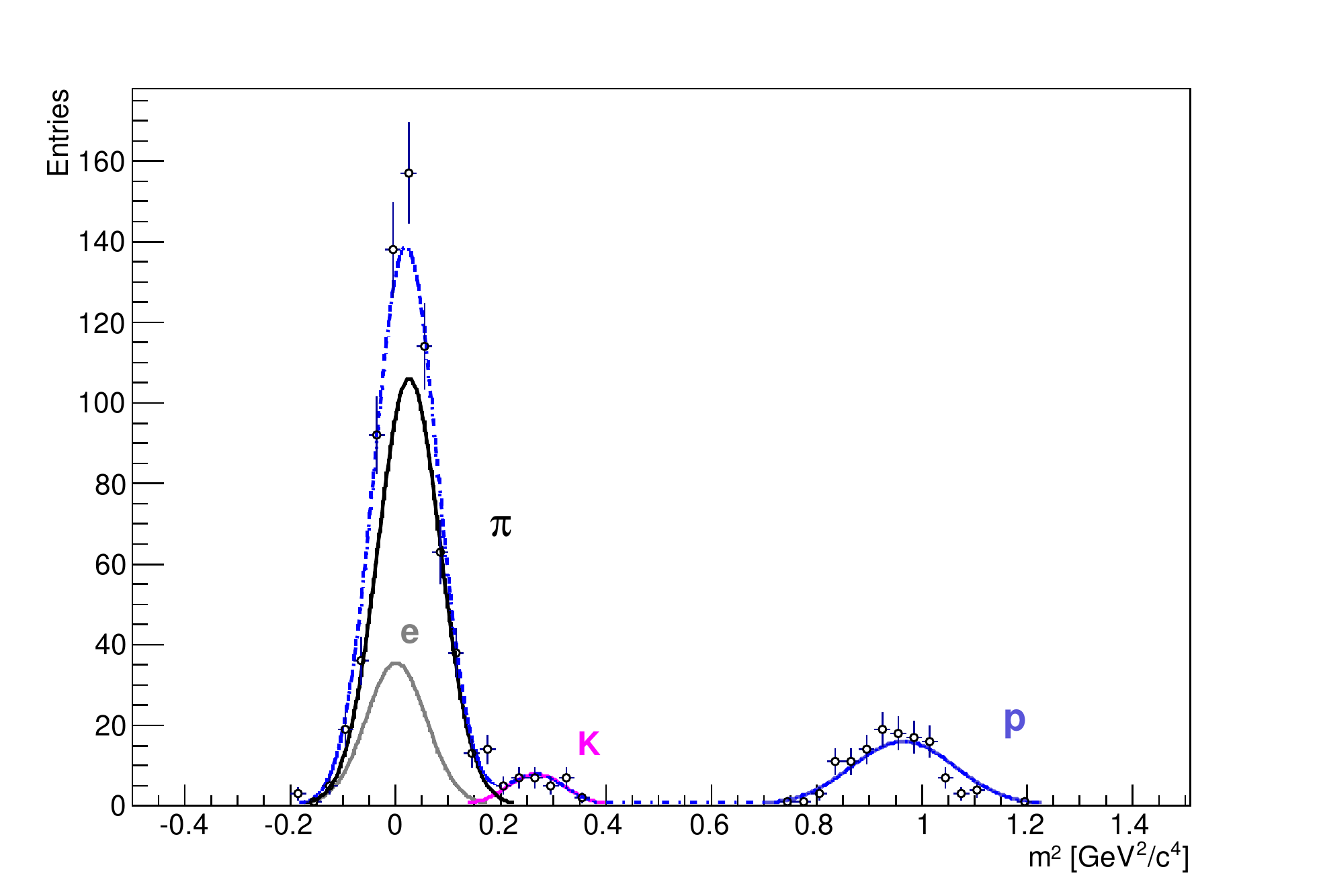}
    \end{subfigure}%
    \begin{subfigure}{.5\textwidth}
	\centering
	\includegraphics[width=.98\linewidth]{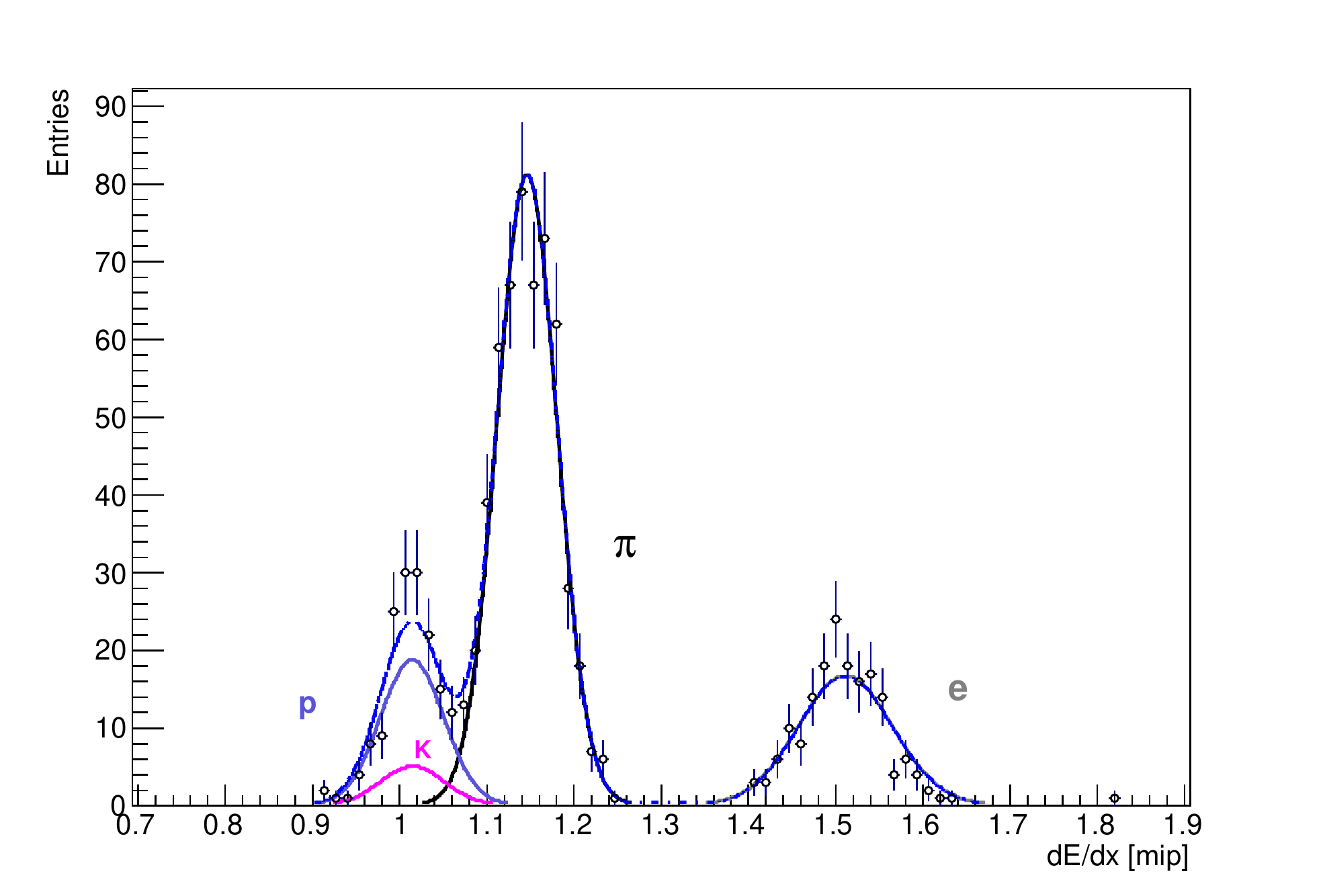}
    \end{subfigure}
    \caption{Distributions of $m^{2}$ and $dE/dx$ for positively charged particles for the second longitudinal bin $z2$ in the intervals $60<\theta<80$~mrad and $2.27< p <2.88$~GeV/$c$.
    The top plot shows the two dimensional spectrum whereas bottom left and bottom right are projections on the $m^{2}$ and $dE/dx$ variables respectively.
    The lines on the top plot indicates the 1$\sigma$, 2$\sigma$ and 3$\sigma$ contours of the fitted Gaussian functions for each particle species.
    The lines on the bottom plot indicate the Gaussian functions for the four different particle species (solid curves) and for the total of the four species (dashed curves).}
    \label{fig:FitProjResults}
\end{figure*}

\begin{figure}[ht]
    \centering
    \includegraphics[width=.98\linewidth]{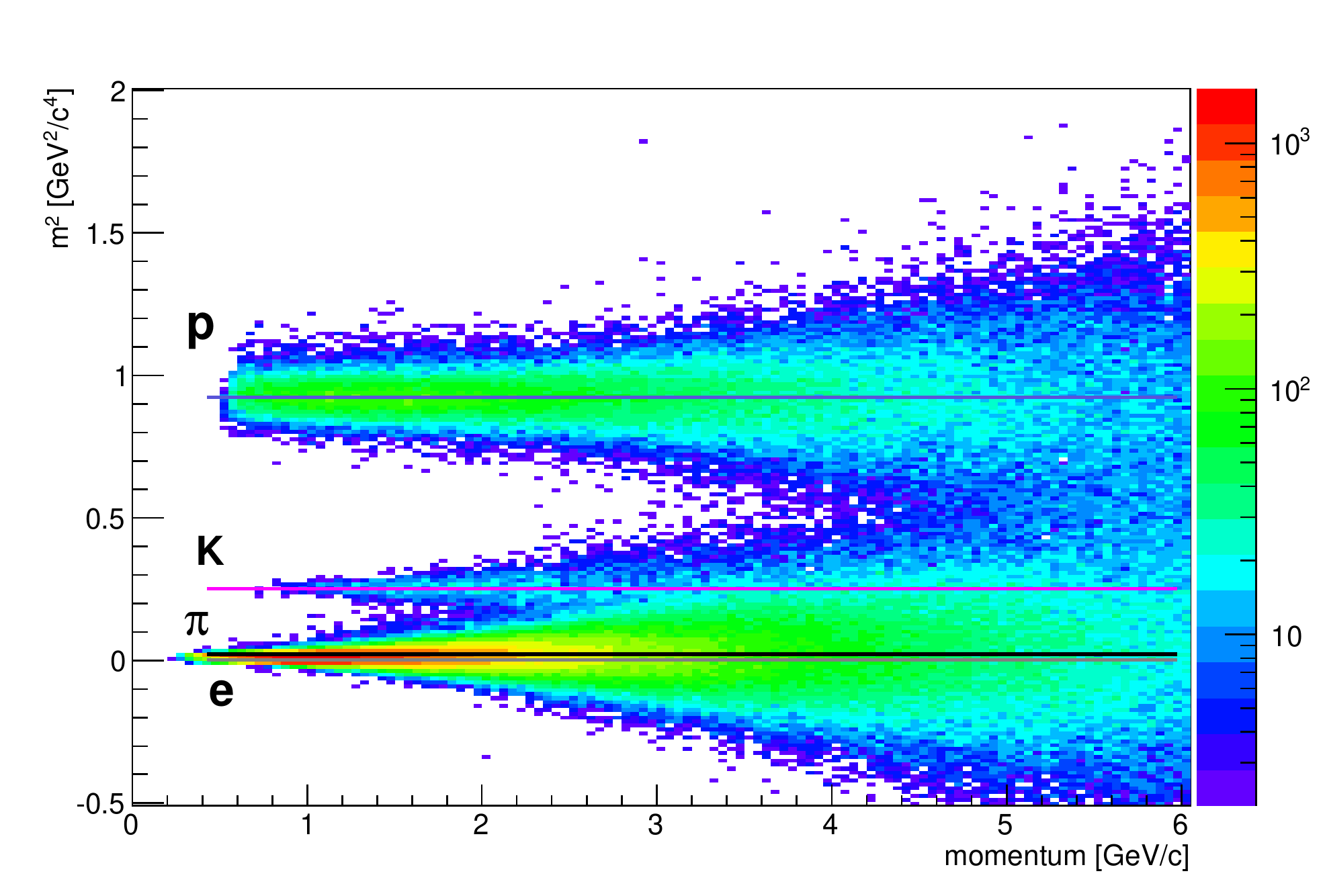}
    \caption{Distribution of particles $m^{2}$ as a function of momentum obtained using the ToF-F measurements.}
    \label{fig:m2Init}
\end{figure}

\begin{figure*}[ht]
    \centering
    \begin{subfigure}{.5\textwidth}
	\centering
	\includegraphics[width=.98\linewidth]{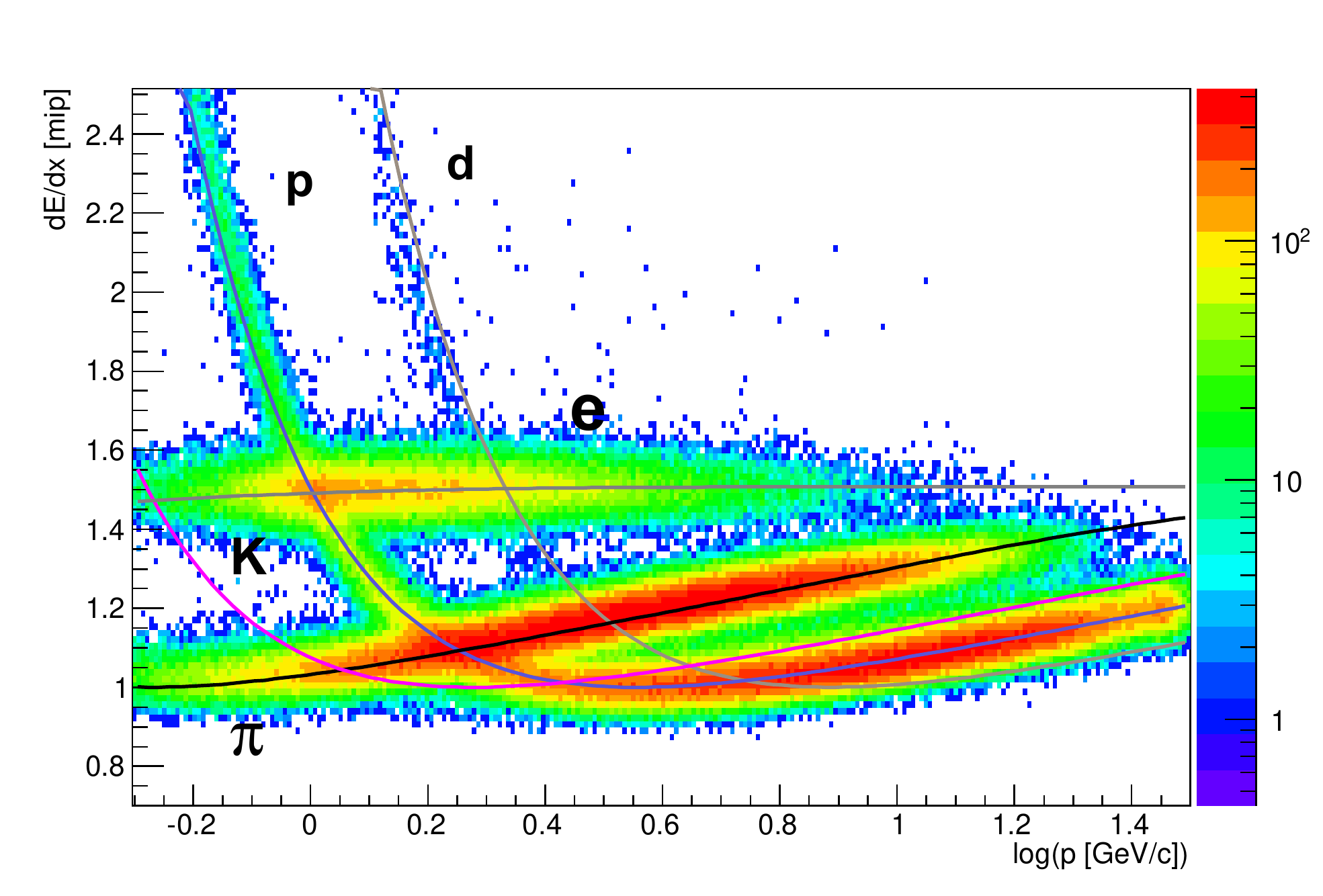}
    \end{subfigure}%
    \begin{subfigure}{.5\textwidth}
	\centering
	\includegraphics[width=.98\linewidth]{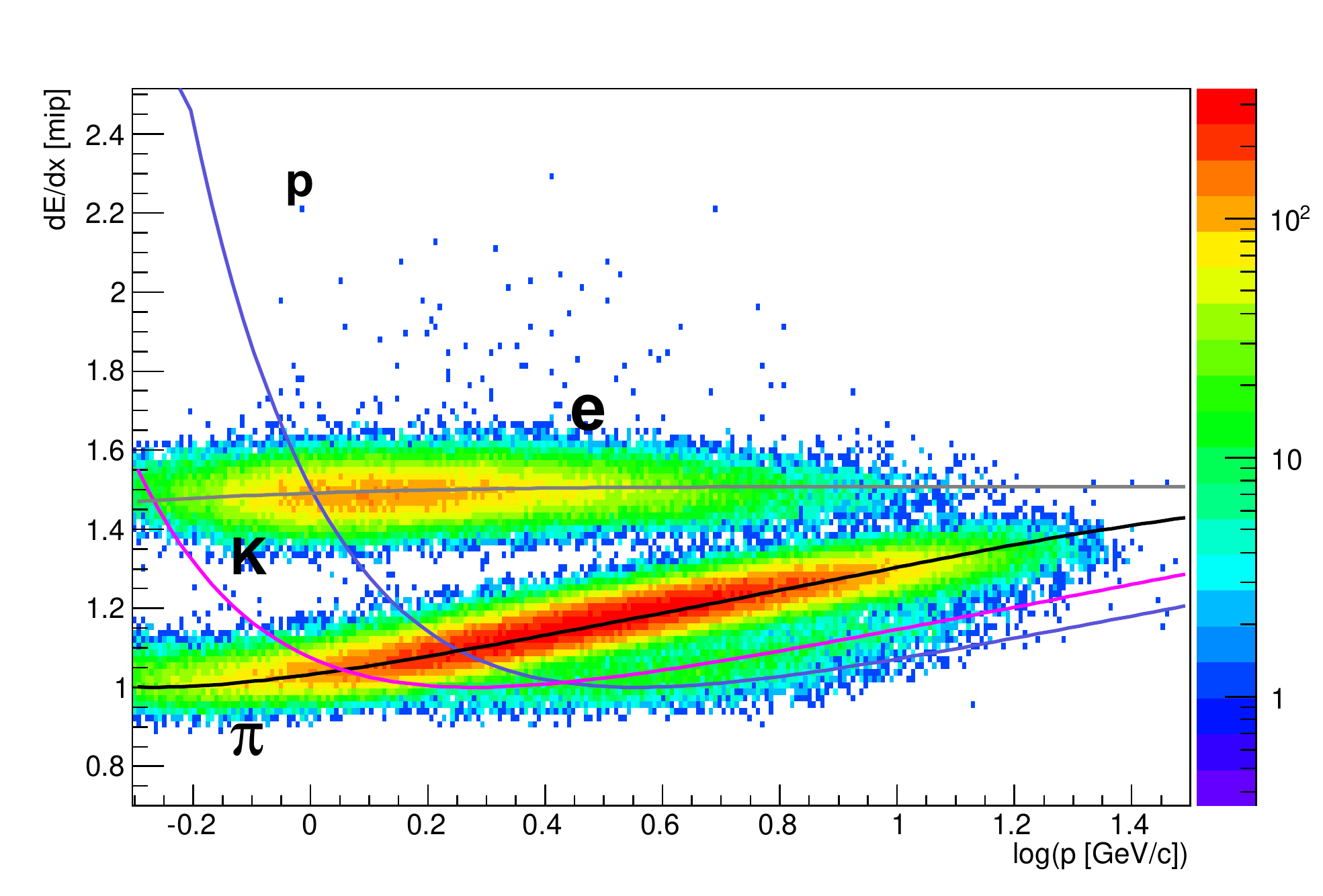}
    \end{subfigure}
    \caption{The $dE/dx$ distributions as a function of momentum for positively ({\it{left}}) and negatively ({\it{right}}) charged particles. The Bethe-Bloch parametrizations are superimposed.}
    \label{fig:BBCurves}
\end{figure*}

Particle identification (PID) in \NASixtyOne relies on measurements of the energy loss $dE/dx$ 
in the TPCs and the time-of-flight that is 
used to compute the particle mass squared, $m^{2}$.
The method is illustrated in Fig.~\ref{fig:FitProjResults} (top panel) which depicts 
how the different particles ($p$, $K$, $\pi$ and $e$) can be separated in the $(m^{2},dE/dx)$ plane.
A $(m^{2},dE/dx)$ distribution, separately for positively and negatively charged tracks, is
obtained for each bin $(p,\theta,z)$ determined at the surface of the 
replica target. The data distributions are then fitted to joint probability density functions (pdf) for the mass squared and the 
energy loss. Due to the independence of the $dE/dx$ and $m^{2}$ 
variables, the joint pdf reduces to the product of the corresponding marginal distributions which are described by Gaussian
distributions. The complete pdf is a sum of two-dimensional 
Gaussian distributions of four particle species, $p$, $K$, $\pi$ and $e$. 
For the initialization of the fit, the resolution of the mass squared 
and the expected energy loss for each particle species is 
obtained from parametrizations of the data distributions shown 
in Figs.~\ref{fig:m2Init} and~\ref{fig:BBCurves} as a function of the track momentum. The resolution of 
the energy loss measurement is a function of the number of reconstructed clusters on the track ($1/\sqrt{N}$). 
For the topology dependent cuts defined in this analysis the $dE/dx$ resolution can be approximated by a constant value of 3$\%$ due to the sufficiently large number of clusters on 
each track. Independent normalization factors are introduced for 
each particle species. Since the individual pdfs are normalized to 
unity, particle yields are given by the normalization factors which 
are obtained from a two-dimensional log-likelihood minimization 
illustrated in Fig.~\ref{fig:FitProjResults}.
The projections on the $m^{2}$ and $dE/dx$ variables better illustrate the quality of the fit results
(see Fig.~\ref{fig:FitProjResults}, bottom panels).


\subsection{Simulation} \label{subsec:Simulation}

Simulations were performed to generate events of 31~GeV/$c$ protons interacting with the T2K replica target.
In order to be consistent with the T2K neutrino beam simulation program \cite{T2Kflux},
the simulation package \mbox{FLUKA2011~\cite{Fluka,Fluka_CERN,Fluka_new}} is used in \NASixtyOne to generate the interactions inside the graphite target.
The GEANT3 \cite{Brun:1987ma} transport code was used to track the particles through the detector and GCALOR~\cite{GCALOR} handled the interactions in the spectrometer.
More details can be found in Ref.~\cite{Nicolas}.

The full simulation chain consists of three parts:
\begin{enumerate}[(i)]

    \item First, a stand-alone FLUKA simulation. The target geometry is described as a 90~cm long graphite rod with aluminum flanges and the S3 counter.
	The target was positioned at the location determined by the alignment procedure applied to the data as explained in Section~\ref{subsec:TargetPosAlign}.
	The incident proton beam profile was simulated following the shape of the distributions for the positions and divergences given by the data.
	The momentum of the beam was set precisely at 30.92~GeV/$c$ to match the beam momentum measured in the data.
	Information on interactions happening inside the target was stored. 
	Position, momentum as well as polar and azimuthal angles of the particles exiting the target were recorded at the surface of the target and saved as output of the FLUKA simulation.
    
    \item A GEANT3-based program used the kinematic parameters of particles produced by FLUKA at 
     the surface of the target and propagated them through the \NASixtyOne experimental setup. 
     The GCALOR model handled all hadronic interactions in the spectrometer.
     Moreover, a detailed simulation of various detector effects was included.
     
    \item The tracks were finally reconstructed following the same reconstruction procedure as the one applied to the data. All information from the FLUKA simulation and the simulated tracks until their reconstruction was stored in the final output files. This allows to get the full history of the simulation and to match the reconstructed to simulated tracks.
\end{enumerate}

The simulation was used to calculate corrections for pions 
resulting from various sources: 
i) weak decay of heavier particles producing additional pions, 
ii) interactions in the detector material, 
iii) track reconstruction efficiency and resolution, 
iv) decay in flight. 
In total, 10 millions of protons on target were simulated.

\section{Systematic Uncertainties} \label{sec:Syst}

Six different sources of systematic uncertainties are considered.
Their contributions to the systematic uncertainty are
described in detail below.
Five of them are similar to the thin target p+C@31~GeV/$c$ analysis~\cite{thin2009paper}, only the last one (backward extrapolation) is specific to the T2K replica target analysis.

\subsection{Particle Identification} \label{Sec:Syst-PID}

The $dE/dx$ distribution for the different particle species in each of 
the $(p,\theta,z)$ bins is approximated by a single Gaussian.
In order to estimate 
the uncertainty related to this approximation 
two Gaussians with the same mean value but different widths were used to fit the $dE/dx$ distributions.
At low momenta, the particle identification is  constrained by the ToF-F information and hence the magnitude of the uncertainty due to describing the energy loss by a single Gaussian is expected to be negligible.
At higher momenta, when the resolution of the time-of-flight measurements does not allow 
to distinguish the different particle species, using two Gaussians instead of 
a single Gaussian in the fitting procedure 
leads to differences of up to $2\%$ at momenta higher than 10~GeV/$c$.

\subsection{Feed-down corrections} \label{Sec:Syst-FeedDown}

Pions not originating from the target but traversing the spectrometer and 
reconstructed as exiting the target surface 
represent the so-called feed-down contribution.
The feed-down correction comes from particles of various origins: (i)
interactions of particles outside the target, (ii) decays in flight of
strange particles.
The correction factor for the feed-down contribution is computed based on simulations produced with FLUKA as primary hadronic generator.
This correction is  model dependent and an uncertainty on this model prediction has to be assigned.
As for the 
thin target analysis, $30\%$ of the correction was assumed 
as the systematic uncertainty of the correction  \cite{pion_paper,thin2009paper}.
This 
uncertainty reaches values as large as $5\%$ of the pion yield for momenta lower than 2~GeV/$c$.
It decreases significantly at higher momenta.

\subsection{Reconstruction efficiency} \label{Sec:Syst-Reco}
Following the 
thin target analysis, a constant $2\%$ uncertainty 
on the efficiency of the reconstruction procedure is assigned \cite{thin2009paper}.

\subsection{ToF-F reconstruction efficiency} \label{Sec:Syst-ToF}

The correction for the ToF-F reconstruction efficiency is computed based on 
the procedure described in Ref.\,\cite{Hasler:2039148}.
The efficiency was estimated on a sample of physics events with 
a strict cut on a time window around the triggered interaction.
In the procedure a track traversing the ToF-F geometrical acceptance 
was required to make a hit in the corresponding slab that can be used 
later to compute a value of $m^{2}$ of the particle. 
The efficiency was parametrized as a function of slab position 
with respect to the beam and the track momentum. This dependence is 
small but not negligible. 
A constant $2\%$ over the entire phase space is hence assigned as the systematic uncertainty of the ToF-F reconstruction efficiency.

\subsection{$\pi$-loss} \label{Sec:Syst-PiLoss}

As mentioned above, the loss of pions can be regarded as tracks being measured in the TPCs
and aiming towards the \mbox{ToF-F} acceptance 
but not having a recorded hit in the \mbox{ToF-F} due to decay 
and due to absorption or interactions of pions with the detector.
The corrections related to the decay are computed via 
the precisely known pion decay which should be model independent.
Hence, when varying the number of requested measured points in the \mbox{MTPCs}, 
one does not expect to see differences in the final spectra.
Any variations would represent an uncertainty 
due to imperfections in the description of the spectrometer 
which can introduce a difference in the acceptance and 
reconstruction efficiency (merging track segments between VTPC-2 and MTPC-L/R)
between simulated and real data. 
This uncertainty decreases fast with increasing particle momentum.
Below 2~GeV/$c$ this contribution can be larger than $5\%$ 
but usually is not larger than $1\%$ at higher momenta.

\subsection{Backward track extrapolation} \label{Sec:Syst-BackExtrap}

The uncertainty due to the backward extrapolation procedure is 
induced by the uncertainty on
relative position of the target and TPCs,
as presented in Section~\ref{subsec:TargetPosAlign}.
The main goal of the backward extrapolation is to 
attribute
to each track a specific longitudinal $z$ bin as well as to determine the momentum and polar angle at the surface of the target.
By shifting the target within the  uncertainties on the different coordinates, the number of tracks exiting from each different $(p,\theta,z)$ bin will vary.
This variation is used as bin-by-bin systematic uncertainty on the final spectra due
to the backward extrapolation.
This contribution is the most important one for the most upstream $z$ bin at low polar angle, and for the most downstream bin at high polar angles.
It can range up to $10\%$ in these two specific
phase-space regions.

\subsection{Summary of systematic uncertainties} 
\label{Sec:Syst-Summary}

\begin{figure*}[ht]
    \centering
    \includegraphics[width=1.\textwidth]{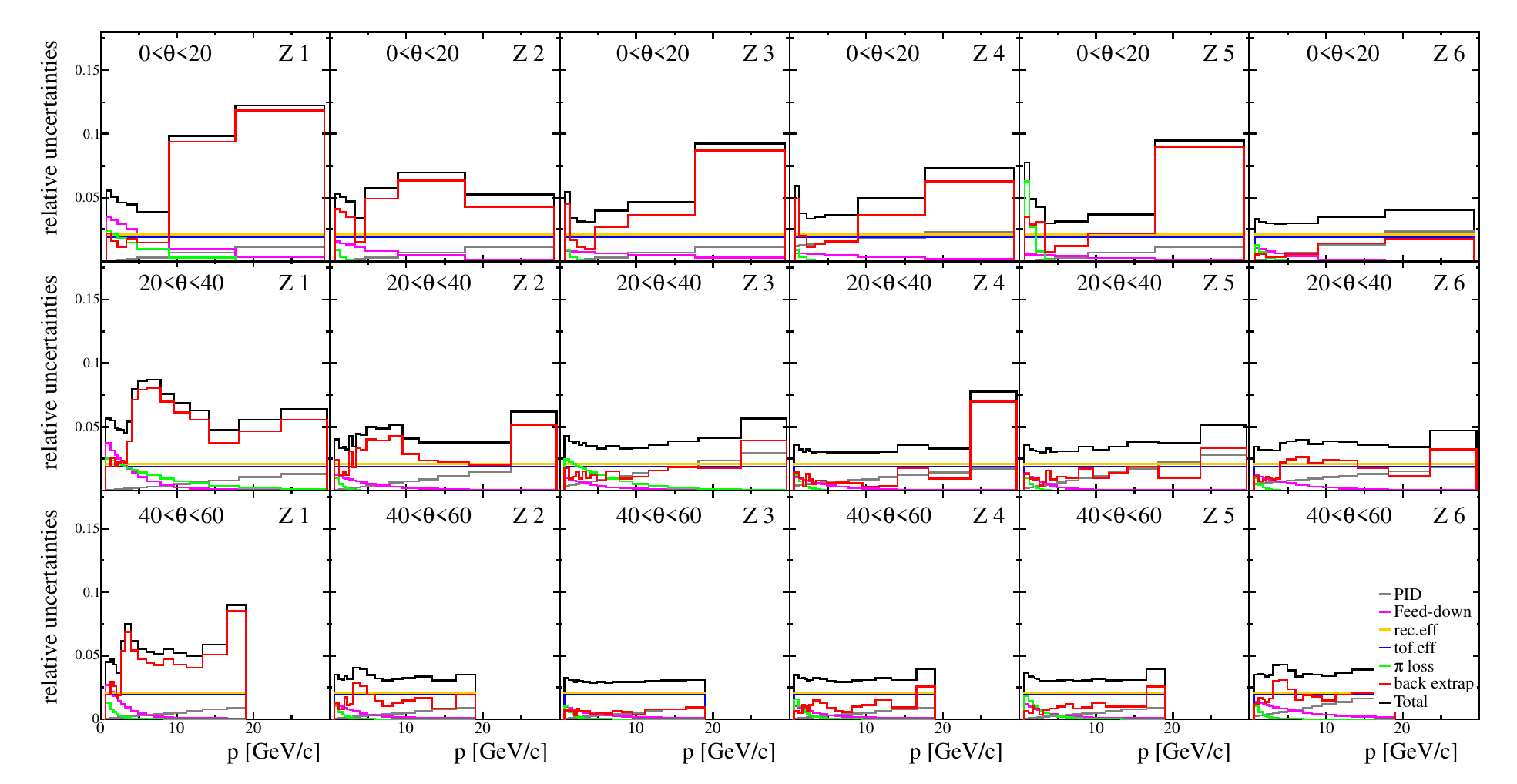}
    \includegraphics[width=1.\textwidth]{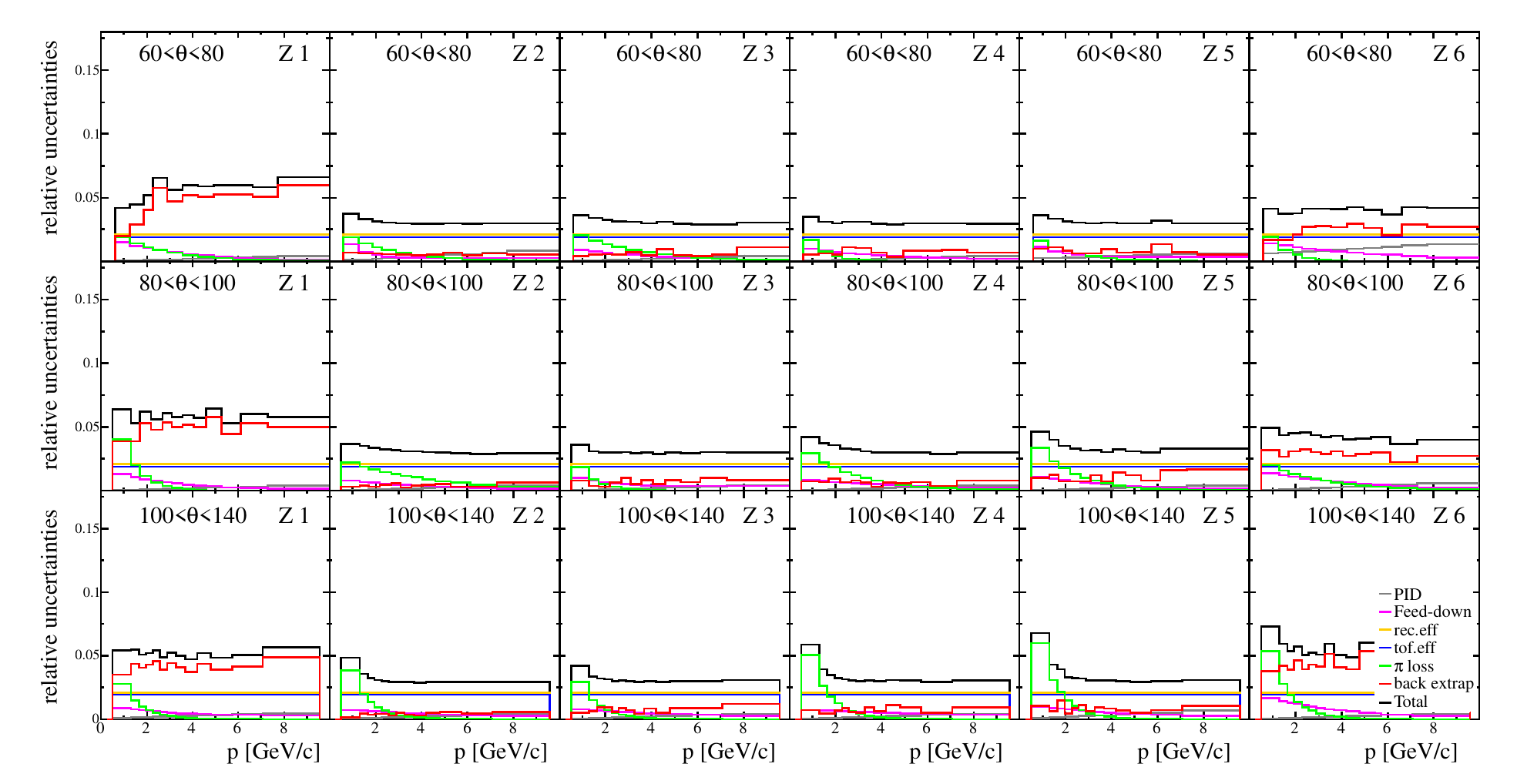}
    \caption{Components of the systematic uncertainties for positively charged pion spectra,
     in the polar angle range from 0 to 140~mrad, and for the six longitudinal bins as a function of momentum.}
    \label{fig:SysPiPosCan1}
\end{figure*}

\begin{figure*}[ht]
    \centering
    \includegraphics[width=1.\textwidth]{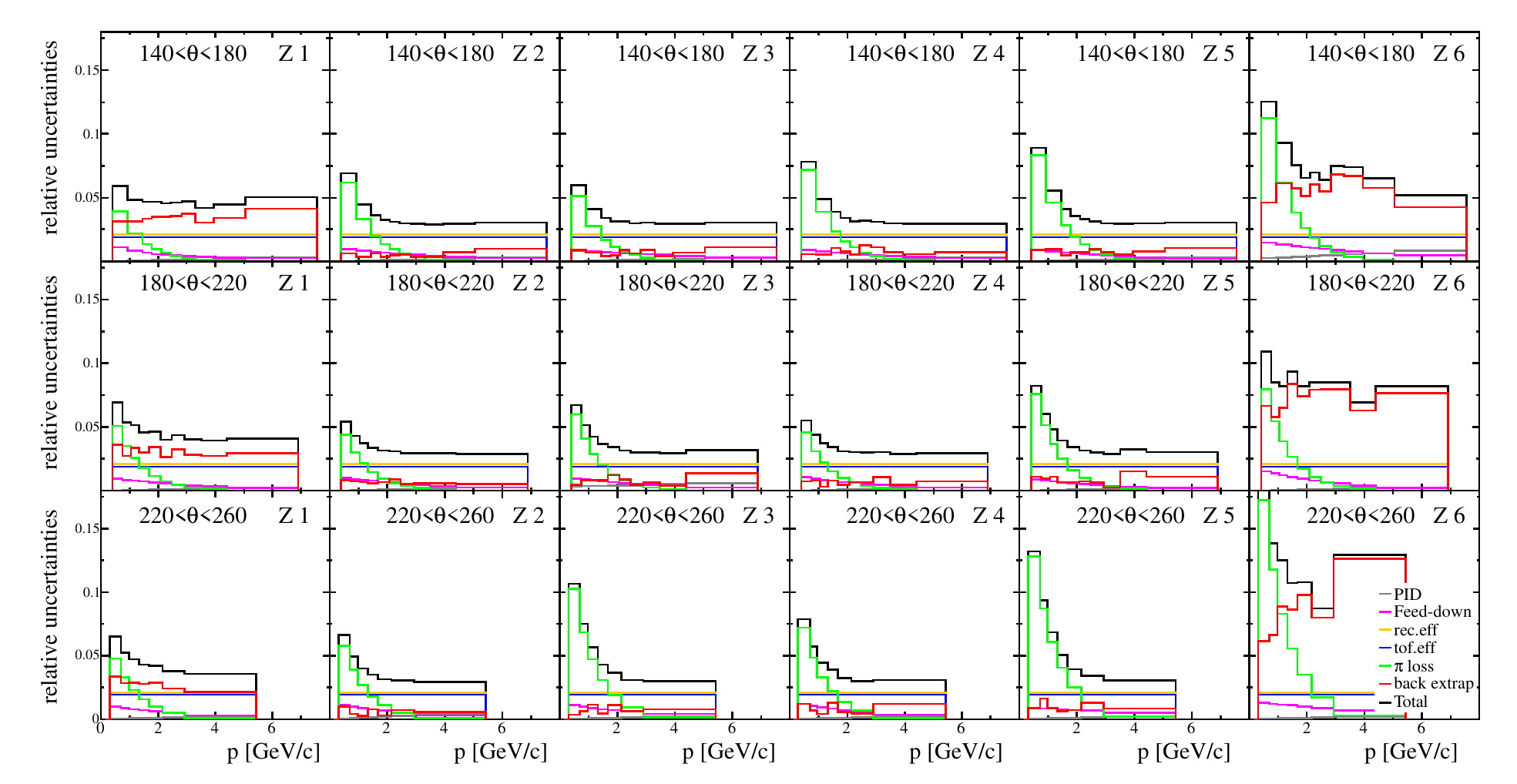}
    \includegraphics[width=1.\textwidth]{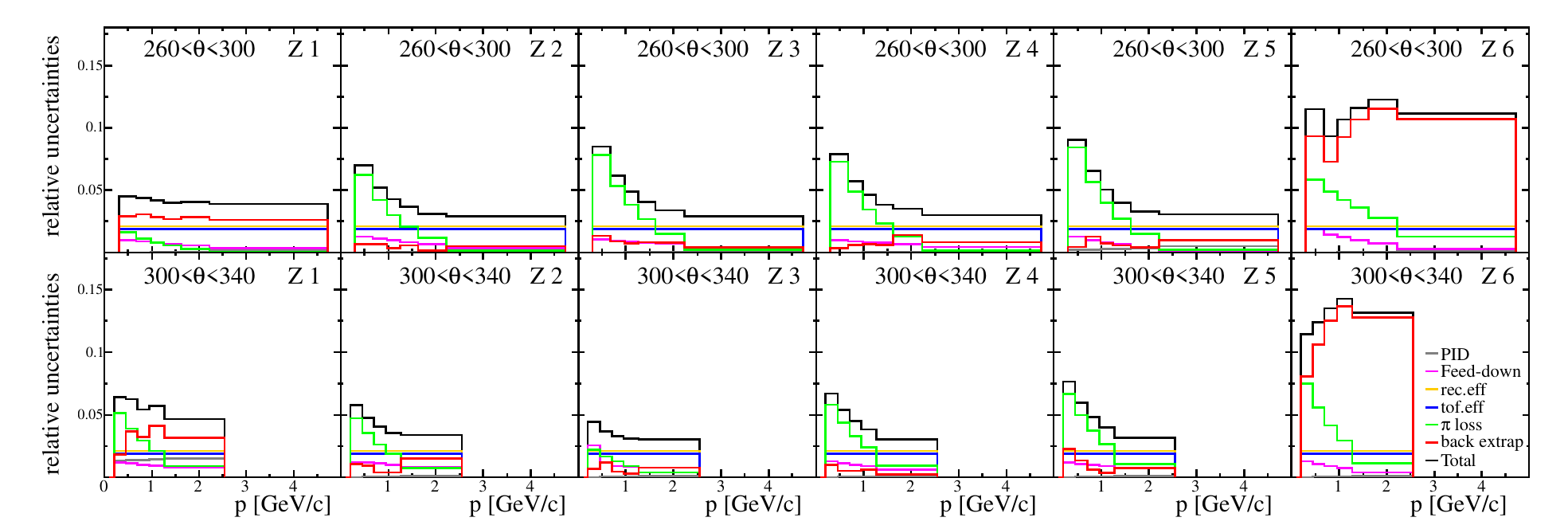}
    \caption{Components of the systematic uncertainties for positively charged pion spectra,
     in the polar angle range from 140 to 340~mrad, and for the six longitudinal bins as a function of momentum.}
    \label{fig:SysPiPosCan2}
\end{figure*}

\begin{figure*}[ht]
    \centering
    \includegraphics[width=1.\textwidth]{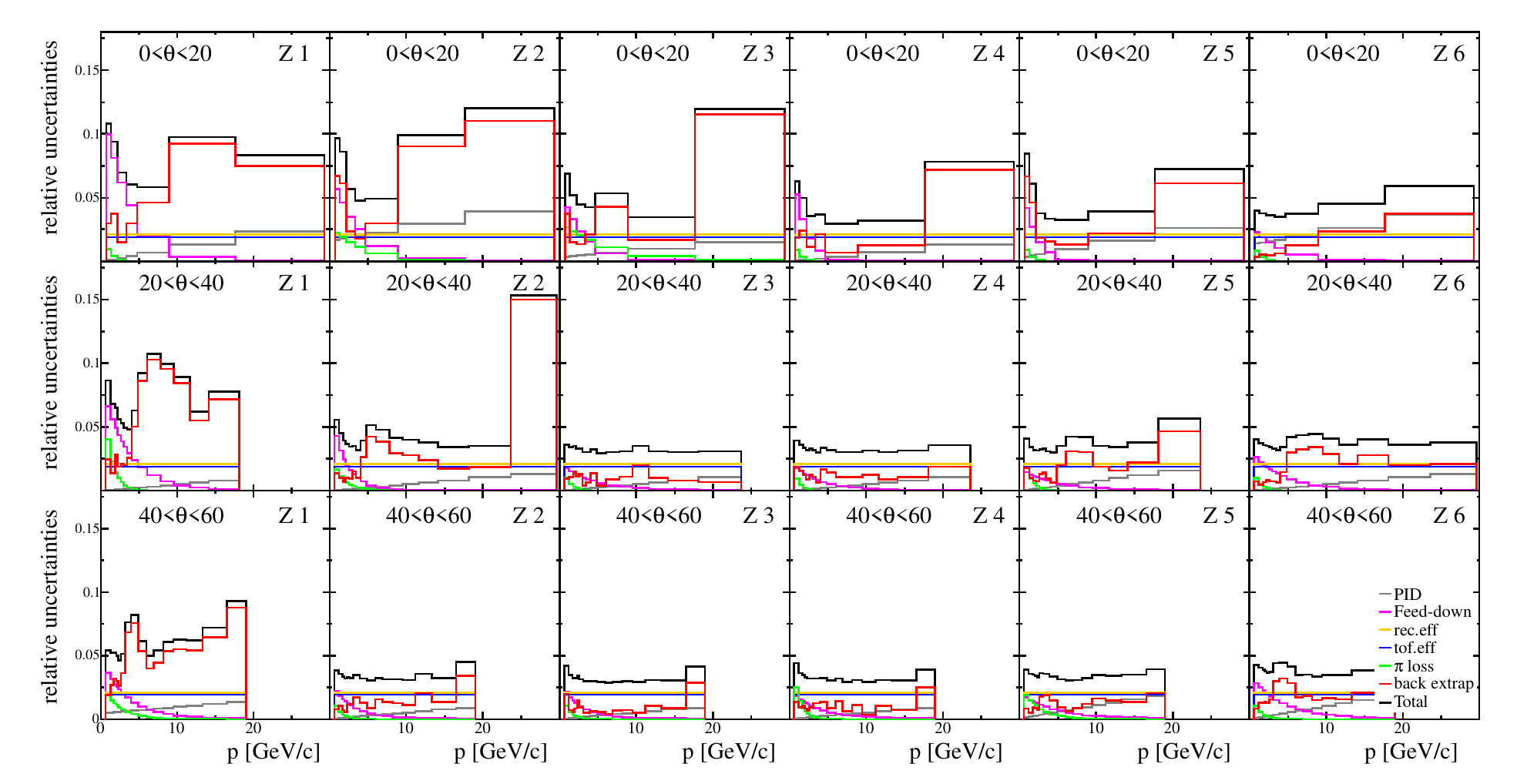}
    \includegraphics[width=1.\textwidth]{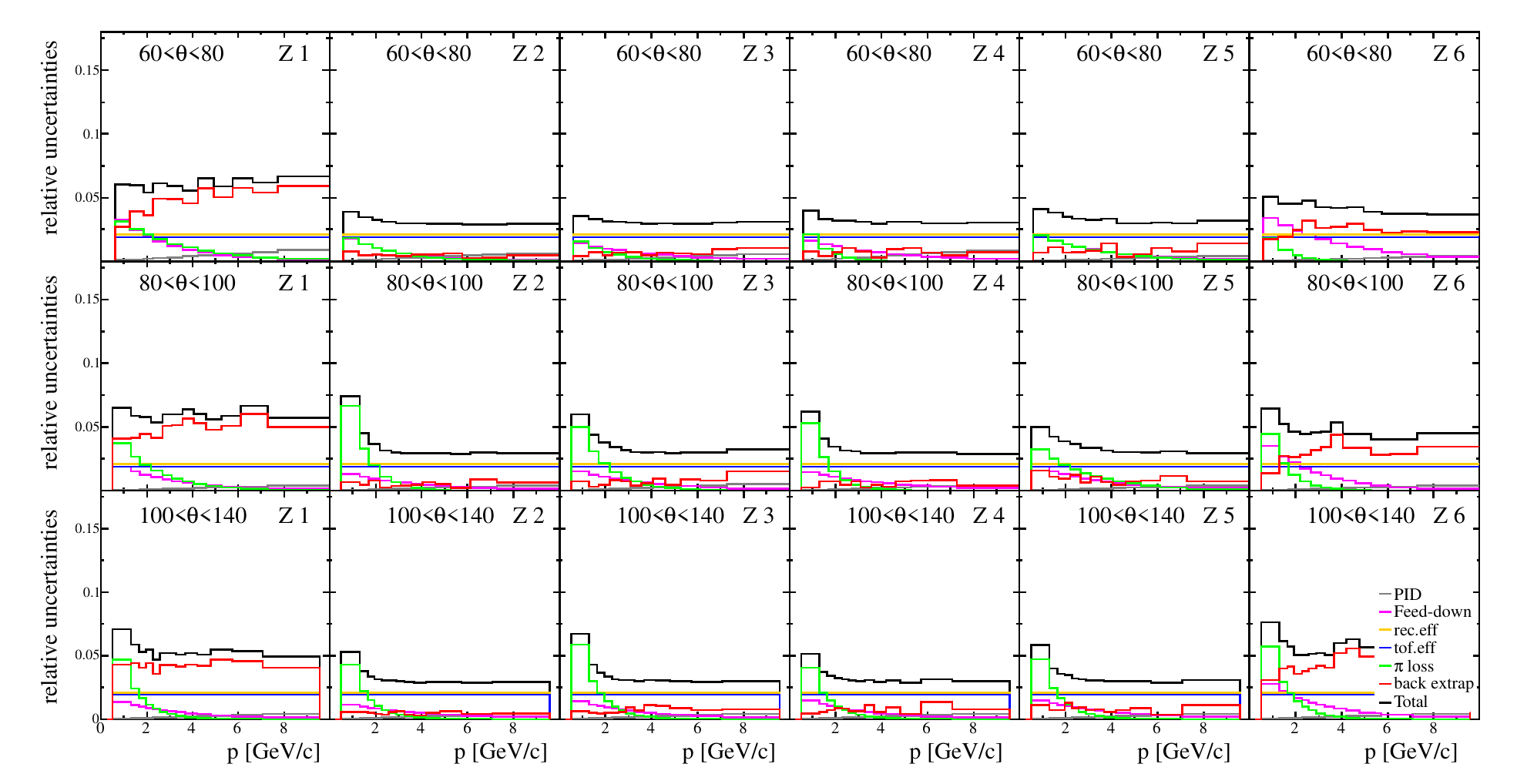}
    \caption{Components of the systematic uncertainties for negatively charged pion spectra,
    in the polar angle range from 0 to 140~mrad, and for the six longitudinal bins as a function of momentum.}
    \label{fig:SysPiNegCan1}
\end{figure*}

\begin{figure*}[ht]
    \centering
    \includegraphics[width=1.\textwidth]{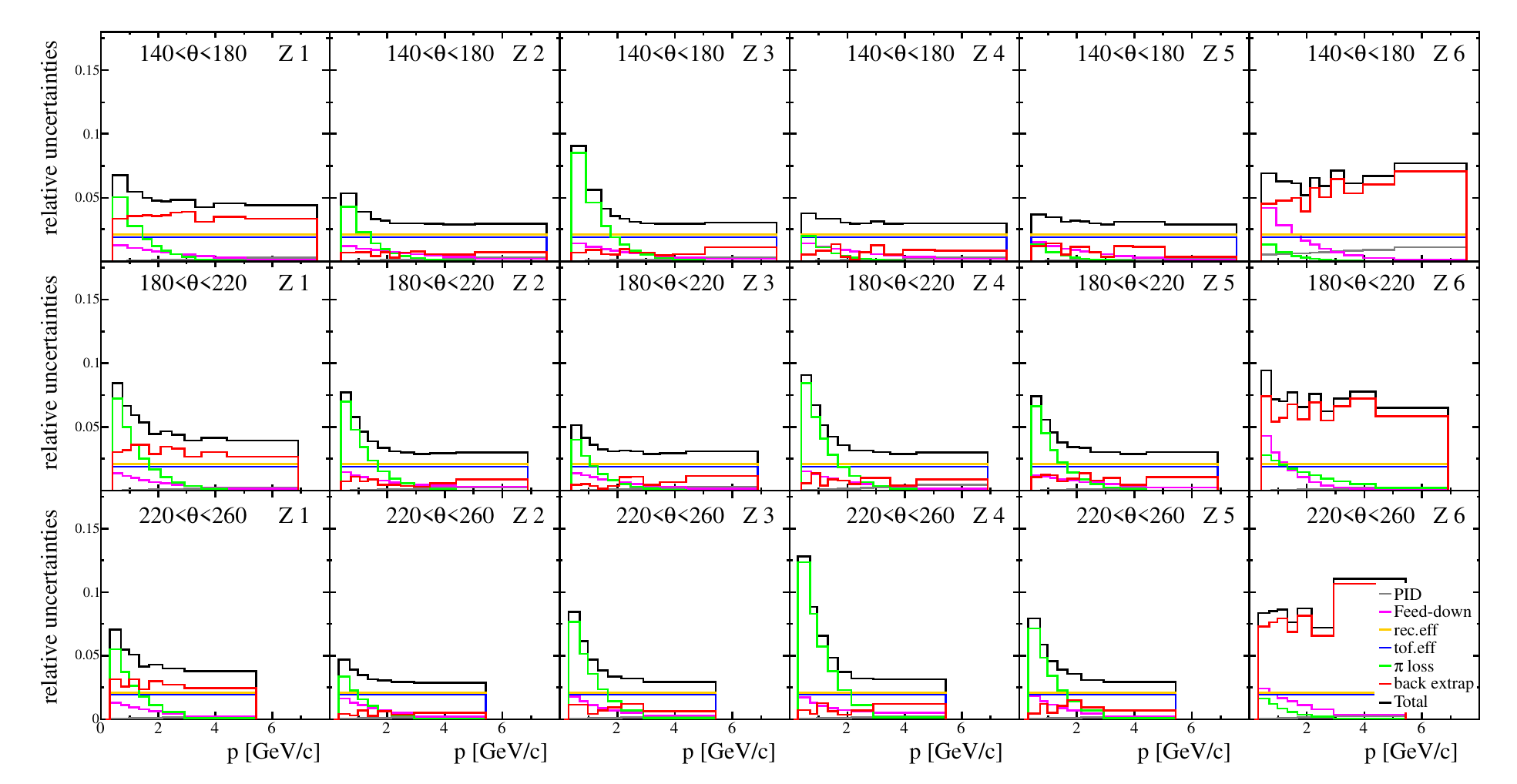}
    \includegraphics[width=1.\textwidth]{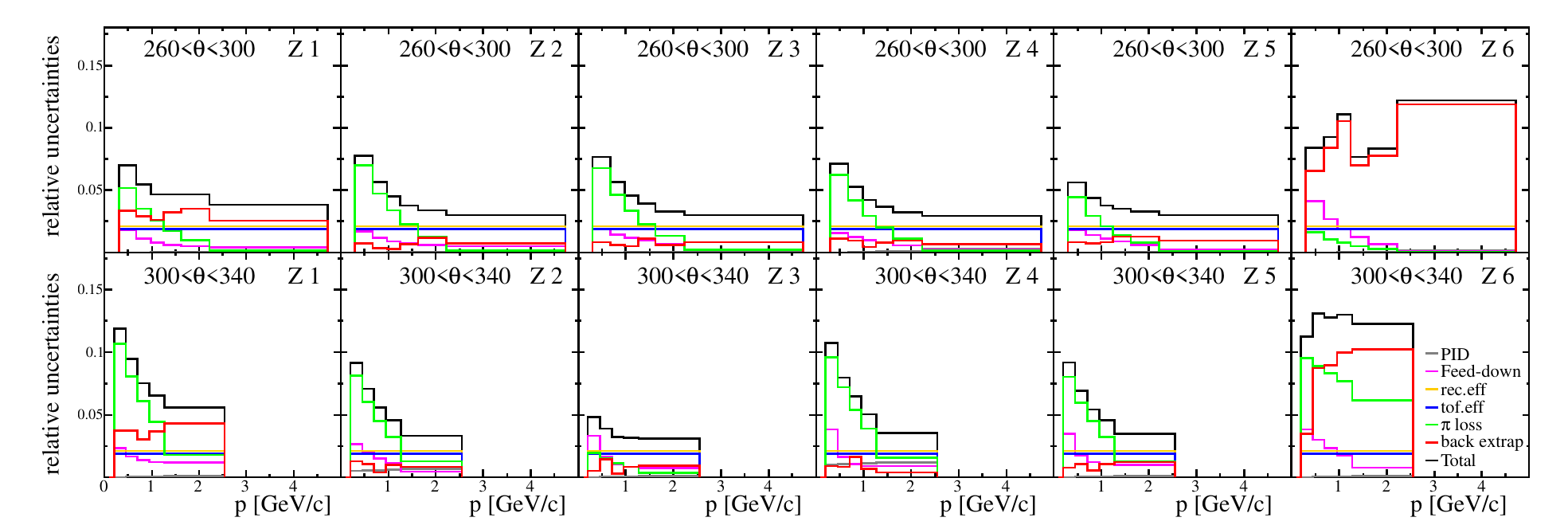}
    \caption{Components of the systematic uncertainties for negatively charged pion spectra,
     in the polar angle range from 140 to 340~mrad, and for the six longitudinal bins as a function of momentum.}
    \label{fig:SysPiNegCan2}
\end{figure*}

The systematic uncertainties are presented in Figs.~\ref{fig:SysPiPosCan1} and~\ref{fig:SysPiPosCan2} for positively charged pions and in Figs.~\ref{fig:SysPiNegCan1} and~\ref{fig:SysPiNegCan2} for negatively charged pions.
They are displayed 
in $z$ and $\theta$ bins as a function of momentum.
The numerical values can be found in Ref.~\cite{Hasler:2039148}.

The momentum and angular resolutions are significantly smaller than the bin
sizes, and bin-to-bin correlations are very small. The only significant
bin-to-bin correlation concerns the first and second longitudinal bins along
the target. These correlations were studied in the context of the
systematic errors on alignment and backward extrapolation.

The dominant contribution for the most upstream and downstream $z$ bins is due to the backward extrapolation.
This can be understood by the fact that an uncertainty of the target position gets translated into a fake track migration between the longitudinal bins at the reconstruction level of the exit point of the tracks from the target surface.
For the central part of the target (longitudinal bins $z1$ to $z5$) this effect gets averaged between the bins.
At low momenta, the uncertainties of the feed-down corrections as well as of the pion loss are significant.
In this region the systematic uncertainty due to particle identification is quite small.
This can be explained by the fact that the $tof-dE/dx$ approach separates well 
the particle species, as seen in Fig.~\ref{fig:FitProjResults}.

\subsection{Statistical uncertainties}
Statistical uncertainties of the final corrected pion spectra receive contributions from 
the finite statistics of the data as well as the simulated events used to obtain the correction factors. 
The dominating contribution is the uncertainty of the normalisation factors returned by the 
maximum likelihood method applied for PID. 
The simulation statistics were higher than the data statistics by more than a factor of 6, hence the related uncertainties are by about a factor of 3 smaller. Added in 
quadrature to the statistical uncertainties of the data, these errors become negligible.

\subsection{Cross-check with the $h^-$ analysis technique}

The primary results obtained with the $tof-dE/dx$ analysis technique
were cross-checked using a complementary $h^-$ analysis method
which was presented in previous publications \cite{pion_paper,thin2009paper}.

The $\pi^-$ differential yields computed with the $h^-$ technique 
are, in general, in good agreement with those obtained with the $tof-dE/dx$ 
approach.
Some small deviations appear only in the upstream part of the target 
for the polar angle bin
$0 < \theta < 20$~mrad and momentum below 7~GeV/$c$,
where the $\pi^-$ differential yields from the $h^-$ analysis 
exceed those obtained with the $tof-dE/dx$ approach
by a few percent, actually remaining at the edge of the estimated 
uncertainties. 
In the other $p-\theta$ bins very good
agreement between the two analysis techniques is observed.
More details on the comparison of $\pi^-$ differential yields 
can be found in Ref.~\cite{Status_Report_2015}.

The agreement between the $h^-$ and $tof-dE/dx$ analysis for the extracted
$\pi^-$ yields gives confidence in the high precision of the results
and in the reliability of the estimate of systematic uncertainties. 



\section{Results and comparison with the hadron production model used by T2K} \label{sec:Results}

\begin{figure*}[ht]
    \centering
    \includegraphics[width=1.\textwidth]{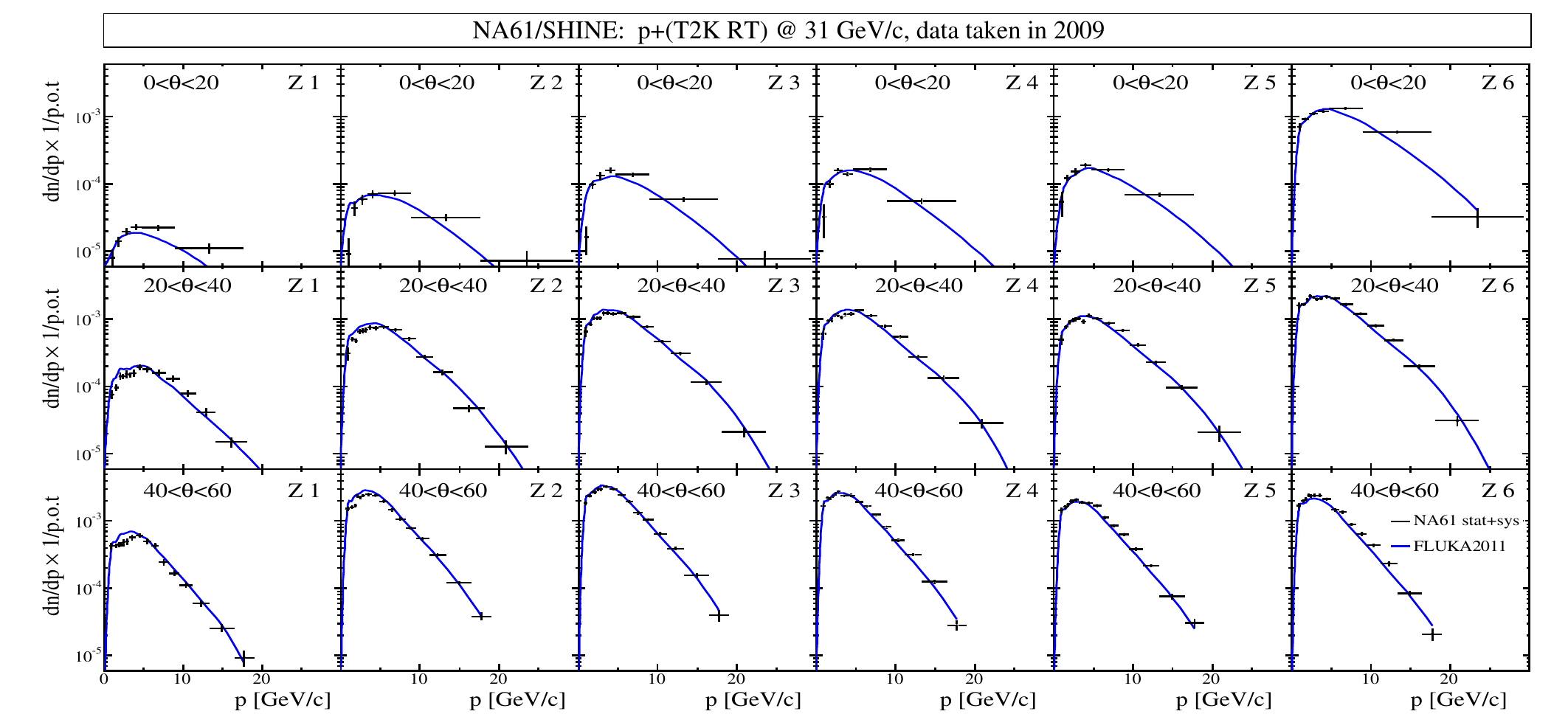}
    \includegraphics[width=1.\textwidth]{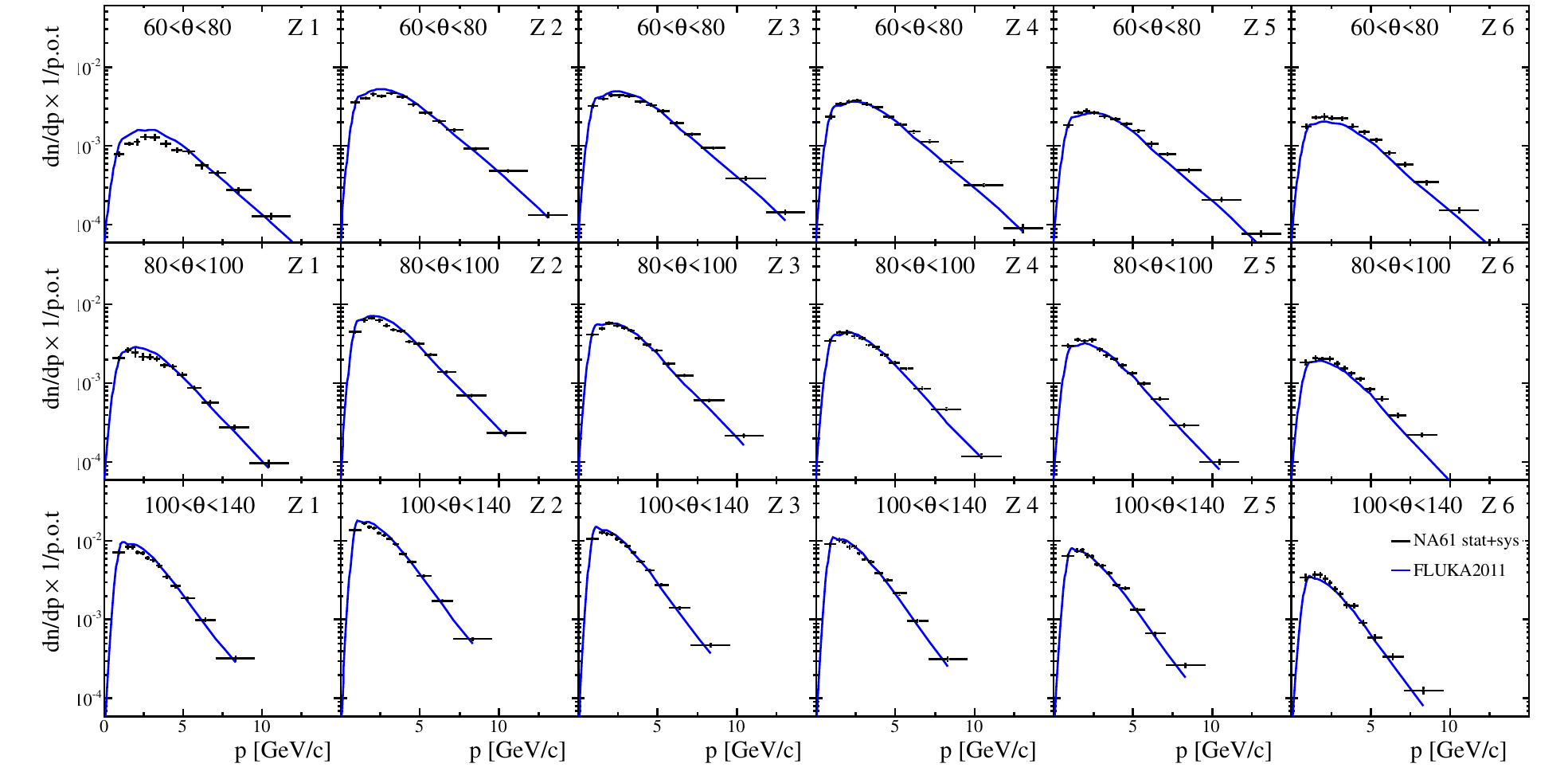}
    \caption{Spectra of positively charged pions at the surface of the T2K replica target,
     in the polar angle range from 0 to 140~mrad, and for the six longitudinal bins
     as a function of momentum.
     The normalization is per proton on target. 
     The prediction from FLUKA2011 is overlaid.}
    \label{fig:ResultsPiPosCan1}
\end{figure*}

\begin{figure*}[ht]
    \centering
    \includegraphics[width=1.\textwidth]{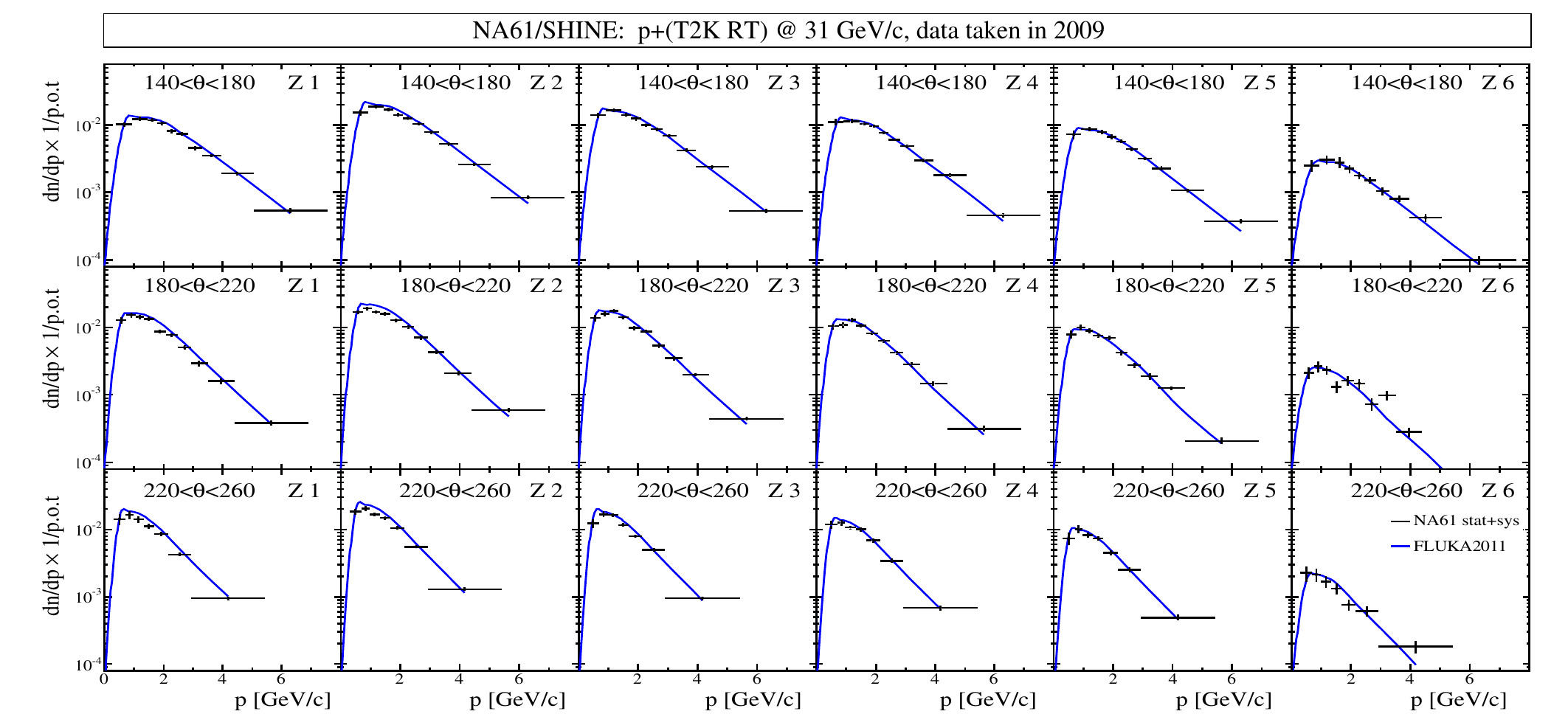}
    \includegraphics[width=1.\textwidth]{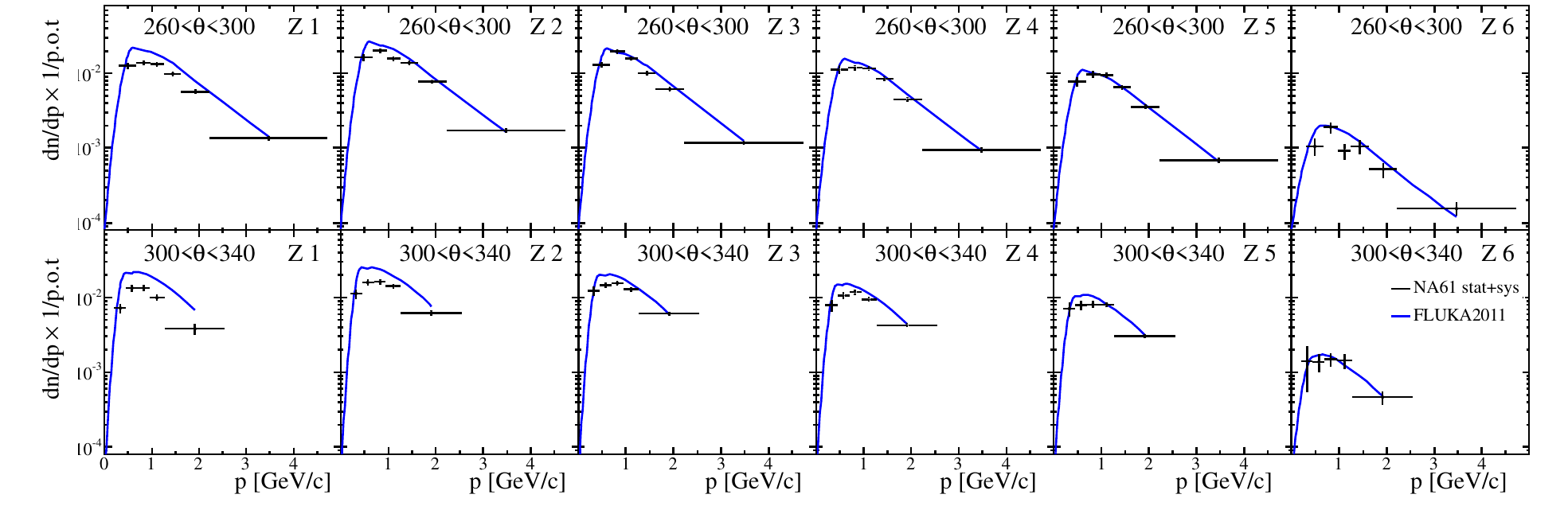}
    \caption{Spectra of positively charged pions at the surface of the T2K replica target, 
     in the polar angle range from 140 to 340~mrad, and for the six longitudinal bins 
     as a function of momentum.
     The normalization is per proton on target. 
     The prediction from FLUKA2011 is overlaid.}
    \label{fig:ResultsPiPosCan2}
\end{figure*}

\begin{figure*}[ht]
    \centering
    \includegraphics[width=1.\textwidth]{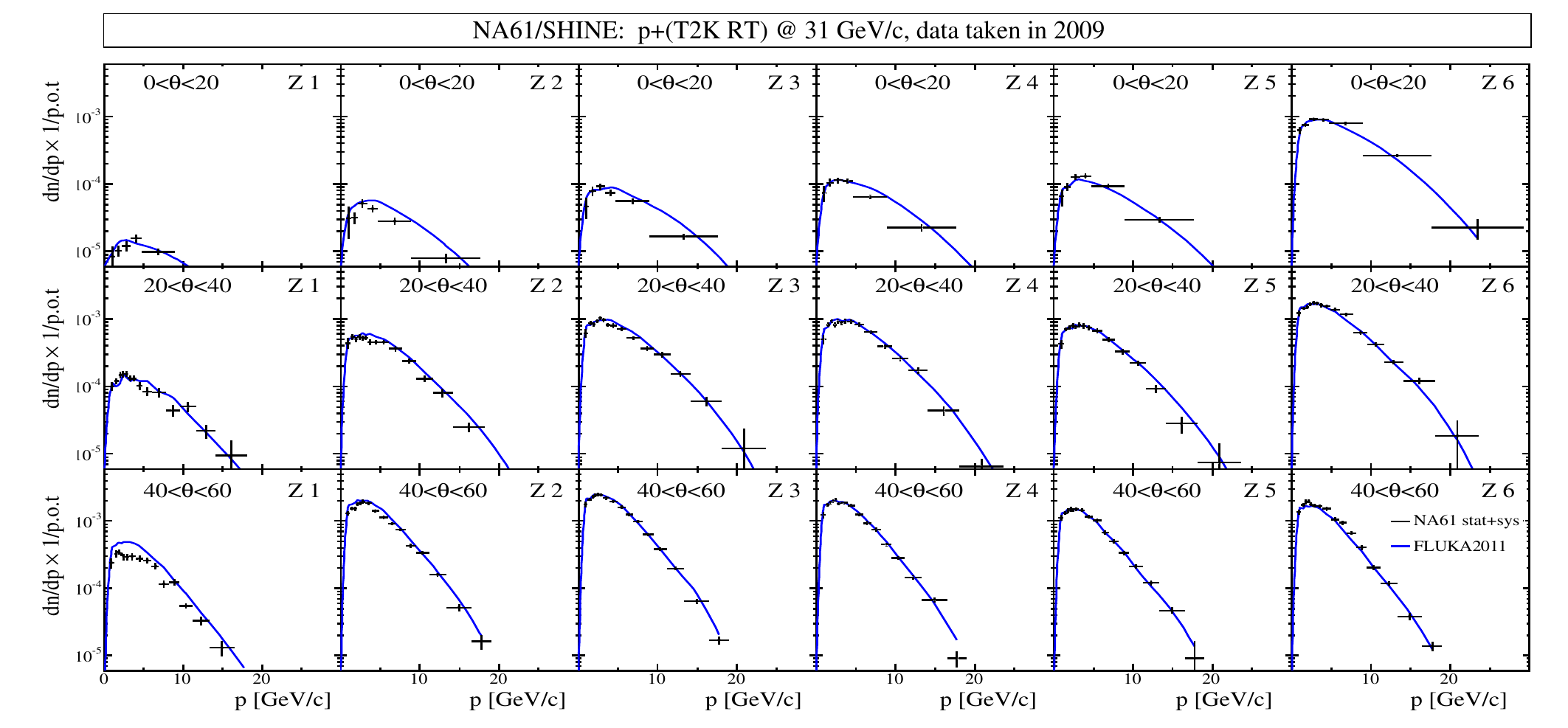}
    \includegraphics[width=1.\textwidth]{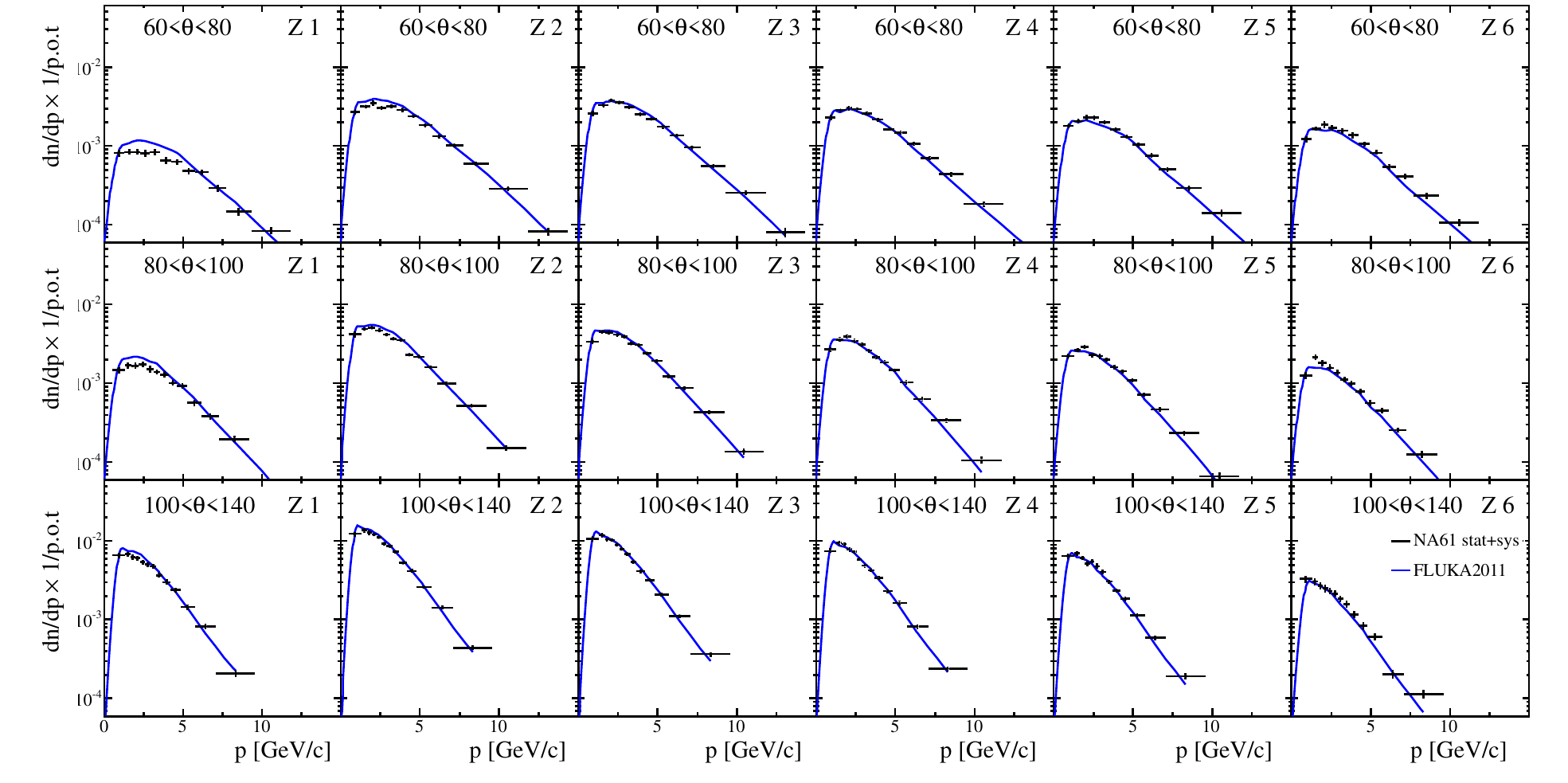}
    \caption{Spectra of negatively charged pions at the surface of the T2K replica target, 
     in the polar angle range from 0 to 140~mrad, and for the six longitudinal bins
     as a function of momentum.
     The normalization is per proton on target. 
     The prediction from FLUKA2011 is overlaid.}
    \label{fig:ResultsPiNegCan1}
\end{figure*}

\begin{figure*}[ht]
    \centering
    \includegraphics[width=1.\textwidth]{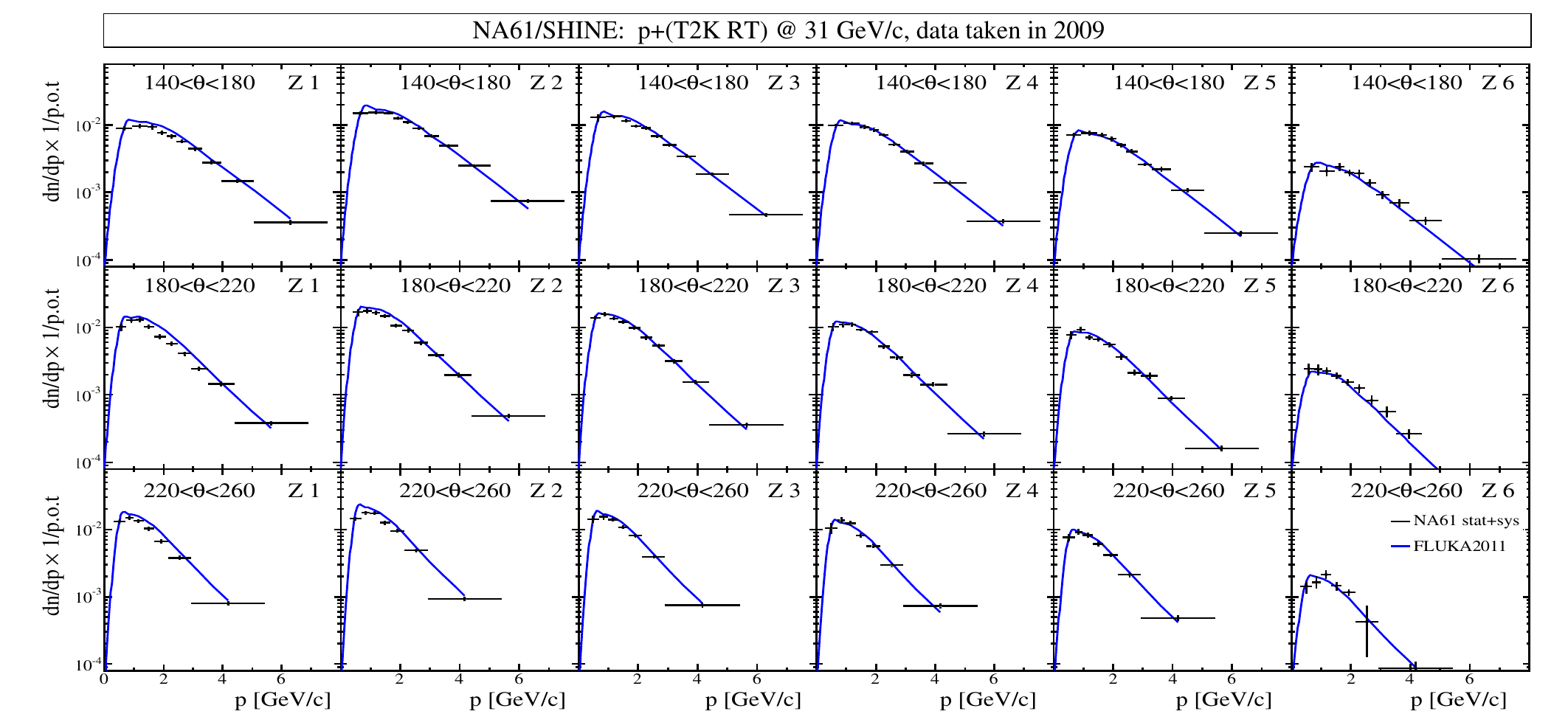}
    \includegraphics[width=1.\textwidth]{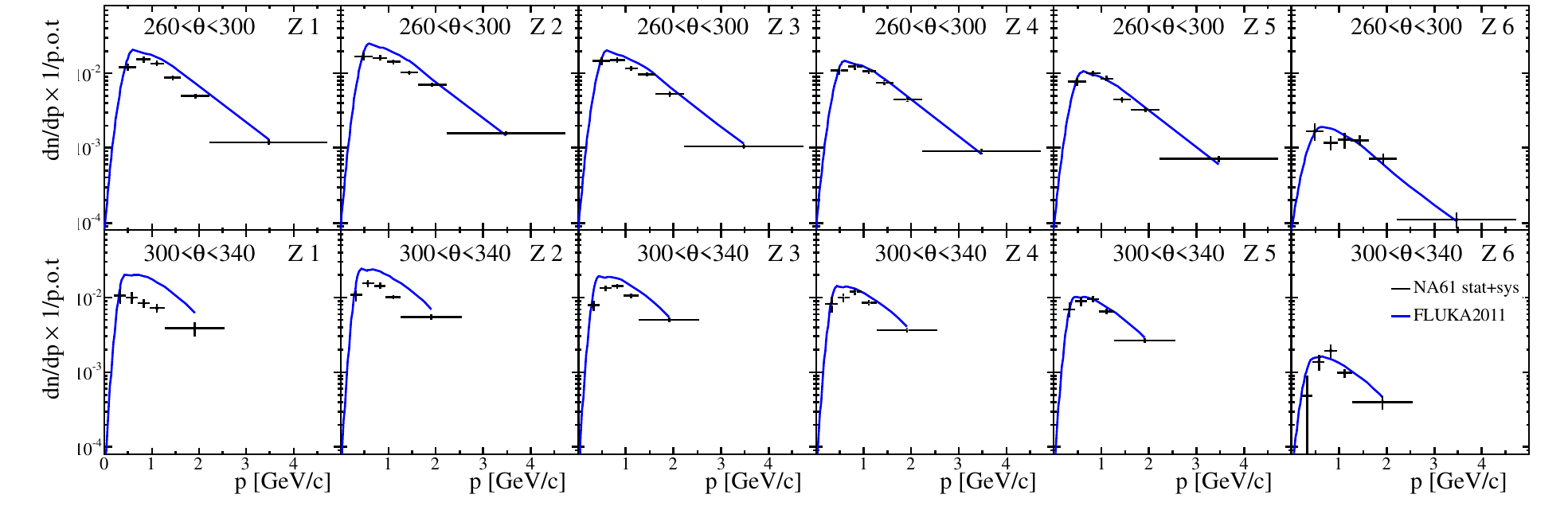}
    \caption{Spectra of negatively charged pions at the surface of the T2K replica target, 
     in the polar angle range from 140 to 340~mrad, and for the six longitudinal bins 
     as a function of momentum.
     The normalization is per proton on target. 
     The prediction from FLUKA2011 is overlaid.}
    \label{fig:ResultsPiNegCan2}
\end{figure*}

Differential multiplicities of positively and negatively charged pions emitted from the T2K replica target exposed to a 31~GeV/$c$ proton beam are presented in Figs.~\ref{fig:ResultsPiPosCan1} to~\ref{fig:ResultsPiNegCan2} 
and in tabular form in Ref.~\cite{EDMS_results}.
The normalization is done to the number of protons on target.



\label{subsec:ComparisonsModelThinTarget}

To compare the 
replica target results with model predictions and with the results of 
thin target measurements~\cite{thin2009paper}, a procedure developed by the T2K beam group is used.
A complete description of this procedure can be found in Ref.\,\cite{T2Kflux}.
This procedure is based on the re-weighting of the model predictions
with experimental measurements\footnote{Re-weighting is performed by assigning a multiplicative positive real
number to each simulated particle, and if appropriate its interaction or
decay products, in such a way that its entry in histograms, probability
functions and flux calculations is multiplied by the product of all the
weights it has received during the simulation.  The weights are
calculated in such a way that the Monte-Carlo simulation reproduces an
input data set.}.
While running a Monte-Carlo simulation of the T2K neutrino flux predictions, the interactions of the incident protons with the 90~cm long graphite target and all subsequent processes leading to the creation of a neutrino are recorded.
By considering only the particles exiting the target surface and re-weighting only the interactions occurring inside the target, it is possible to directly use this procedure to compare model predictions with the T2K replica target measurements.
Furthermore, the interactions taking place inside the target can be constrained by 
the available hadron 
production
data. 
Those data are predominantly the \NASixtyOne measurements. Other external data are used as well 
as extrapolations to lower 
incident hadron energies~\cite{T2Kflux} in order to constrain re-interactions inside the target.
The constraint is done by re-weighting the produced Monte-Carlo particles based on two physics quantities:
\begin{enumerate}[(i)]
    \item the hadron differential multiplicities:\\
	at each interaction, the predicted hadron spectra normalized to mean multiplicities can be re-weighted with respect to the corresponding spectra measured by \NASixtyOne in p+C interactions at 31~GeV/$c$.
    \item the production interaction rate:\\
	at each interaction, the rate at which hadrons interact and produce new particles can be re-weighted to the production cross section extracted from the \NASixtyOne thin target measurements \cite{thin2009paper};
	this production cross section is defined as $\sigma_{prod} = \sigma_{inel} - \sigma_{qe}$, where $\sigma_{inel}$ is the inelastic cross section and 
$\sigma_{qe}$ is the quasi-elastic cross section which 
      characterizes the break up of the carbon nucleus without production of new hadrons.
\end{enumerate}
The weights are computed as the ratio between the measured values and the model predictions.
The procedure developed by the T2K beam group allows to apply the two above-mentioned weighting schemes independently and hence to evaluate the effect of each of them.

\begin{figure*}
    \centering
    \includegraphics[width=0.9\textwidth]{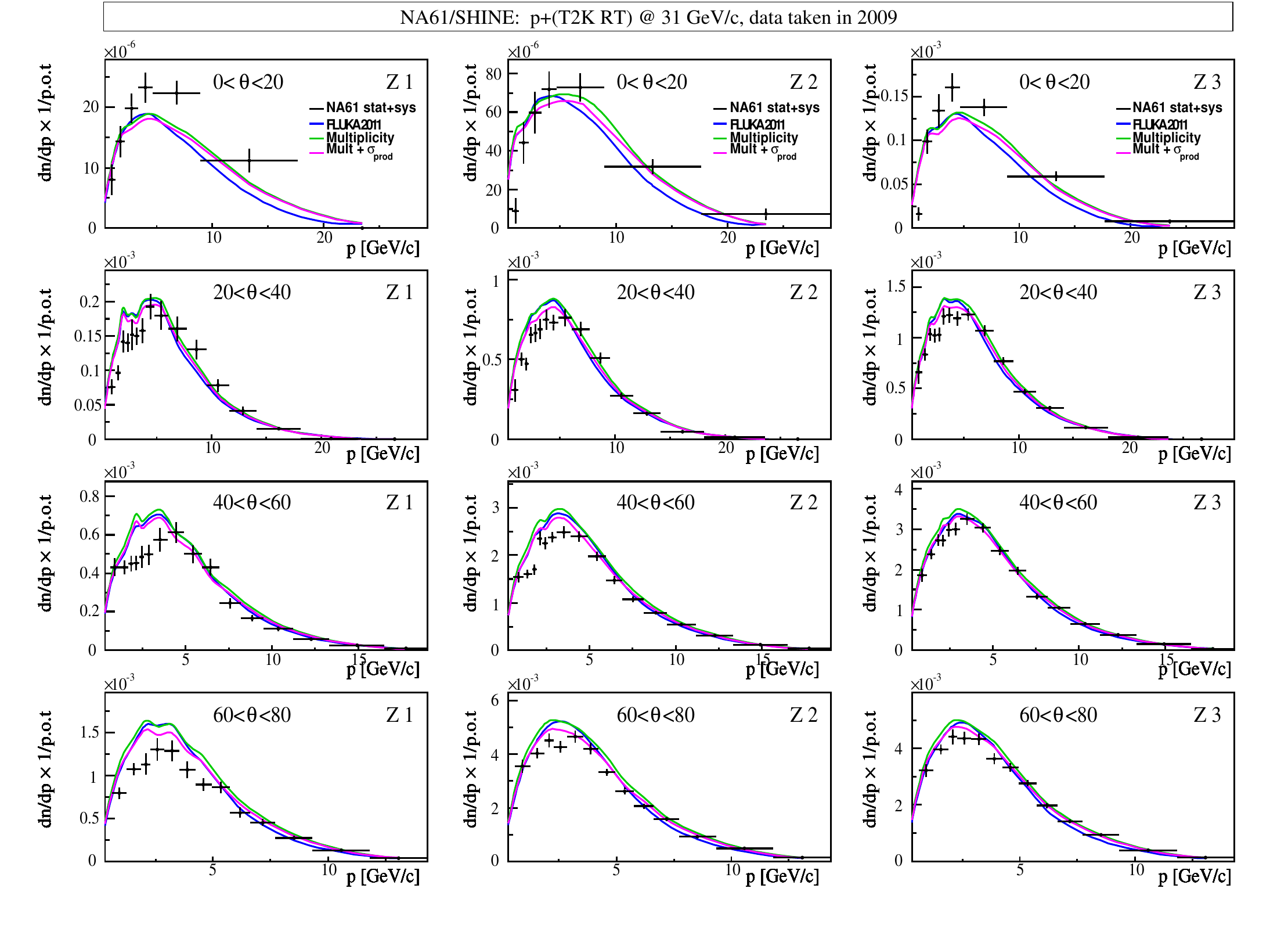}
    \caption{Spectra of positively charged pions overlaid with nominal FLUKA predictions (blue), FLUKA re-weighted for the multiplicities (green) and FLUKA re-weighted for multiplicities and production cross section $\sigma_{prod}$ (magenta) for the three upstream longitudinal bins, Z1--Z3, and in the polar angles between 0 and 80~mrad plotted as a function of momentum.}
    \label{fig:DataFlukaReWeightPos1}
\end{figure*}

\begin{figure*}
    \centering
    \includegraphics[width=0.9\textwidth]{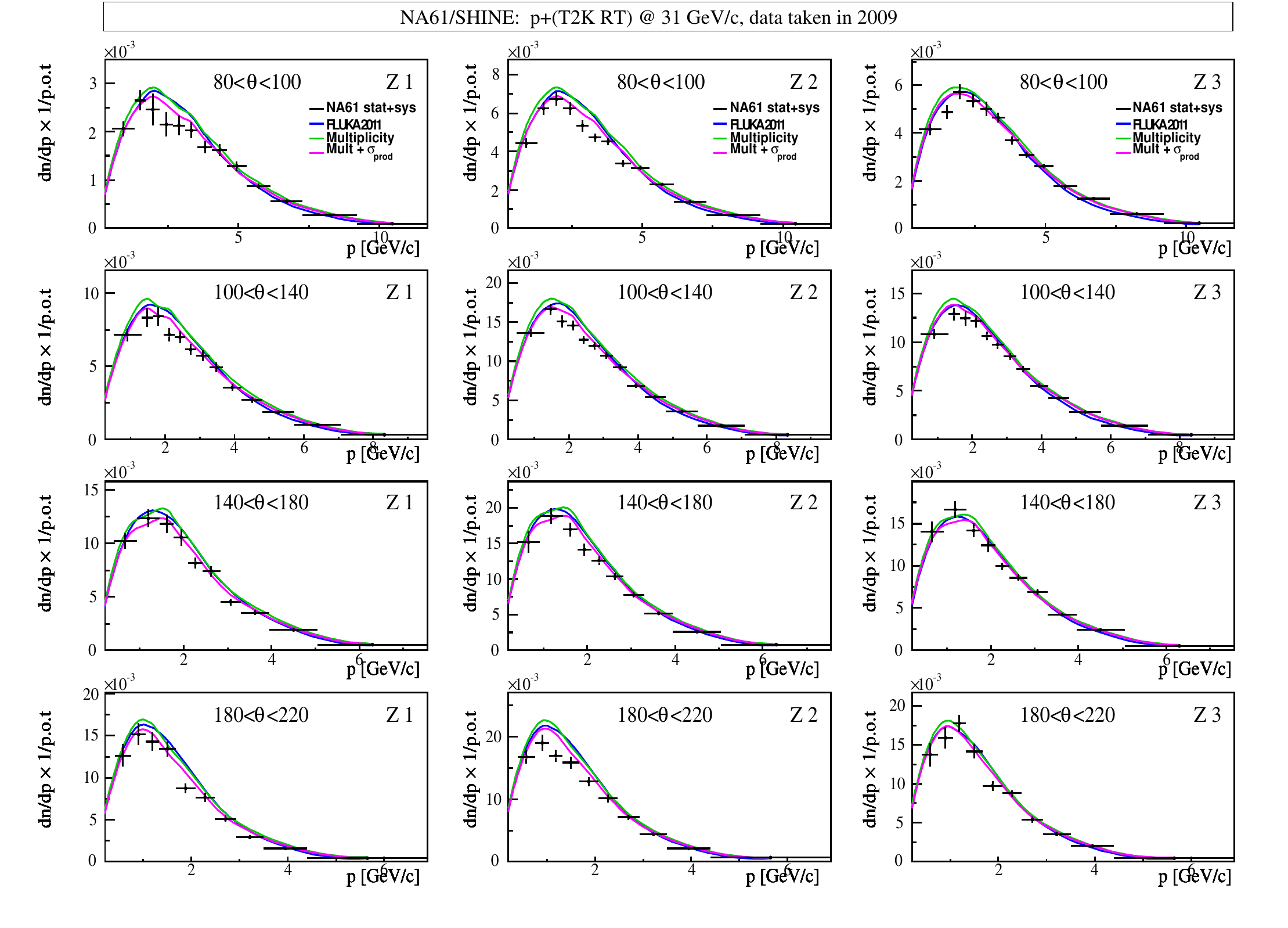}
    \caption{Spectra of positively charged pions overlaid with nominal FLUKA predictions (blue), FLUKA re-weighted for the multiplicities (green) and FLUKA re-weighted for multiplicities and production cross section $\sigma_{prod}$ (magenta) for the three upstream longitudinal bins, Z1--Z3, and in the polar angles between 80 and 220~mrad plotted as a function of momentum.}
    \label{fig:DataFlukaReWeightPos2}
\end{figure*}

\begin{figure*}
    \centering
    \includegraphics[width=0.9\textwidth]{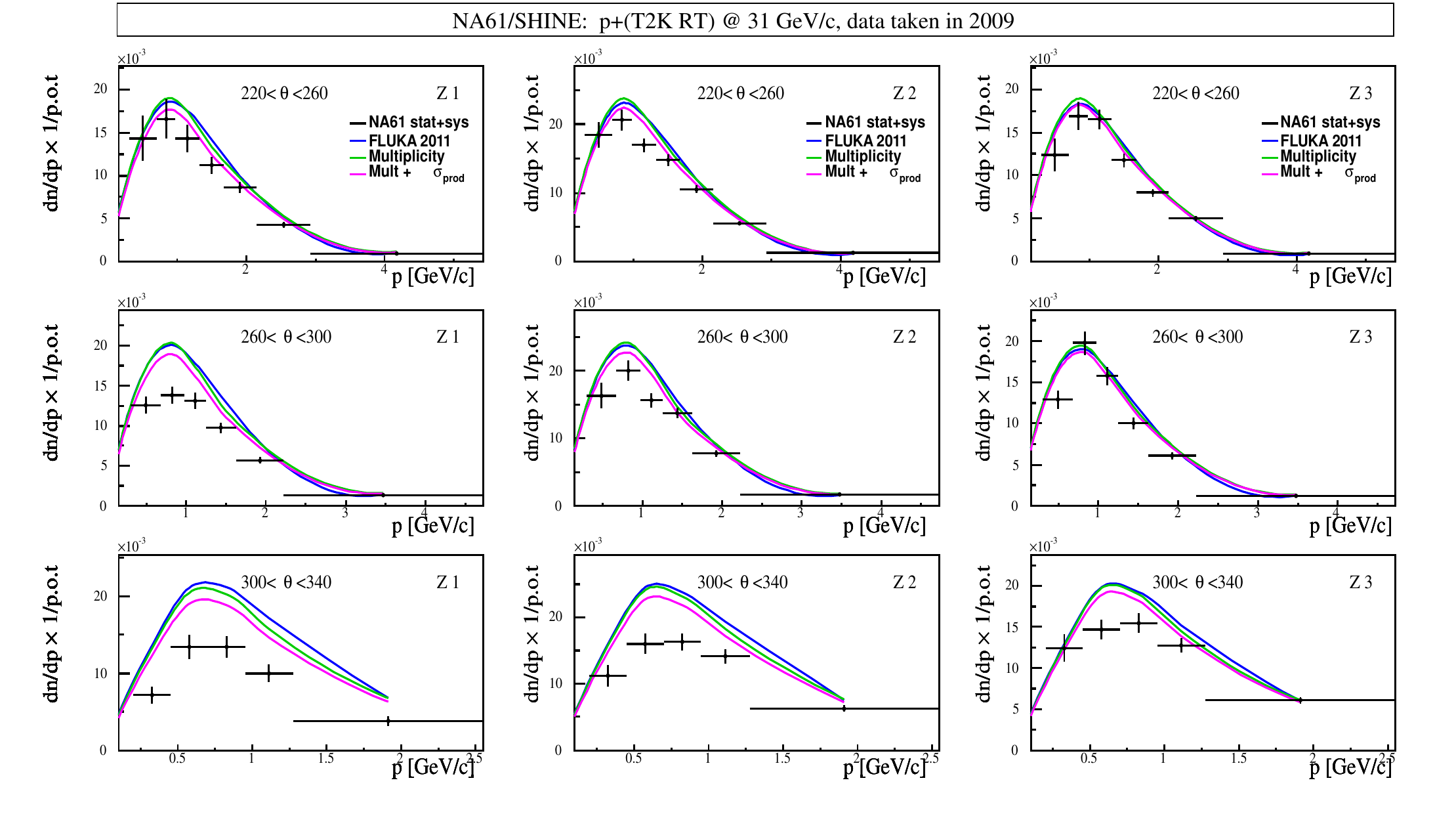}
    \caption{Spectra of positively charged pions overlaid with nominal FLUKA predictions (blue), FLUKA re-weighted for the multiplicities (green) and FLUKA re-weighted for multiplicities and production cross section $\sigma_{prod}$ (magenta) for the three upstream longitudinal bins, Z1--Z3, and in the polar angles between 220 and 340~mrad plotted as a function of momentum.}
    \label{fig:DataFlukaReWeightPos3}
\end{figure*}

\begin{figure*}
    \centering
    \includegraphics[width=0.9\textwidth]{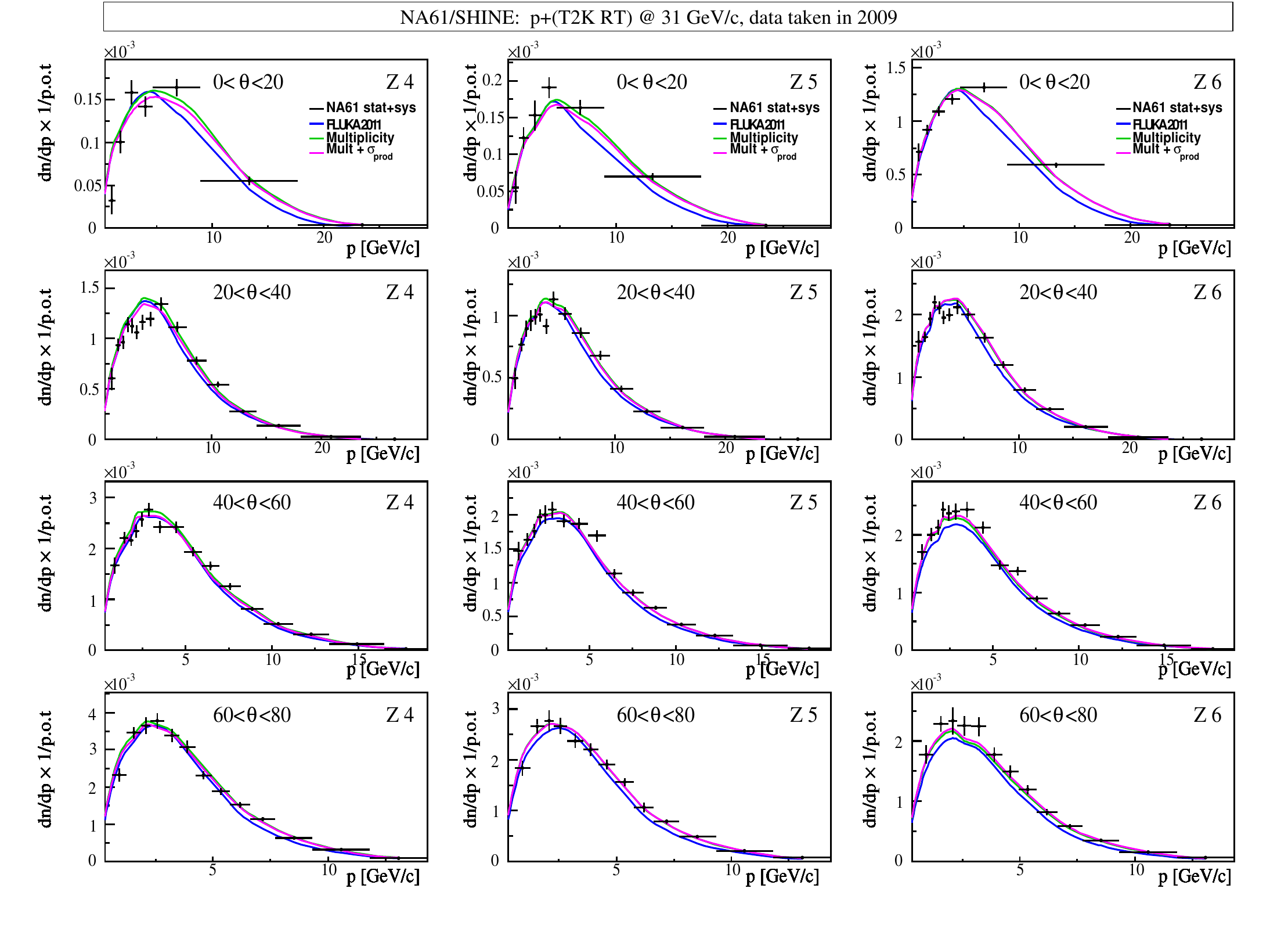}
    \caption{Spectra of positively charged pions overlaid with nominal FLUKA predictions (blue), FLUKA re-weighted for the multiplicities (green) and FLUKA re-weighted for multiplicities and production cross section $\sigma_{prod}$ (magenta) for the three downstream longitudinal bins, Z4--Z6, and in the polar angles between 0 and 80~mrad plotted as a function of momentum.}
    \label{fig:DataFlukaReWeightPos4}
\end{figure*}

\begin{figure*}
    \centering
    \includegraphics[width=0.9\textwidth]{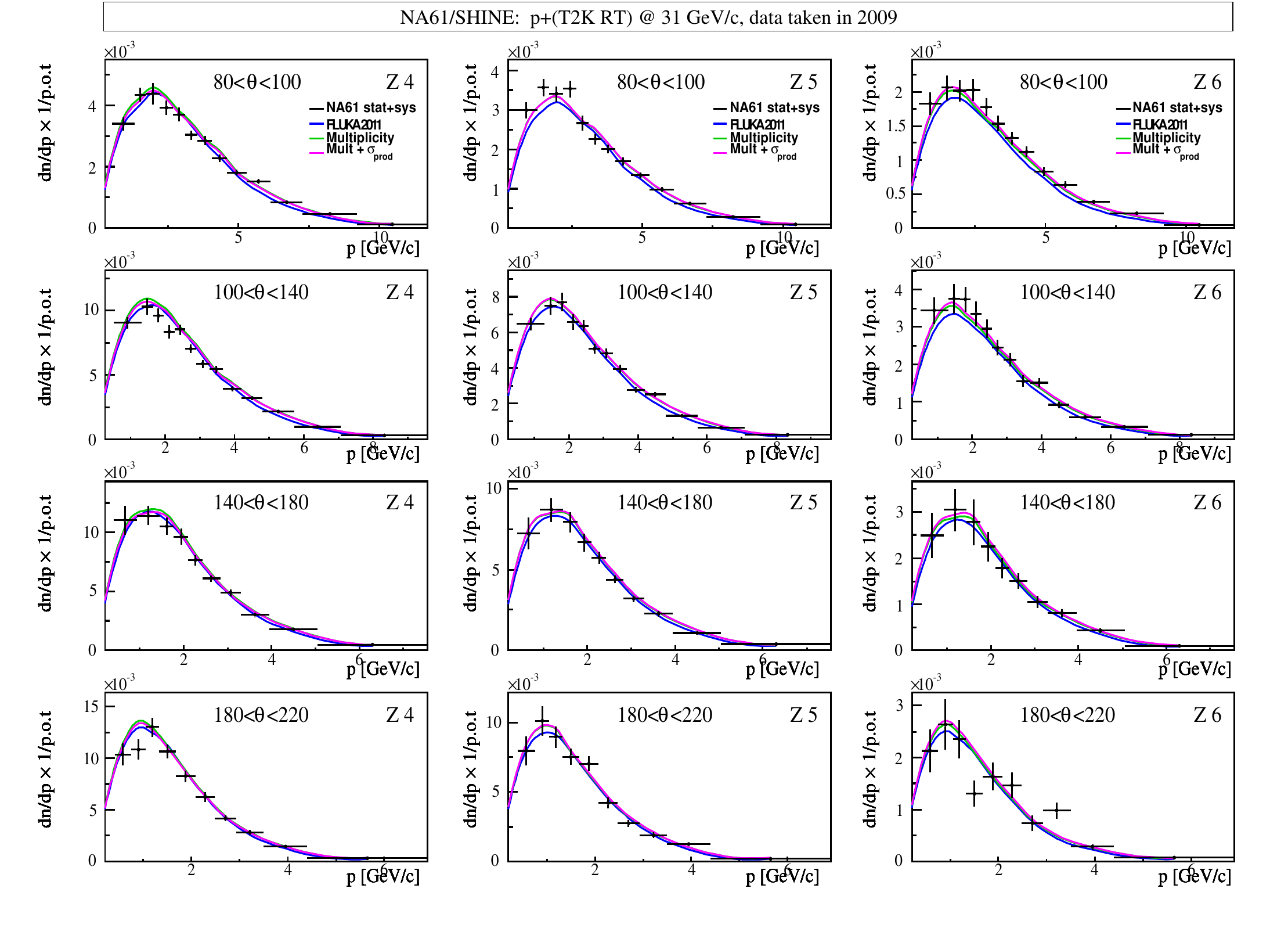}
    \caption{Spectra of positively charged pions overlaid with nominal FLUKA predictions (blue), FLUKA re-weighted for the multiplicities (green) and FLUKA re-weighted for multiplicities and production cross section $\sigma_{prod}$ (magenta) for the three downstream longitudinal bins, Z4--Z6, and in the polar angles between 80 and 220~mrad plotted as a function of momentum.}
    \label{fig:DataFlukaReWeightPos5}
\end{figure*}

\begin{figure*}
    \centering
    \includegraphics[width=0.9\textwidth]{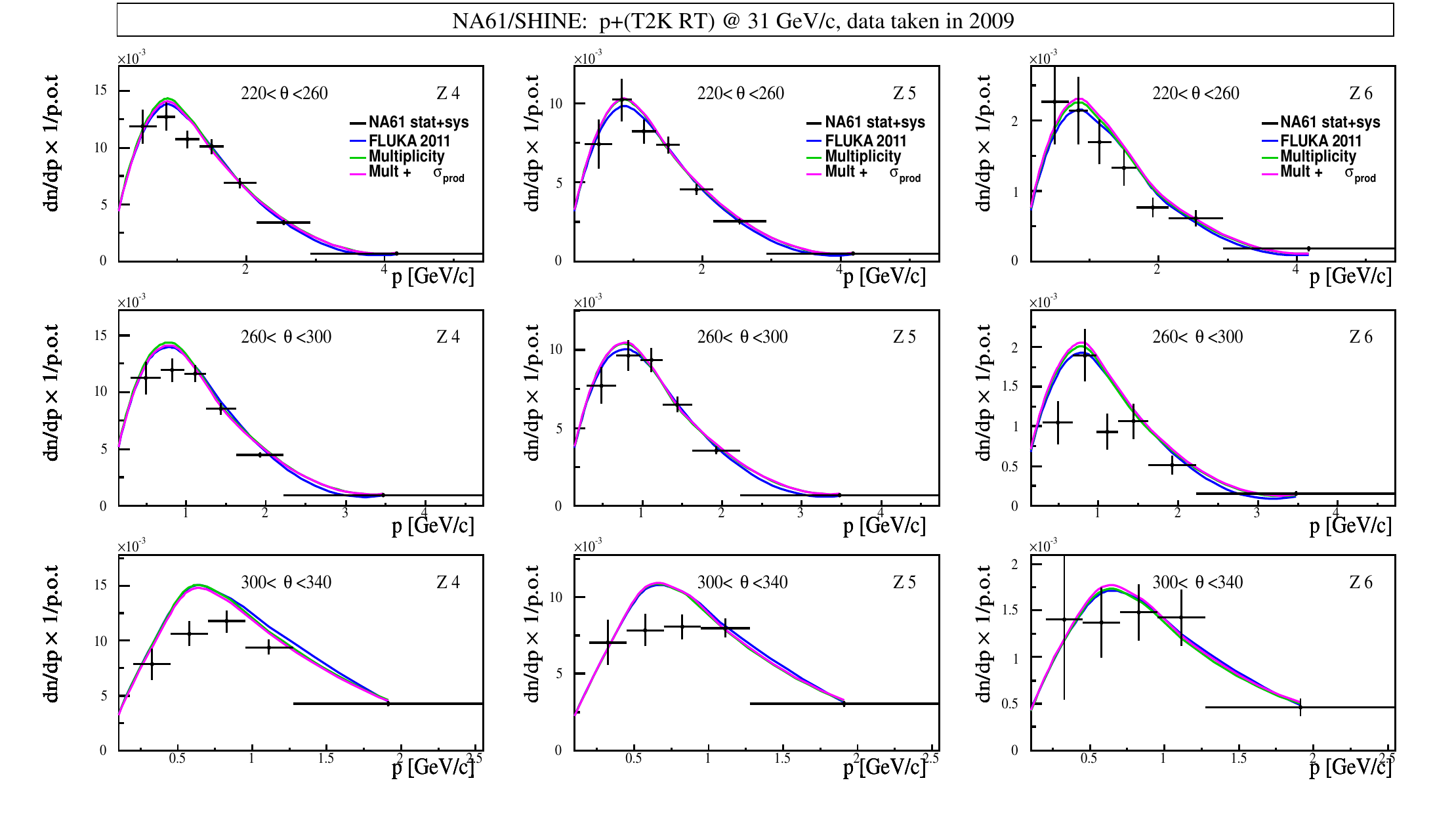}
    \caption{Spectra of positively charged pions overlaid with nominal FLUKA predictions (blue), FLUKA re-weighted for the multiplicities (green) and FLUKA re-weighted for multiplicities and production cross section $\sigma_{prod}$ (magenta) for the three downstream longitudinal bins, Z4--Z6, and in the polar angles between 220 and 340~mrad plotted as a function of momentum.}
    \label{fig:DataFlukaReWeightPos6}
\end{figure*}

In Figures~\ref{fig:DataFlukaReWeightPos1} to \ref{fig:DataFlukaReWeightPos6} three comparisons are presented:
\begin{enumerate}[(i)]

    \item the T2K replica target results are compared with the nominal FLUKA \cite{Fluka,Fluka_CERN,Fluka_new} predictions.

    \item the T2K replica target results are compared with the FLUKA predictions re-weighted
     for the hadron multiplicities \cite{thin2009paper}. As can be seen, the weighting of the hadron 
     multiplicities has a small effect which in general does not exceed a few percent.
     This is expected as the FLUKA predictions reproduce rather well the differential
     cross sections of pions measured by \NASixtyOne in p+C interactions at 31~GeV/$c$.
     
    \item the T2K replica target results are compared with the FLUKA predictions
     which are re-weighted for hadron multiplicities and production cross section \cite{thin2009paper}.
    In order to illustrate the sensitivity of the model predictions to the production 
    cross section, 
    a re-weighting procedure was applied such that it effectively decreases by 20 mb
    the production cross section in FLUKA \footnote{At 31~GeV/$c$ the modification
    corresponds to lowering the value of the FLUKA prediction from 241~mb to 221~mb.
    Both values are within two standard deviations agreement with the $\sigma_{prod}$ 
    measured by \NASixtyOne using a thin carbon target: 
    $230.7 \pm 2.7 {\rm (stat)} \pm 1.2 \rm{(det)} ^{+6.3}_{-3.4} {\rm (mod)}$ mb 
    \cite{thin2009paper}.}.
    As shown in Figs \ref{fig:DataFlukaReWeightPos1}-\ref{fig:DataFlukaReWeightPos6},
    this has a visible effect on the pion yields along the target.
    For the upstream part of the replica target and at lower pion momentum, 
    the predictions are thus lowered by 5-10 \%, bringing them closer to 
    the replica target measurements. At higher momenta the predicted
    spectra stay unchanged. 
    For the downstream part of the target, the agreement lies always
    within the uncertainty of the replica target data points.

\end{enumerate}

The  model whereby total cross sections are described by a sum of elastic scattering, inelastic scattering and production cross section, on which FLUKA is based, does not seem to reproduce  well the longitudinal distribution of particles along the replica target.
This point justifies further studies, first with the higher statistics data taken in 2010, and possibly with an upgraded experimental set-up in the future.

\section{Towards T2K flux predictions} \label{sec:T2KFluxPredictions}

\subsection{Monte-Carlo re-weighting with replica target pion spectra}
\label{subsec:MCReweightingLT}

The implementation of the 
T2K replica target results within the T2K neutrino flux prediction can essentially follow the one for the 
thin target re-weighting procedure~\cite{T2Kflux}.
The major modification to be introduced is to save the relevant parameters of the particles exiting the surface of the  target in the simulation process.
This specific information is added to the already existing history chain for each produced neutrino.
Recording these parameters allows, once the full neutrino beam simulation is completed, to come back to each pion exiting the surface of the target and being a parent of neutrinos.
The neutrinos are then assigned a weight which is dependent on the $(p,\theta,z)$ parameters of the pions at the surface of the target and  computed as 
\begin{equation}
    w(p,\theta,z) = N^{meas}_{NA61}(p,\theta,z) / N^{sim}_{T2K}(p,\theta,z) \,,
    \label{eq:PiWeight}
\end{equation}
where $N^{meas}_{NA61}$ are the hadron yields at the surface of 
the target measured by \NASixtyOne
and $N^{sim}_{T2K}$ are the yields of emitted particles at the surface of the target 
calculated within the T2K beam simulation program.
In both cases the yields are normalized to the number of incoming protons.

Other particle species exiting the target surface are re-weighted following the already established 
thin target procedure.
Re-inte\-ractions outside of the target (in the focusing horns and along the beam line) are also tuned with the nominal T2K neutrino flux calculation based on 
the thin target data.

In order to compare, in a consistent way, the neutrino flux prediction re-weighted with the 
thin target procedure and with the 
T2K replica target 
results, it is important to use the same settings for the entire flux computation.
As the weights related to the 
T2K replica target dataset were computed with the \NASixtyOne beam profile, this same beam profile is used to run a full neutrino flux simulation within the T2K neutrino flux simulation code~\cite{T2Kflux}.
Using this code a comparison between the 
thin target and 
T2K replica target 
re-weighting procedures can then be performed.

In order to study  different effects of the 
flux tuning with the thin target procedure, 
the neutrino flux at SK is first re-weighted for the multiplicity only 
and then for the multiplicity and production cross section for proton-Carbon interactions.
For the latter, as in the case of the pion spectra, 
the production cross section is reduced by 20 mb as described in Section~\ref{sec:Results}.

\begin{figure*}[tbh]
    \includegraphics[width=0.5\textwidth]{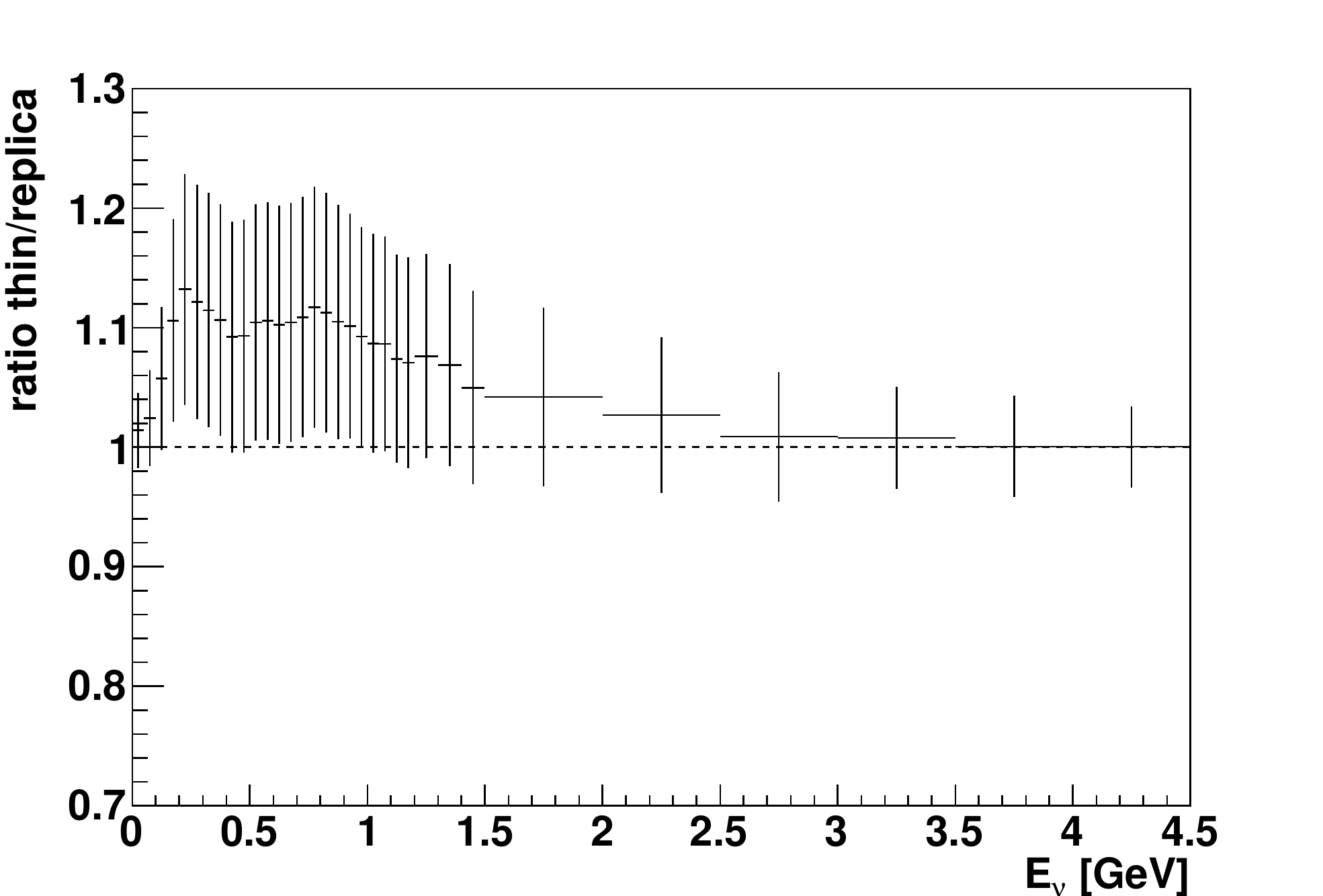}
    \includegraphics[width=0.5\textwidth]{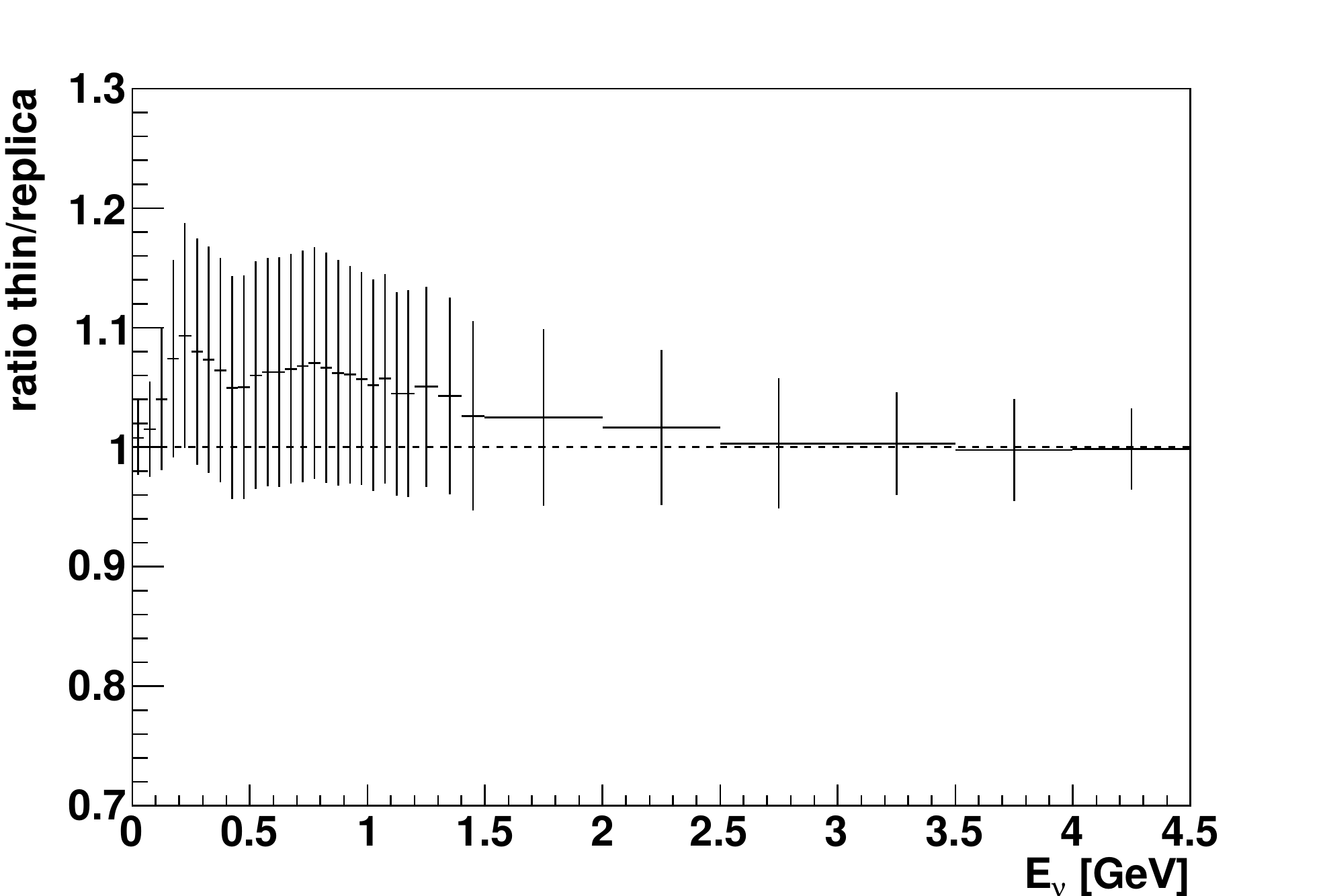}
    \caption{Ratio of the $\nu_{\mu}$ flux at SK re-weighted with the thin target procedure
      to the p+(T2K RT)@31~GeV/$c$ flux.
      For the latter pion spectra presented in this paper have been used in the re-weighting.
      The dominant part of the $\nu_{\mu}$ spectra at high $E_\nu$ comes from kaons and thus
      is not affected by the re-weighting of the pion spectra.
      The left plot shows the ratio calculated using the FLUKA production cross section,
      whereas the FLUKA cross section reduced by 20 mb was used to obtain 
      the ratio presented in the right plot.
      Vertical error bars show the full uncertainties on the ratio which are dominated 
      by systematical uncertainties.
    }
    \label{fig:RatioNumuFluxes}
\end{figure*}

The ratio of $\nu_{\mu}$ flux re-weighted with the thin target procedure over 
the T2K replica target flux at SK is shown in Fig.~\ref{fig:RatioNumuFluxes}.
At low momentum the thin target procedure predicts a 10\% larger value of the flux
which is however consistent with the replica target prediction within one standard deviation.
At high neutrino energy, the ratio tends to one.
This is  due to the fact that within this energy range the dominant part of the $\nu_{\mu}$ spectra comes from kaons 
and 
thus is not affected by the re-weighting procedure on pion spectra.
At lower energy, where the pion parent particles dominate, 
the thin target tuning predicts a higher neutrino flux at SK.
By lowering the proton-carbon production cross section, 
reduced spectra are obtained for the predictions of the number of pions 
exiting the surface of the replica target.
This results in a lower $\nu_{\mu}$ flux prediction  at SK.

The beam profile of \NASixtyOne (space and angular distribution of
incoming protons) were used for the neutrino flux calculations
presented above. The profile of the T2K beam is about 50\% narrower and
better centered. Given the knowledge, on an event-by-event basis, of the
incoming proton parameters, it is possible to re-weight the \NASixtyOne 
data to reproduce more precisely the situation of the T2K beam. Such 
an analysis was performed in  Ref.\,\cite{Hasler:2039148}: it showed  
only a minor impact on the resulting neutrino fluxes that is small compared with
the systematic errors.


\subsection{Propagation of $\pi$ uncertainties to the $\nu_\mu$ flux}
\label{subsec:ErrorAnalysis}

The propagation of pion spectra obtained with the replica target measurements 
to the final neutrino flux predictions is done via the weights computed 
with Eq.\,(\ref{eq:PiWeight}).
In this section we focus on the propagation of uncertainties to the
systematic uncertainties of the neutrino flux.

For each neutrino energy bin $i$, one can write the contribution of the re-weighting factors, for each $(p,\theta,z)$ bin $j$, as a linear combination

\begin{equation}
    E_{\nu_i} = \sum_{j=1}^{N} a_{ij}\cdot \omega_{j} ~,~~ \text{where} ~~~~ \omega_{j}=\frac{n^{NA61}_{j}}{n^{FLUKA}_{j}}
\end{equation}
and the coefficients $a_{ij}$ are related to the contribution of each $(p,\theta,z)$ bin to each neutrino energy bin.
Propagating the T2K replica target uncertainties means propagating the uncertainties on the weights $\omega_{j}$.


The propagation is done via computation of covariance matrices for each source of uncertainties
\begin{equation}
    C_Y = F_X \cdot C_X \cdot F^{T}_X
\end{equation}
with
\begin{equation}
    F_X =
    \begin{pmatrix}
	a_{11} & \ldots & a_{1n} \\
	\vdots & \ddots & \vdots \\
	a_{p1} & \ldots & a_{pn}
    \end{pmatrix}
    \qquad \text{and} \qquad
    C_X =
    \begin{pmatrix}
	\sigma^{2}_{x_1} & \ldots & \sigma_{x_1,x_n} \\
	\sigma_{x_2,x_1} & \ldots & \sigma_{x_2,x_n} \\
	\vdots           & \ddots & \vdots              \\
	\sigma_{x_n,x_1} & \ldots & \sigma^{2}_{x_n}
    \end{pmatrix}.
\end{equation}
The off-diagonal elements $\sigma_{x_i,x_j} , i\neq j$ of the matrix $C_X$ are the correlation coefficients for the 
T2K replica target results.
The diagonal elements of the matrix $C_Y$ give the uncertainty on the final neutrino flux predictions with respect to the propagation of the 
replica target uncertainties.
For instance, the first diagonal element of the matrix $C_Y$ can be written 
with the notation, $Y_1 = f_1 (X)$, as:
\begin{equation}
    \sigma^{2}_{y_1} = \sum_j \left( a_{1j} \right)^{2} \sigma^{2}_{x_j}
    + \sum_{i\neq j} \sum \frac{\partial f_1}{\partial x_i} \frac{\partial f_1}{\partial x_j} \sigma_{x_i,x_j}
~.
\end{equation}

\begin{figure*}[htb]
    \centering
    \includegraphics[width=0.49\textwidth]{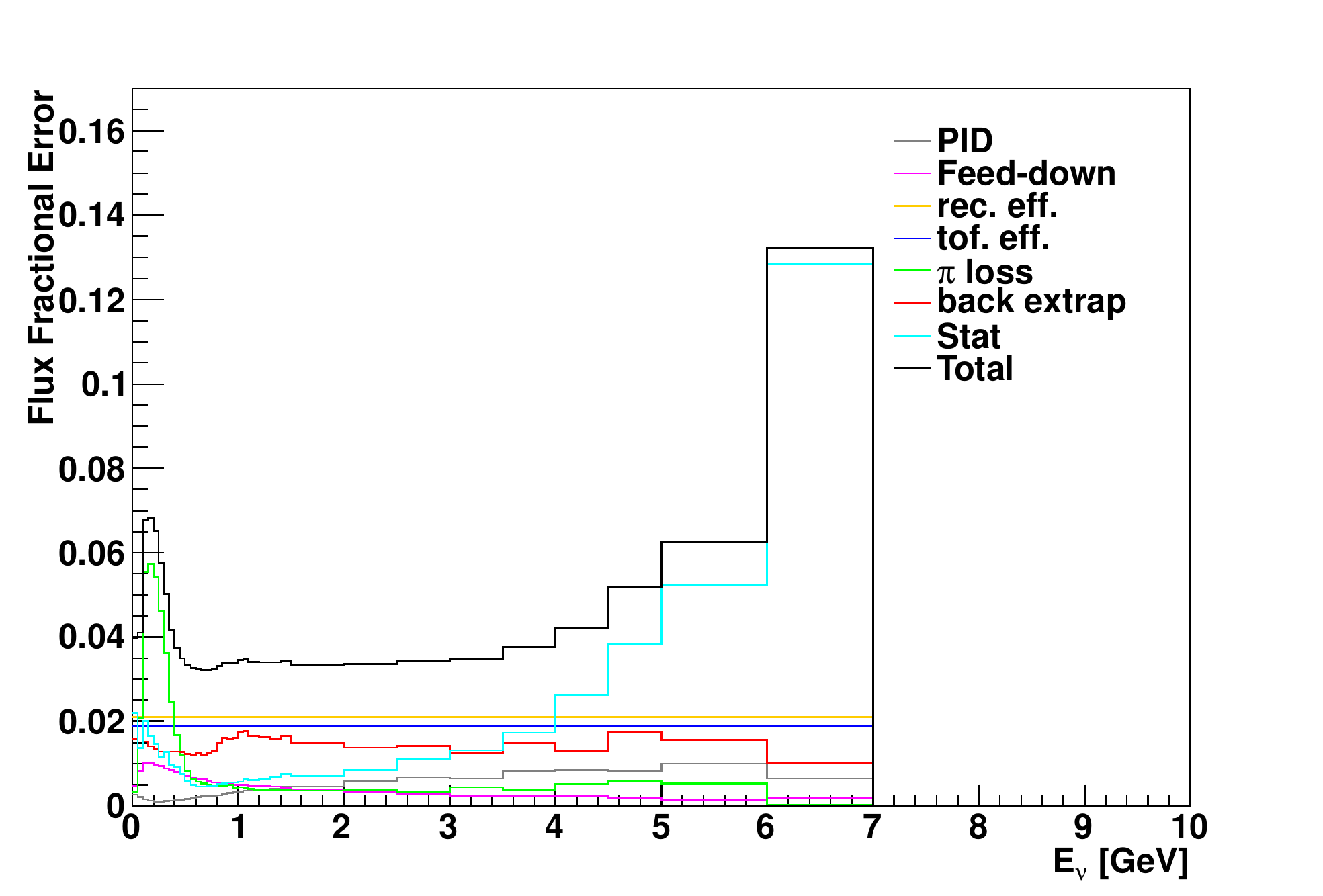}
    \caption{Propagated statistical and systematic uncertainties for the p+(T2K RT)@31~GeV/$c$ replica target results for the $\nu_{\mu}$ flux at SK. The statistical uncertainties are considered as not being correlated between the different \NASixtyOne analysis bins, while for each component of the systematic uncertainties the correlation is considered to be maximal. The different components of the systematic uncertainties are considered as not being correlated between each other.}
    \label{fig:SysPropagationToNumuSK}
\end{figure*}

For each component of the systematic uncertainties related to the 
T2K replica target results, a full correlation between the analysis bins is considered, but no correlation is taken into account between the different contributions to the  systematic uncertainties.
The statistical uncertainties are treated as uncorrelated between the different analysis bins as well as uncorrelated with the different components of the systematic uncertainties.
Figure~\ref{fig:SysPropagationToNumuSK} shows the result of the propagation of the systematic and statistical uncertainties to the $\nu_{\mu}$ flux at SK.
Each line for the systematic uncertainties corresponds to a component described in Section~\ref{sec:Syst}.
It is important to note that these uncertainties correspond only to the component of the $\nu_{\mu}$ flux produced by pions exiting the target surface.
Hence it does not represent the full uncertainties.
As presented in Fig.~\ref{fig:FractionNonTargetSK}, it covers about $87\%$ of the flux at the most probable neutrino energy but only $10\%$ at 4~GeV.
Hence, with the 
T2K replica target results around $87\%$ of the $\nu_{\mu}$ flux at SK can be predicted with $3.5\%$ uncertainty
at the most probable energy, while at 4~GeV only $10\%$ of the flux can be predicted with a $4\%$ uncertainty.
The uncertainty on the remaining part of the flux will have to be estimated from the production of kaons off the surface of the target and/or the re-interactions along the beam line.
These estimations are out of the scope of this paper.

The fraction of the SK neutrino flux that is coming from pions originating from
the long target surface becomes very small at low energies below 400~MeV, this
explains the result that the relative flux uncertainty related to
measurements made in NA61/SHINE is small. Probably the flux at these low
energies originates from muon decays and tertiary pions.

\section{Conclusions} \label{sec:Conclusions}

The data taken in 2009 with 30 GeV protons impinging on a 90~cm long, 2.6~cm diameter carbon rod, were analysed and fully-corrected $\pi^{+}$ and $\pi^{-}$ spectra at the surface of the  target were obtained.
Using the beamline simulation program of T2K,
these spectra were compared with the 
FLUKA2011 prediction re-weighted with cross section measurements 
obtained by \NASixtyOne with a thin target.
A reasonable agreement was found, although 
an even better description 
was obtained
when lowering the production cross section by 9\%.
A method allowing the direct implementation of the T2K replica target results, as well as the propagation of their uncertainties, in the T2K beam simulation was demonstrated.
Further results will be obtained from a higher statistics dataset taken by \NASixtyOne in 2010.

\begin{acknowledgements}

We would like to thank the CERN PH, BE and EN Departments for the
strong support of \NASixtyOne.

This work was supported by
the Hungarian Scientific Research Fund (grants OTKA 68506 and 71989),
the J\'anos Bolyai Research Scholarship of
the Hungarian Academy of Sciences,
the Polish Ministry of Science and Higher Education (grants 667\slash N-CERN\slash2010\slash0, NN\,202\,48\,4339 and NN\,202\,23\,1837),
the Polish National Center for Science (grants~2011\slash03\slash N\slash ST2\slash03691, 2012\slash04\slash M\slash ST2\slash00816 and 
2013\slash11\slash N\slash ST2\slash03879),
the Foundation for Polish Science --- MPD program, co-financed by the European Union within the European Regional Development Fund,
the Federal Agency of Education of the Ministry of Education and Science of the
Russian Federation (SPbSU research grant 11.38.193.2014),
the Russian Academy of Science and the Russian Foundation for Basic Research (grants 08-02-00018, 09-02-00664 and 12-02-91503-CERN),
the Ministry of Education, Culture, Sports, Science and Tech\-no\-lo\-gy, Japan, Grant-in-Aid for Sci\-en\-ti\-fic Research (grants 18071005, 19034011, 19740162, 20740160 and 20039012),
the German Research Foundation (grant GA\,1480/2-2),
the U. S. Department of Energy,
the EU-funded Marie Curie Outgoing Fellowship,
Grant PIOF-GA-2013-624803,
the Bulgarian Nuclear Regulatory Agency and the Joint Institute for
Nuclear Research, Dubna (bilateral contract No. 4418-1-15\slash 17),
Ministry of Education and Science of the Republic of Serbia (grant OI171002),
Swiss Nationalfonds Foundation 
(grants 206621\_117734 and 20FL20\_154223) 
and ETH Research Grant TH-01\,07-3.

\end{acknowledgements}


\bibliographystyle{spphys} 

\bibliography{submit}


\end{document}